%
%
%
\def\unredoffs{} \def\redoffs{\voffset=-.31truein\hoffset=-.48truein}
\def\speclscape{}
%
%
%
%
%
\newbox\leftpage \newdimen\fullhsize \newdimen\hstitle \newdimen\hsbody
\tolerance=1000\hfuzz=2pt
\catcode`\@=11 
\ifx\hyperdef\UNd@FiNeD\def\hyperdef#1#2#3#4{#4}\def\hyperref#1#2#3#4{#4}\fi
\def\bigans{b }
\def\answ{b }
%
\ifx\answ\bigans\message{(This will come out unreduced.}
\magnification=1200\unredoffs\baselineskip=16pt plus 2pt minus 1pt
\hsbody=\hsize \hstitle=\hsize 
\else\message{(This will be reduced.} \let\l@r=L
\magnification=1000\baselineskip=16pt plus 2pt minus 1pt \vsize=7truein
\redoffs \hstitle=8truein\hsbody=4.75truein\fullhsize=10truein\hsize=\hsbody
\output={\ifnum\pageno=0 
  \shipout\vbox{\speclscape{\hsize\fullhsize\makeheadline}
    \hbox to \fullhsize{\hfill\pagebody\hfill}}\advancepageno
  \else
  \almostshipout{\leftline{\vbox{\pagebody\makefootline}}}\advancepageno
  \fi}
\def\almostshipout#1{\if L\l@r \count1=1 \message{[\the\count0.\the\count1]}
      \global\setbox\leftpage=#1 \global\let\l@r=R
 \else \count1=2
  \shipout\vbox{\speclscape{\hsize\fullhsize\makeheadline}
      \hbox to\fullhsize{\box\leftpage\hfil#1}}  \global\let\l@r=L\fi}
\fi
%
\newcount\yearltd\yearltd=\year\advance\yearltd by -2000

\def\Title#1#2{\nopagenumbers\abstractfont\hsize=\hstitle\rightline{#1}%
\vskip 1in\centerline{\titlefont #2}\abstractfont\vskip .5in\pageno=0}
\def\Date#1{\vfill\leftline{#1}\tenpoint\supereject\global\hsize=\hsbody%
\footline={\hss\tenrm\hyperdef\hypernoname{page}\folio\folio\hss}}%
%

\def\draftmode{\message{ DRAFTMODE }\def\draftdate{{\rm preliminary draft:
\number\month/\number\day/{0}\number\yearltd\ \ \hourmin}}%
\headline={\hfil\draftdate}\writelabels\baselineskip=20pt plus 2pt minus 2pt
 {\count255=\time\divide\count255 by 60 \xdef\hourmin{\number\count255}
  \multiply\count255 by-60\advance\count255 by\time
  \xdef\hourmin{\hourmin:\ifnum\count255<10 0\fi\the\count255}}}
\def\nolabels{\def\wrlabeL##1{}\def\eqlabeL##1{}\def\reflabeL##1{}}
\def\writelabels{\def\wrlabeL##1{\leavevmode\vadjust{\rlap{\smash%
{\line{{\escapechar=` \hfill\rlap{\sevenrm\hskip.03in\string##1}}}}}}}%
\def\eqlabeL##1{{\escapechar-1\rlap{\sevenrm\hskip.05in\string##1}}}%
\def\reflabeL##1{\noexpand\llap{\noexpand\sevenrm\string\string\string##1}}}
\nolabels
%
\global\newcount\secno \global\secno=0
\global\newcount\meqno \global\meqno=1
\def\s@csym{}
\def\newsec#1{\global\advance\secno by1%
{\toks0{#1}\message{(\the\secno. \the\toks0)}}%
\global\subsecno=0\eqnres@t\let\s@csym\secsym\xdef\secn@m{\the\secno}\noindent
{\bf\hyperdef\hypernoname{section}{\the\secno}{\the\secno.} #1}%
\writetoca{{\string\hyperref{}{section}{\the\secno}{\it\the\secno.}} {{\it #1} }}%
\par\nobreak\medskip\nobreak}
\def\eqnres@t{\xdef\secsym{\the\secno.}\global\meqno=1\bigbreak\bigskip}
\def\sequentialequations{\def\eqnres@t{\bigbreak}}\xdef\secsym{}
\global\newcount\subsecno \global\subsecno=0
\def\subsec#1{\global\advance\subsecno by1%
{\toks0{#1}\message{(\s@csym\the\subsecno. \the\toks0)}}%
\ifnum\lastpenalty>9000\else\bigbreak\fi       \global\subsubsecno=0
\noindent{\it\hyperdef\hypernoname{subsection}{\secn@m.\the\subsecno}%
{\secn@m.\the\subsecno.} #1}\writetoca{\string\quad
{\string\hyperref{}{subsection}{\secn@m.\the\subsecno}{\secn@m.\the\subsecno.}}
{#1}}\par\nobreak\medskip\nobreak}
\def\appendix#1#2{\global\meqno=1\global\subsecno=0\xdef\secsym{\hbox{#1.}}%
\bigbreak\bigskip\noindent{\bf Appendix \hyperdef\hypernoname{appendix}{#1}%
{#1.} #2}{\toks0{(#1. #2)}\message{\the\toks0}}%
\xdef\s@csym{#1.}\xdef\secn@m{#1}%
\writetoca{\string\hyperref{}{appendix}{#1}{{\it Appendix} {\it #1.}} {\it #2}}%
\par\nobreak\medskip\nobreak}
%
%
\def\checkm@de#1#2{\ifmmode{\def\f@rst##1{##1}\hyperdef\hypernoname{equation}%
{#1}{#2}}\else\hyperref{}{equation}{#1}{#2}\fi}
\def\eqnn#1{\DefWarn#1\xdef #1{(\noexpand\relax\noexpand\checkm@de%
{\s@csym\the\meqno}{\secsym\the\meqno})}%
\wrlabeL#1\writedef{#1\leftbracket#1}\global\advance\meqno by1}
\def\f@rst#1{\c@t#1a\em@ark}\def\c@t#1#2\em@ark{#1}
\def\eqna#1{\DefWarn#1\wrlabeL{#1$\{\}$}%
\xdef #1##1{(\noexpand\relax\noexpand\checkm@de%
{\s@csym\the\meqno\noexpand\f@rst{##1}}{\hbox{$\secsym\the\meqno##1$}})}
\writedef{#1\numbersign1\leftbracket#1{\numbersign1}}\global\advance\meqno by1}
\def\eqn#1#2{\DefWarn#1%
\xdef #1{(\noexpand\hyperref{}{equation}{\s@csym\the\meqno}%
{\secsym\the\meqno})}$$#2\eqno(\hyperdef\hypernoname{equation}%
{\s@csym\the\meqno}{\secsym\the\meqno})\eqlabeL#1$$%
\writedef{#1\leftbracket#1}\global\advance\meqno by1}
\def\xeqn{\expandafter\xe@n}\def\xe@n(#1){#1}
\def\xeqna#1{\expandafter\xe@n#1}
\def\eqns#1{(\e@ns #1{\hbox{}})}
\def\e@ns#1{\ifx\UNd@FiNeD#1\message{eqnlabel \string#1 is undefined.}%
\xdef#1{(?.?)}\fi{\let\hyperref=\relax\xdef\next{#1}}%
\ifx\next\em@rk\def\next{}\else%
\ifx\next#1\xeqn#1\else\def\n@xt{#1}\ifx\n@xt\next#1\else\xeqna#1\fi
\fi\let\next=\e@ns\fi\next}

\def\DefWarn#1{\ifx\UNd@FiNeD#1\else
\immediate\write16{*** WARNING: the label \string#1 is already defined ***}\fi}
%
\newskip\footskip\footskip14pt plus 1pt minus 1pt 
\def\footnotefont{\ninepoint}\def\f@t#1{\footnotefont #1\@foot}
\def\f@@t{\baselineskip\footskip\bgroup\footnotefont\aftergroup\@foot\let\next}
\setbox\strutbox=\hbox{\vrule height9.5pt depth4.5pt width0pt}
\global\newcount\ftno \global\ftno=0
\def\foot{\global\advance\ftno by1\def\foot@rg{\hyperref{}{footnote}%
{\the\ftno}{\the\ftno}\xdef\foot@rg{\noexpand\hyperdef\noexpand\hypernoname%
{footnote}{\the\ftno}{\the\ftno}}}\footnote{$^{\foot@rg}$}}
%
\newwrite\ftfile
\def\footend{\def\foot{\global\advance\ftno by1\chardef\wfile=\ftfile
\hyperref{}{footnote}{\the\ftno}{$^{\the\ftno}$}%
\ifnum\ftno=1\immediate\openout\ftfile=\jobname.fts\fi%
\immediate\write\ftfile{\noexpand\smallskip%
\noexpand\item{\noexpand\hyperdef\noexpand\hypernoname{footnote}
{\the\ftno}{f\the\ftno}:\ }\pctsign}\findarg}%
\def\footatend{\vfill\eject\immediate\closeout\ftfile{\parindent=20pt
\centerline{\bf Footnotes}\nobreak\bigskip\input \jobname.fts }}}
\def\footatend{}
%
%
\global\newcount\refno \global\refno=1
\newwrite\rfile
\def\ref{[\hyperref{}{reference}{\the\refno}{\the\refno}]\nref}
\def\nref#1{\DefWarn#1%
\xdef#1{[\noexpand\hyperref{}{reference}{\the\refno}{\the\refno}]}%
\writedef{#1\leftbracket#1}%
\ifnum\refno=1\immediate\openout\rfile=\jobname.refs\fi
\chardef\wfile=\rfile\immediate\write\rfile{\noexpand\item{[\noexpand\hyperdef%
\noexpand\hypernoname{reference}{\the\refno}{\the\refno}]\ }%
\reflabeL{#1\hskip.31in}\pctsign}\global\advance\refno by1\findarg}
\def\findarg#1#{\begingroup\obeylines\newlinechar=`\^^M\pass@rg}
{\obeylines\gdef\pass@rg#1{\writ@line\relax #1^^M\hbox{}^^M}%
\gdef\writ@line#1^^M{\expandafter\toks0\expandafter{\striprel@x #1}%
\edef\next{\the\toks0}\ifx\next\em@rk\let\next=\endgroup\else\ifx\next\empty%
\else\immediate\write\wfile{\the\toks0}\fi\let\next=\writ@line\fi\next\relax}}
\def\striprel@x#1{} \def\em@rk{\hbox{}}
\def\lref{\begingroup\obeylines\lr@f}
\def\lr@f#1#2{\DefWarn#1\gdef#1{\let#1=\UNd@FiNeD\ref#1{#2}}\endgroup\unskip}

\def\addref#1{\immediate\write\rfile{\noexpand\item{}#1}} 
\def\listrefs{\footatend\vfill\supereject\immediate\closeout\rfile\writestoppt
\baselineskip=\footskip\centerline{{\bf References}}\bigskip{\parindent=20pt%
\frenchspacing\escapechar=` \input \jobname.refs\vfill\eject}\nonfrenchspacing}
\def\startrefs#1{\immediate\openout\rfile=\jobname.refs\refno=#1}
\def\xref{\expandafter\xr@f}\def\xr@f[#1]{#1}
\def\refs#1{\count255=1[\r@fs #1{\hbox{}}]}
\def\r@fs#1{\ifx\UNd@FiNeD#1\message{reflabel \string#1 is undefined.}%
\nref#1{need to supply reference \string#1.}\fi%
\vphantom{\hphantom{#1}}{\let\hyperref=\relax\xdef\next{#1}}%
\ifx\next\em@rk\def\next{}%
\else\ifx\next#1\ifodd\count255\relax\xref#1\count255=0\fi%
\else#1\count255=1\fi\let\next=\r@fs\fi\next}
%

%
\newwrite\ffile\global\newcount\figno \global\figno=1
\def\fig{fig.~\hyperref{}{figure}{\the\figno}{\the\figno}\nfig}
\def\nfig#1{\DefWarn#1%
\xdef#1{fig.~\noexpand\hyperref{}{figure}{\the\figno}{\the\figno}}%
\writedef{#1\leftbracket fig.\noexpand~\xfig#1}%
\ifnum\figno=1\immediate\openout\ffile=\jobname.figs\fi\chardef\wfile=\ffile%
{\let\hyperref=\relax
\immediate\write\ffile{\noexpand\medskip\noexpand\item{Fig.\ %
\noexpand\hyperdef\noexpand\hypernoname{figure}{\the\figno}{\the\figno}. }
\reflabeL{#1\hskip.55in}\pctsign}}\global\advance\figno by1\findarg}
\def\listfigs{\vfill\eject\immediate\closeout\ffile{\parindent40pt
\baselineskip14pt\centerline{{\bf Figure Captions}}\nobreak\medskip
\escapechar=` \input \jobname.figs\vfill\eject}}
\def\xfig{\expandafter\xf@g}\def\xf@g fig.\penalty\@M\ {}
\def\figs#1{figs.~\f@gs #1{\hbox{}}}
\def\f@gs#1{{\let\hyperref=\relax\xdef\next{#1}}\ifx\next\em@rk\def\next{}\else
\ifx\next#1\xfig #1\else#1\fi\let\next=\f@gs\fi\next}
\def\figin{\epsfcheck\figin}\def\figins{\epsfcheck\figins}
\def\epsfcheck{\ifx\epsfbox\UNd@FiNeD
\message{(NO epsf.tex, FIGURES WILL BE IGNORED)}
\gdef\figin##1{\vskip2in}\gdef\figins##1{\hskip.5in}
\else\message{(FIGURES WILL BE INCLUDED)}%
\gdef\figin##1{##1}\gdef\figins##1{##1}\fi}
\def\DefWarn#1{}
\def\figinsert{\goodbreak\midinsert}
\def\ifig#1#2#3{\DefWarn#1\xdef#1{Fig.~\noexpand\hyperref{}{figure}%
{\the\figno}{\the\figno}}\writedef{#1\leftbracket fig.\noexpand~\xfig#1}%
\figinsert\figin{\centerline{#3}}\medskip\centerline{\vbox{\baselineskip12pt
\advance\hsize by -1truein\noindent\wrlabeL{#1=#1}\footnotefont%
{\bf Fig.~\hyperdef\hypernoname{figure}{\the\figno}{\the\figno}:} #2}}
\bigskip\endinsert\global\advance\figno by1}
\newwrite\lfile
{\escapechar-1\xdef\pctsign{\string\%}\xdef\leftbracket{\string\{}
\xdef\rightbracket{\string\}}\xdef\numbersign{\string\#}}
\def\writedefs{\immediate\openout\lfile=\jobname.defs \def\writedef##1{%
{\let\hyperref=\relax\let\hyperdef=\relax\let\hypernoname=\relax
 \immediate\write\lfile{\string\def\string##1\rightbracket}}}}%
\def\writestop{\def\writestoppt{\immediate\write\lfile{\string\pageno
 \the\pageno\string\startrefs\leftbracket\the\refno\rightbracket
 \string\def\string\secsym\leftbracket\secsym\rightbracket
 \string\secno\the\secno\string\meqno\the\meqno}\immediate\closeout\lfile}}
\def\writestoppt{}\def\writedef#1{}
\def\seclab#1{\DefWarn#1%
\xdef #1{\noexpand\hyperref{}{section}{\the\secno}{\the\secno}}%
\writedef{#1\leftbracket#1}\wrlabeL{#1=#1}}
\def\subseclab#1{\DefWarn#1%
\xdef #1{\noexpand\hyperref{}{subsection}{\secn@m.\the\subsecno}%
{\secn@m.\the\subsecno}}\writedef{#1\leftbracket#1}\wrlabeL{#1=#1}}
\def\applab#1{\DefWarn#1%
\xdef #1{\noexpand\hyperref{}{appendix}{\secn@m}{\secn@m}}%
\writedef{#1\leftbracket#1}\wrlabeL{#1=#1}}
\newwrite\tfile \def\writetoca#1{}
\def\leaderfill{\leaders\hbox to 1em{\hss.\hss}\hfill}
\def\writetoc{\immediate\openout\tfile=\jobname.toc
   \def\writetoca##1{{\edef\next{\write\tfile{\noindent ##1
   \string\leaderfill {\string\hyperref{}{page}{\noexpand\number\pageno}%
                       {\noexpand\number\pageno}} \par}}\next}}}
\newread\ch@ckfile
\def\listtoc{\immediate\closeout\tfile\immediate\openin\ch@ckfile=\jobname.toc
\ifeof\ch@ckfile\message{no file \jobname.toc, no table of contents this pass}%
\else\closein\ch@ckfile\centerline{\bf Contents}\nobreak\medskip%
{\baselineskip=18.5pt  \footnotefont
\parskip=2pt\catcode`\@=12\input\jobname.toc
\catcode`\@=12\bigbreak\bigskip}\fi}
\catcode`\@=12 
%
\edef\tfontsize{\ifx\answ\bigans scaled\magstep3\else scaled\magstep4\fi}
\font\titlerm=cmr10 \tfontsize \font\titlerms=cmr7 \tfontsize
\font\titlermss=cmr5 \tfontsize \font\titlei=cmmi10 \tfontsize
\font\titleis=cmmi7 \tfontsize \font\titleiss=cmmi5 \tfontsize
\font\titlesy=cmsy10 \tfontsize \font\titlesys=cmsy7 \tfontsize
\font\titlesyss=cmsy5 \tfontsize \font\titleit=cmti10 \tfontsize
\skewchar\titlei='177 \skewchar\titleis='177 \skewchar\titleiss='177
\skewchar\titlesy='60 \skewchar\titlesys='60 \skewchar\titlesyss='60
\def\titlefont{\def\rm{\fam0\titlerm}
\textfont0=\titlerm \scriptfont0=\titlerms \scriptscriptfont0=\titlermss
\textfont1=\titlei \scriptfont1=\titleis \scriptscriptfont1=\titleiss
\textfont2=\titlesy \scriptfont2=\titlesys \scriptscriptfont2=\titlesyss
\textfont\itfam=\titleit \def\it{\fam\itfam\titleit}\rm}
 \ifx\answ\bigans\else scaled\magstep1\fi
\ifx\answ\bigans\def\abstractfont{\tenpoint}\else
\font\absit=cmti10 scaled \magstep1
\font\abssl=cmsl10 scaled \magstep1
\font\absrm=cmr10 scaled\magstep1 \font\absrms=cmr7 scaled\magstep1
\font\absrmss=cmr5 scaled\magstep1 \font\absi=cmmi10 scaled\magstep1
\font\absis=cmmi7 scaled\magstep1 \font\absiss=cmmi5 scaled\magstep1
\font\abssy=cmsy10 scaled\magstep1 \font\abssys=cmsy7 scaled\magstep1
\font\abssyss=cmsy5 scaled\magstep1 \font\absbf=cmbx10 scaled\magstep1
\skewchar\absi='177 \skewchar\absis='177 \skewchar\absiss='177
\skewchar\abssy='60 \skewchar\abssys='60 \skewchar\abssyss='60
\def\abstractfont{\def\rm{\fam0\absrm}
\textfont0=\absrm \scriptfont0=\absrms \scriptscriptfont0=\absrmss
\textfont1=\absi \scriptfont1=\absis \scriptscriptfont1=\absiss
\textfont2=\abssy \scriptfont2=\abssys \scriptscriptfont2=\abssyss
\textfont\itfam=\absit \def\it{\fam\itfam\absit}\def\footnotefont{\tenpoint}%
\textfont\slfam=\abssl \def\sl{\fam\slfam\abssl}%
\textfont\bffam=\absbf \def\bf{\fam\bffam\absbf}\rm}\fi
\def\tenpoint{\def\rm{\fam0\tenrm}
\textfont0=\tenrm \scriptfont0=\sevenrm \scriptscriptfont0=\fiverm
\textfont1=\teni  \scriptfont1=\seveni  \scriptscriptfont1=\fivei
\textfont2=\tensy \scriptfont2=\sevensy \scriptscriptfont2=\fivesy
\textfont\itfam=\tenit \def\it{\fam\itfam\tenit}\def\footnotefont{\ninepoint}%
\textfont\bffam=\tenbf \def\bf{\fam\bffam\tenbf}\def\sl{\fam\slfam\tensl}\rm}
\font\ninerm=cmr9 \font\sixrm=cmr6 \font\ninei=cmmi9 \font\sixi=cmmi6
\font\ninesy=cmsy9 \font\sixsy=cmsy6 \font\ninebf=cmbx9
\font\nineit=cmti9 \font\ninesl=cmsl9 \skewchar\ninei='177
\skewchar\sixi='177 \skewchar\ninesy='60 \skewchar\sixsy='60
\def\ninepoint{\def\rm{\fam0\ninerm}
\textfont0=\ninerm \scriptfont0=\sixrm \scriptscriptfont0=\fiverm
\textfont1=\ninei \scriptfont1=\sixi \scriptscriptfont1=\fivei
\textfont2=\ninesy \scriptfont2=\sixsy \scriptscriptfont2=\fivesy
\textfont\itfam=\ninei \def\it{\fam\itfam\nineit}\def\sl{\fam\slfam\ninesl}%
\textfont\bffam=\ninebf \def\bf{\fam\bffam\ninebf}\rm}
%
%
\def\noblackbox{\overfullrule=0pt}
\hyphenation{anom-aly anom-alies coun-ter-term coun-ter-terms}
\def\inv{^{\raise.15ex\hbox{${\scriptscriptstyle -}$}\kern-.05em 1}}

\def\Dsl{\,\raise.15ex\hbox{/}\mkern-13.5mu D} 
\def\dsl{\raise.15ex\hbox{/}\kern-.57em\partial}

\def\lspace{\ifx\answ\bigans{}\else\qquad\fi}
\def\lbspace{\ifx\answ\bigans{}\else\hskip-.2in\fi} 
\def\boxeqn#1{\vcenter{\vbox{\hrule\hbox{\vrule\kern3pt\vbox{\kern3pt
	\hbox{${\displaystyle #1}$}\kern3pt}\kern3pt\vrule}\hrule}}}
\def\mbox#1#2{\vcenter{\hrule \hbox{\vrule height#2in
		\kern#1in \vrule} \hrule}}  
%

\def\vev#1{\langle #1 \rangle}

\def\darr#1{\raise1.5ex\hbox{$\leftrightarrow$}\mkern-16.5mu #1}

\def\roughly#1{\raise.3ex\hbox{$#1$\kern-.75em\lower1ex\hbox{$\sim$}}}

\global\newcount\subsubsecno \global\subsubsecno=0
\def\subsubsec#1{\global\advance\subsubsecno by1%
{\toks0{#1}\message{(\the\secno\the\subsecno\the\subsubsecno. \the\toks0)}}%
\ifnum\lastpenalty>9000\else\bigbreak\fi
\noindent{\it\hyperdef\hypernoname{subsubsection}{\the\secno.\the\subsecno\the\subsubsecno}%
{\the\secno.\the\subsecno.\the\subsubsecno.} #1}
\par\nobreak\medskip\nobreak}
\def\boxit#1{\vbox{\hrule\hbox{\vrule\kern8pt
\vbox{\hbox{\kern8pt}\hbox{\vbox{#1}}\hbox{\kern8pt}}
\kern8pt\vrule}\hrule}}
\def\mathboxit#1{\vbox{\hrule\hbox{\vrule\kern8pt\vbox{\kern8pt
\hbox{$\displaystyle #1$}\kern8pt}\kern8pt\vrule}\hrule}}
\def\slashchar#1{\setbox0=\hbox{$#1$}           
   \dimen0=\wd0                                 
   \setbox1=\hbox{/} \dimen1=\wd1               
   \ifdim\dimen0>\dimen1                        
      \rlap{\hbox to \dimen0{\hfil/\hfil}}      
      #1                                        
   \else                                        
      \rlap{\hbox to \dimen1{\hfil$#1$\hfil}}   
      /                                         
   \fi}
\def\sqr#1#2{{\vcenter{\vbox{\hrule height.#2pt
         \hbox{\vrule width.#2pt height#1pt \kern#1pt
            \vrule width.#2pt}
         \hrule height.#2pt}}}}



\noblackbox
\baselineskip=14.5pt
\def\crr{\noalign{\vskip5pt}}

\def\comment#1{{}}
\def\ss#1{{\scriptstyle{#1}}}

\def\z{{\zeta}}
\def\ap{\alpha'}

\def\cf{{\it cf.\ }}
\def\ie{{\it i.e.\ }}
\def\eg{{\it e.g.\ }}
\def\eqq{{\it Eq.\ }}
\def\eqqs{{\it Eqs.\ }}

\def\eps{\epsilon}

\def\de{\delta}
\def\si{\sigma}\def\Si{{\Sigma}}

\def\Hc{{\cal H}}
\def\Uc{{\cal U}}
\def\Qc{{\cal Q}}

\newif\ifnref

\nreffalse

\input epsf

\def\figin{\epsfcheck\figin}\def\figins{\epsfcheck\figins}
\def\epsfcheck{\ifx\epsfbox\UnDeFiNeD
\message{(NO epsf.tex, FIGURES WILL BE IGNORED)}
\gdef\figin##1{\vskip2in}\gdef\figins##1{\hskip.5in}
\else\message{(FIGURES WILL BE INCLUDED)}%
\gdef\figin##1{##1}\gdef\figins##1{##1}\fi}
\def\DefWarn#1{}
\def\figinsert{\goodbreak\midinsert}  
\def\ifig#1#2#3{\DefWarn#1\xdef#1{Fig.~\the\figno}
\writedef{#1\leftbracket fig.\noexpand~\the\figno}%
\figinsert\figin{\centerline{#3}}\medskip\centerline{\vbox{\baselineskip12pt
\advance\hsize by -1truein\noindent\footnotefont\centerline{{\bf
Fig.~\the\figno}\ \sl #2}}}
\bigskip\endinsert\global\advance\figno by1}

\def\iifig#1#2#3#4{\DefWarn#1\xdef#1{Fig.~\the\figno}
\writedef{#1\leftbracket fig.\noexpand~\the\figno}%
\figinsert\figin{\centerline{#4}}\medskip\centerline{\vbox{\baselineskip12pt
\advance\hsize by -1truein\noindent\footnotefont\centerline{{\bf
Fig.~\the\figno}\ \ \sl #2}}}\smallskip\centerline{\vbox{\baselineskip12pt
\advance\hsize by -1truein\noindent\footnotefont\centerline{\ \ \ \sl #3}}}
\bigskip\endinsert\global\advance\figno by1}


\def\tilde{\widetilde}

\def\h {{1\over 2}}

\def\ov {\overline}
\def\o {\over}
\def\fc#1#2{{#1 \o #2}}

\def\IZ{ {\bf Z}}\def\IC{{\bf C}}\def\IN{ {\bf N}}
\def\IQ{ {\bf Q}}


\def\br{\hfill\break}

\def\lf {\left}
\def\ri {\right}
\def\ra {\rightarrow}
\def\lra {\longrightarrow}

\def\p {\partial}

\def\Zc {{\cal Z}}
 
 \def\Oc {{\cal O}}
\def\Lc {{\cal L}} 
\def\Mc {{\cal M}} \def\Ac {{\cal A}}
 
 \def\Uc {{\cal U}}

\def\shuffle{{\hskip0.10cm \vrule height 0pt width 8pt depth 0.75pt  \hskip-0.3cm\ss{\rm III}\hskip0.05cm}}
\lref\SpradlinWP{
  M.~Spradlin and A.~Volovich,
``Symbols of One-Loop Integrals From Mixed Tate Motives,''
JHEP {\bf 1111}, 084 (2011).
[arXiv:1105.2024 [hep-th]].
}

\lref\DT{
J.~Fleischer, A.V.~Kotikov and O.L.~Veretin,
``Applications of the large mass expansion,''
Acta Phys.\ Polon.\ B {\bf 29}, 2611 (1998).
[hep-ph/9808243];
``Analytic two loop results for selfenergy type and vertex type diagrams with one nonzero mass,''
Nucl.\ Phys.\ B {\bf 547}, 343 (1999).
[hep-ph/9808242];\br
A.V.~Kotikov, L.N.~Lipatov, A.I.~Onishchenko and V.N.~Velizhanin,
``Three loop universal anomalous dimension of the Wilson operators in N=4 SUSY Yang-Mills model,''
Phys.\ Lett.\ B {\bf 595}, 521 (2004), [Erratum-ibid.\ B {\bf 632}, 754 (2006)].
[hep-th/0404092].
}

\lref\Zagier{
D. Zagier, ``Values of zeta functions and their applications,'' in First European Congress of Mathematics (Paris, 1992), Vol. II, A. Joseph et. al. (eds.), Birkh{\"{a}}user, Basel, 1994, pp. 497-512.}

\lref\Brown{
  F.~Brown,
``On the decomposition of motivic multiple zeta values,''
[arXiv:1102.1310 [math.NT]], to appear in `Galois-Teichm\"uller theory and Arithmetic Geometry', Advanced Studies in Pure Mathematics.
}

\lref\Goncharov{
A.B. Goncharov, 
``Galois symmetries of fundamental groupoids and noncommutative geometry,''
 Duke Math. J. 128 (2005) 209-284. [arXiv:math/0208144v4 [math.AG]].
}

\lref\goncharov{
A.B. Goncharov,
``Multiple polylogarithms and mixed Tate motives,''
	[arXiv:math/ 0103059v4 [math.AG]].
}

\lref\goncharovi{
A.B. Goncharov,
``Multiple polylogarithms, cyclotomy and modular complexes,''
Math. Res. Letters 5, (1998) 497-516  [arXiv:1105.2076v1 [math.AG]]
}

\lref\TATE{
F.~Brown, ``Mixed Tate Motives over $\IZ$,'' Ann. Math. 175 (2012) 949--976.}

\lref\MSSi{C.R.~Mafra, O.~Schlotterer and S.~Stieberger, 
 ``Complete N-Point Superstring Disk Amplitude I. Pure Spinor Computation,''
[arXiv:1106.2645 [hep-th]].
}

\lref\MSSii{C.R.~Mafra, O.~Schlotterer and S.~Stieberger,
``Complete N-Point Superstring Disk Amplitude II. Amplitude and Hypergeometric Function Structure,''
[arXiv:1106.2646 [hep-th]].
}

\lref\GRAV{
  S.~Stieberger,
 ``Constraints on Tree-Level Higher Order Gravitational Couplings in Superstring Theory,''
Phys.\ Rev.\ Lett.\  {\bf 106}, 111601 (2011).
[arXiv:0910.0180 [hep-th]].
}

\lref\StiebergerTE{
  S.~Stieberger and T.R.~Taylor,
``Multi-Gluon Scattering in Open Superstring Theory,''
Phys.\ Rev.\ D {\bf 74}, 126007 (2006).
[hep-th/0609175].
}

\lref\DataMine{
J.~Bl\"umlein, D.J.~Broadhurst and J.A.M.~Vermaseren,
``The Multiple Zeta Value Data Mine,''
Comput.\ Phys.\ Commun.\  {\bf 181}, 582 (2010).
[arXiv:0907.2557 [math-ph]].
}

\lref\Broadi{
  D.J.~Broadhurst,
``On the enumeration of irreducible k fold Euler sums and their roles in knot theory and field theory,''
[hep-th/9604128].
}

\lref\Broadii{
  D.J.~Broadhurst and D.~Kreimer,
``Association of multiple zeta values with positive knots via Feynman diagrams up to 9 loops,''
Phys.\ Lett.\ B {\bf 393}, 403 (1997).
[hep-th/9609128].
}

\lref\Harmpol{E.~Remiddi and J.A.M.~Vermaseren,
``Harmonic polylogarithms,''
Int.\ J.\ Mod.\ Phys.\ A {\bf 15}, 725 (2000).
[hep-ph/9905237].
}

\lref\HuberYG{
  T.~Huber and D.~Ma\^{\i}tre,
``HypExp: A Mathematica package for expanding hypergeometric functions around integer-valued parameters,''
Comput.\ Phys.\ Commun.\  {\bf 175}, 122 (2006).
[hep-ph/0507094].
``HypExp 2, Expanding Hypergeometric Functions about Half-Integer Parameters,''
Comput.\ Phys.\ Commun.\  {\bf 178}, 755 (2008).
[arXiv:0708.2443 [hep-ph]].
}

\lref\MochZR{
 S.~Moch, P.~Uwer and S.~Weinzierl,
``Nested sums, expansion of transcendental functions and multiscale multiloop integrals,''
J.\ Math.\ Phys.\  {\bf 43}, 3363 (2002).
[hep-ph/0110083].
}

\lref\MochUC{
  S.~Moch and P.~Uwer,
``XSummer: Transcendental functions and symbolic summation in form,''
Comput.\ Phys.\ Commun.\  {\bf 174}, 759 (2006).
[math-ph/0508008].
}

\lref\BSS{
J. Br\"odel, O.~Schlotterer and S.~Stieberger, 
 ``Structure of Open and Closed Superstring Amplitudes:
Motivic Multiple Zeta Values and $\ap$--Expansions,''  to appear.}

\lref\KawaiXQ{
  H.~Kawai, D.C.~Lewellen and S.H.H.~Tye,
``A Relation Between Tree Amplitudes Of Closed And Open Strings,''
  Nucl.\ Phys.\  B {\bf 269}, 1 (1986).
}

\lref\BCJ{
  Z.~Bern, J.J.M.~Carrasco and H.~Johansson,
  ``New Relations for Gauge-Theory Amplitudes,''
  Phys.\ Rev.\  D {\bf 78}, 085011 (2008)
  [arXiv:0805.3993 [hep-ph]].
}

\lref\BjerrumBohrRD{
  N.E.J.~Bjerrum-Bohr, P.H.~Damgaard and P.~Vanhove,
  ``Minimal Basis for Gauge Theory Amplitudes,''
  Phys.\ Rev.\ Lett.\  {\bf 103}, 161602 (2009)
  [arXiv:0907.1425 [hep-th]].
}

\lref\StiebergerHQ{
  S.~Stieberger,
  ``Open \& Closed vs. Pure Open String Disk Amplitudes,''
  arXiv:0907.2211 [hep-th].
}

\lref\BernFY{
Z.~Bern, J.J.M.~Carrasco, H.~Johansson,
``The Structure of Multiloop Amplitudes in Gauge and Gravity Theories,''
Nucl.\ Phys.\ Proc.\ Suppl.\  {\bf 205-206}, 54-60 (2010).
[arXiv:1007.4297 [hep-th]];\br
Z.~Bern, J.J.M. Carrasco, L.J.~Dixon, H.~Johansson, R.~Roiban,
 ``Amplitudes and Ultraviolet Behavior of N=8 Supergravity,''
[arXiv:1103.1848 [hep-th]];\br
L.J.~Dixon,
``Scattering amplitudes: the most perfect microscopic structures in the universe,''
J.\ Phys.\ A {\bf 44}, 454001 (2011).
[arXiv:1105.0771 [hep-th]].
}

\lref\AMHV{
S.~Stieberger and T.R.~Taylor,
 ``Maximally Helicity Violating Disk Amplitudes, Twistors and Transcendental Integrals,''
[arXiv:1204.3848 [hep-th]].
}

\lref\Dan{
D.~Oprisa and S.~Stieberger,
``Six gluon open superstring disk amplitude, multiple hypergeometric series and Euler-Zagier sums,''
[hep-th/0509042].
}

\lref\Deligne{
P. Deligne and  A. Goncharov, 
``Groupe fondamentaux motivique de Tate mixte,'' 
Ann. Sci. Ecole Norm. Sup. (4) 38 (2005), 1Ð56.
}

\lref\Smirnov{
V.A. Smirnov, 
``Evaluating Feynman Integrals,''
Springer Berlin Heidelberg, November 2010.}

\lref\GSVV{
A.B.~Goncharov, M.~Spradlin, C.~Vergu and A.~Volovich,
 ``Classical Polylogarithms for Amplitudes and Wilson Loops,''
Phys.\ Rev.\ Lett.\  {\bf 105}, 151605 (2010).
[arXiv:1006.5703 [hep-th]].
}

\lref\Claude{
C.~Duhr,``Hopf algebras, coproducts and symbols: an application to Higgs boson amplitudes,''
JHEP {\bf 1208}, 043 (2012).
[arXiv:1203.0454 [hep-ph]].
}

\lref\Symbols{
A.B. Goncharov,
``A simple construction of Grassmannian polylogarithms,'' arXiv: 0908.2238v2 [math.AG].
}

\lref\Browni{
F. Brown, 
``Multiple zeta values and periods of moduli spaces $\Mc_{0,n}$,'' arXiv: math/0606419v1 [math.AG].
}

\lref\Brownii{
F. Brown, 
``P\'eriodes des espaces des modules $\Mc_{0,n}$ et valeurs z\^etas multiples'', 
C.R. Acad. Sci. Paris, Ser. I  342 (2006),  949-954.}

\lref\FB{F. Brown, private communication.}

\lref\ABG{
A.B. Goncharov,
``Multiple zeta-values, Galois groups, and geometry of modular varieties'',
	arXiv:math/0005069v2 [math.AG].}

\lref\Schneps{
F. Brown, S. Carr, and L. Schneps,
``The algebra of cell--zeta values'', 
Compos. Math. Vol. 146 Iss. 3 (2010) 731, [arXiv:0910.0122 [math.NT]].
}


\Title{\vbox{\rightline{AEI--2012--039}
\rightline{MPP--2012--85}
}}
{\vbox{\centerline{Motivic Multiple Zeta Values and Superstring Amplitudes}}}
\medskip
\centerline{O. Schlotterer$^a$ and  S. Stieberger$^b$}
\bigskip
\centerline{\it $^a$ Max--Planck--Institut f\"ur Gravitationsphysik} 
\centerline{\it Albert--Einstein--Institut, 14476 Potsdam, Germany}
\bigskip\medskip
\centerline{\it $^b$Max--Planck--Institut f\"ur Physik}
\centerline{\it Werner--Heisenberg--Institut, 80805 M\"unchen, Germany}

\vskip15pt

\medskip
\bigskip\bigskip\bigskip
\centerline{\bf Abstract}
\vskip .2in
\noindent

The structure of tree--level  open and closed superstring amplitudes is analyzed.
For the open superstring amplitude we find a striking and elegant form, which 
allows to disentangle
its $\ap$--expansion into several contributions accounting for different classes 
of multiple zeta values.
This form is bolstered by the decomposition of motivic 
multiple zeta values, i.e. the latter  encapsulate the $\ap$--expansion
of the superstring amplitude. 
Moreover,  a morphism induced by the coproduct maps the $\ap$--expansion onto a 
non--commutative Hopf algebra. This map  represents a generalization of the symbol of a transcendental function.
In terms of elements of this Hopf algebra the $\ap$--expansion assumes a very simple and symmetric form, which  carries all the relevant information.
Equipped with these results we can also cast the closed superstring amplitude into 
a very elegant form.

\Date{}
\noindent
\goodbreak
\listtoc 
\writetoc
\break
\newsec{Introduction}

One important question in quantum field theory is finding a simple principle to easily compute physical quantities such as Feynman integrals describing higher--order quantum
corrections. Analytic results for Feynman integrals are encoded by transcendental functions 
such as multiple polylogarithms or elliptic functions \Smirnov.
These functions, which depend on the kinematic invariants, have a rich algebraic structure and obey a variety of different classes of relations among each other. Although these equations may allow to 
obtain a short and simple answer in practice it is not straightforward how to concretely 
apply and disentangle these relations to arrive at this simple answer. 
Hence, a guiding principle to get a grip on these relations is important.

A recent step towards an implicit application of these relations, which also leads to quite remarkable simplifications \GSVV, is the concept
of the symbol of a transcendental function, which maps the combinatorics of relations among different multiple 
polylogarithms to the combinatorics of a tensor algebra  \Symbols. All 
the functional identities between the polylogarithms are mapped to simple algebraic relations in the tensor algebra over the group of rational functions.
A generalization of the symbol approach is the coproduct structure of multiple polylogarithms \refs{\Goncharov,\Brown}.
The advantage of the coproduct structure is, that it also keeps track of 
multiple zeta values (MZVs) in contrast to the symbol $S$, for which we have $S(\pi), S(\zeta)=0$. Recently, in Ref.  \Claude\  the coproduct structure has been applied for a concrete physical amplitude.

The properties of scattering amplitudes in both gauge and gravity theories 
suggest a deeper understanding from string theory, \cf Ref.~\BernFY\ 
for a recent review.
Many field theory objects and relations such as 
Bern--Carrasco--Johansson (BCJ) \BCJ\ or Kawai--Lewellen--Tye (KLT) \KawaiXQ\ 
relations can be easily derived from and understood in 
string theory  by tracing these identities back to the monodromy 
properties of the string world--sheet \refs{\StiebergerHQ,\BjerrumBohrRD}.
In this context we also like to mention the question of transcendentality of a Feynman 
integral \DT, which can be related to superstring tree--level amplitudes given by generalized Euler integrals \AMHV.
Moreover, the concept of symbols 
and coproduct structure for Feynman integrals might have  a natural appearance in string theory. In fact, in this work we shall demonstrate, that the aforementioned coproduct structure allows to cast the $\ap$--expansion of the tree--level open and closed superstring amplitude into a short and symmetric form.
 
Generically, the string amplitudes are given by integrals over vertex operator positions
on the Riemann surface describing the interacting string world--sheet.
At higher loops there is also an integral over the moduli space of this manifold.
At tree--level  such integrals over positions boil down to generalized 
Euler integrals  \Dan. 
Expanding the latter w.r.t. to powers in the string tension $\ap$ yields higher--order string corrections to Yang--Mills (YM) theory. 
Their expansion coefficients are given by MZVs multiplying some polynomials in the kinematic invariants: at each order  in $\ap$ only a set of MZVs of a fixed  transcendentality degree (transcendentality level \DT) appears. 
In practice  extracting these orders 
from the integrals \refs{\Dan,\StiebergerTE}, which boils down to computing  generalized Euler--Zagier sums, 
is both cumbersome and provides quite complicated expressions: the appearance of various MZVs of different depth seems to lack any sorted structure.
Furthermore, there is no selection principle to choose the right basis of MZVs
in the $\ap$--expansion.
Just as  computing amplitudes in field theory a lot of their simplicity 
and symmetry structure is lost by using not the most appropriate approach.
In other words, though the final result may have a simple structure, the actual computation might not be able to reproduce this simplicity and yield a difficult answer.

In fact, by passing from the MZVs  to their motivic 
versions \refs{\Goncharov,\Brown} 
and then mapping the latter to elements  of a Hopf algebra 
endows the superstring amplitude with its motivic structure. More precisely,
the isomorphism $\phi$, which is induced by the coproduct, maps the $\ap$--expansion of the  open superstring amplitude $\Ac$ into the very short and intriguing form 
in terms of elements $f_i$ of a non--commutative Hopf algebra:
\eqn\summ{
\Ac\buildrel \phi \over \lra\lf
( \sum_{k=0}^{\infty} f_2^k\ P_{2k} \ri) \ \lf\{\  \sum_{p=0}^{\infty} 
\ \sum_{ i_1,\ldots, i_p \atop\in 2 \IN^+ + 1}
f_{i_1} f_{i_2}\ldots f_{i_p}\ M_{i_p} \ldots M_{i_2} M_{i_1}\  \ri\}\  A\ .}
In Eq. \summ\ the vector $A$ encompasses a basis of YM subamplitudes, the matrices $P_{2k}$ and 
$M_{2n+1}$ encode polynomials of degree $2k$ and $2n+1$, respectively in $\ap$ and the kinematic invariants.
As the vector $A$ the string amplitude $\Ac$ represents a vector of the 
same dimension, \cf section 3 for further notational details.
All the relevant information of the 
$\ap$--expansion of the open superstring amplitude is encapsulated in \summ\ 
without further specifying the latter explicitly in terms of MZVs. 
This way all relations between MZVs are automatically built in as simple algebraic 
relations following from the coalgebra structure. 
Furthermore, the result is independent on any particular selection of a basis of MZVs.
Finally, in contrast to the symbol the map $\phi$, which is invertible, does not lose any information on the amplitude.

The organization of the present  work is as follows. In section 2 we review those aspects
of MZVs, which will be needed in the sequel. In section 3 we present our findings
for the $\ap$--expansion of the $N$--point open superstring amplitude. 
After some short exhibition on the work \Brown\ of F. Brown on motivic MZVs in section 4
we compute the decompositions  of motivic MZVs from weight $11$ until weight $16$ and compare the result with the structure of the open superstring amplitude.
Equipped with these results in section 5 we investigate the motivic structure of
the open superstring amplitude and derive \summ. In section 6 we use our open superstring 
results to also cast the closed string amplitude into a compact form.
In Appendix A we present some more results on the decomposition of motivic MZVs.

\newsec{Aspects of multiple zeta values}

One prime object in both quantum field theory and string theory are 
multiple zeta values (MZVs):
\eqn\MZV{
\zeta_{n_1,\ldots,n_r}:=\zeta(n_1,\ldots,n_r)=
\sum\limits_{0<k_1<\ldots<k_r}\ \prod\limits_{l=1}^r k_l^{-n_l}\ \ \ ,\ \ \ n_l\in\IN^+\ ,\ n_r\geq2\ .}
In this section we review some of their aspects. They can be written as 
special cases~\goncharov
\eqn\mzvs{
\zeta_{n_1,\ldots,n_r}=(-1)^r\ G(\underbrace{0,\ldots,0}_{n_r-1},1\ldots,\underbrace{0,\ldots,0}_{n_1-1},1;1)}
of multiple polylogarithms \refs{\goncharov,\goncharovi}
\eqn\MPOLY{
G(a_1,\ldots,a_n;z)=\int_0^z\fc{dt}{t-a_1}\ G(a_2,\ldots,a_n;t)\ ,}
with $G(z)=1$ and $a_i,z\in\IC$.
In \MZV\ the sum $w=\sum_{l=1}^rn_l$ is called the transcendentality degree or weight of \MZV\  and  $r$ its depth.
The integral representation \mzvs\ is useful to  establish various properties and relations of \MZV.
The set of integral linear combinations of MZVs \MZV\ is a ring, since the product of any two values can be expressed by a
(positive) integer linear combination of the other MZVs \Zagier,  \eg 
\eqn\SHUFFLE{
\zeta_m\ \zeta_n=\zeta_{m,n}+\zeta_{n,m}+\zeta_{m+n}\ .}
This relation is known as quasi--shuffle or stuffle relation.
There are many relations over $\IQ$ among MZVs, \eg 
$\zeta_{1,4}= 2\zeta_5-\zeta_2\zeta_3$. 
We define the (commutative) $\IQ$--algebra $\Zc$ spanned by all MZVs over $\IQ$. The latter is the 
(conjecturally direct) sum over the  
$\IQ$--vector spaces $\Zc_N$ spanned by the set of MZVs \MZV\ of total weight $w=N$, with $n_r\geq2$,
\ie $\Zc=\bigoplus_{k\geq 0}\Zc_k$. 
For a given weight $w\in\IN$ the dimension $\dim_\IQ(\Zc_N)$ of the space $\Zc_N$  is conjecturally given
by $\dim_\IQ(\Zc_N)=d_N$, with $d_N=d_{N-2}+d_{N-3},\ N\geq 3$ and $d_0=1,\ d_1=0,\ d_2=1$ \Zagier. Starting at weight $w=8$ MZVs of depth greater than one $r>1$ appear
in the basis.
By applying stuffle, shuffle, doubling, generalized doubling relations and duality it is possible to reduce the MZVs of a given weight to a minimal set.
Strictly speaking this is explicitly proven only up to weight $26$ \DataMine.  
For $D_{w,r}$ being the number of independent MZVs at weight $w>2$ and depth $r$, which cannot be 
reduced to primitive MZVs of smaller depth and their products,  
it is believed, that $D_{8,2}=1, 
D_{10,2}=1,\ D_{11,3}=1,\ D_{12,2}=1$ and $D_{12,4}=1$~\Broadii. 
For $Z=\fc{\Zc_{>0}}{\Zc_{>0}\Zc_{>0}}$ the graded space of irreducible MZVs we have:
$dim(Z_w)\equiv\sum_r D_{w,r}=1,0,1,0,1,1,1,1,2,2,3,3,4,5$ for $w=3,\ldots,16$, 
respectively \refs{\Broadii,\DataMine}.

The selection of a basis of MZVs can be performed by following some principles. 
For instance a minimal depth representation may be preferable. In addition, one may write as many 
elements of the basis as possible with positive odd indices $n_l$ only. 
However, it is not possible to achieve this for the whole basis, \ie a number of basis 
elements needs at least two even entries \DataMine.
Up to weight $w=16$, one can choose the following basis elements, displayed in 
the following three tables, \cf Tables 1--3.

{\vbox{\ninepoint{$$
\vbox{\offinterlineskip\tabskip=0pt
\halign{\strut\vrule#
&~$#$~\hfil 
&\vrule$#$ 
&~$#$~\hfil 
&\vrule$#$ 
&~$#$~\hfil 
&\vrule$#$ 
&~$#$~\hfil 
&\vrule$#$
&~$#$~\hfil 
&\vrule$#$ 
&~$#$~\hfil 
&\vrule$#$
&~$#$~\hfil 
&\vrule$#$
&~$#$~\hfil 
&\vrule$#$
&~$#$~\hfil 
&\vrule$#$
&~$#$~\hfil 
&\vrule$#$
&~$#$~\hfil 
&~$#$~\hfil 
&\vrule$#$
&~$#$~\hfil 
&~$#$~\hfil 
&\vrule$#$\cr
\noalign{\hrule}
& w &&2 &&3 && 4  && 5 && 6&& 7 && 8 && 9 && 10 && 11& &&12& &\cr
\noalign{\hrule}
&\Zc_w&&\zeta_2  &&\zeta_3  && \zeta_2^2 &&\zeta_5 && \zeta_3^2 &&\zeta_7  &&\zeta_{3,5} &&\zeta_9 &&\zeta_{3,7} && \zeta_{3,3,5}&\zeta_2\  \zeta_3^3 && \zeta_{1,1,4,6}&\zeta_2\ \zeta_{3,7}& \cr
&\ &&\  &&\  && \  &&\zeta_2\ \zeta_3 && \zeta_2 ^3 &&\zeta_2\ \zeta_5  &&\zeta_3\ \zeta_5 &&\zeta_3^3&&\zeta_3\ \zeta_7 &&\zeta_{3,5}\ \zeta_3&\zeta_2\ \zeta_9  
&&\zeta_{3,9}&\zeta_2^2\ \zeta_{3,5}&\cr
&\ &&\  &&\  && \  &&\ &&\ && \zeta_2^2\ \zeta_3  &&\zeta_2\ \zeta_3 ^2&&\zeta_2\ \zeta_7 &&\zeta_5^2&&\zeta_{11}& \zeta_2^2\ \zeta_7  &&\zeta_3\ \zeta_9&
\zeta_2\ \zeta_5^2&\cr
&\ &&\  &&\  && \  &&\ &&\ &&\  &&\zeta_2^4&&\zeta_2^2\ \zeta_5 &&\zeta_2\ \zeta_{3,5} &&\zeta_3^2\ \zeta_5&\zeta_2^3\ \zeta_5   &&\zeta_5\ \zeta_7&\zeta_2\ \zeta_3\ \zeta_7&\cr
&\ &&\  &&\  && \  &&\ &&\ &&\  &&\ &&\zeta_2^3\ \zeta_3 &&\zeta_2\ \zeta_3\ \zeta_5  &&   &\zeta_2^4\ \zeta_3   &&\zeta_3^4&\zeta_2^2\ \zeta_3\ \zeta_5&\cr
&\ &&\  &&\  && \  &&\ &&\ &&\  &&\ &&\ &&\zeta_2^2\ \zeta_3^2 && &  && &\zeta_2^3\ \zeta_3^2 &\cr
&\ &&\  &&\  && \  &&\ &&\ &&\  &&\ &&\ &&\zeta_2^5 && &  && & \zeta_2^6 &\cr
\noalign{\hrule}
&d_w &&1 &&1  &&1 &&2 &&2 &&3  &&4 &&5 &&7 && 9& &&12 & &\cr
\noalign{\hrule}}}$$
\vskip-6pt
\centerline{\noindent{\bf Table 1:}
{\sl Basis elements for $\Zc_w$, with $2\leq w\leq 12$.}}
\vskip10pt}}}

{\vbox{\ninepoint{$$
\vbox{\offinterlineskip\tabskip=0pt
\halign{\strut\vrule#
&~$#$~\hfil 
&\vrule$#$ 
&~$#$~\hfil 
&~$#$~\hfil 
&\vrule$#$ 
&~$#$~\hfil 
&~$#$~\hfil 
&\vrule$#$ 
&~$#$~\hfil 
&~$#$~\hfil 
&~$#$~\hfil
&\vrule$#$\cr
\noalign{\hrule}
& w &&13 & && 14  &  && 15 & &   &\cr
\noalign{\hrule}
&\Zc_w&&\zeta_{3,3,7} & \zeta_2\ \zeta_{3,3,5} &&\zeta_{3,3,3,5} &\zeta_2\ \zeta_{1,1,4,6} 
&& \zeta_{1,1,3,4,6}&\zeta_2\ \zeta_{3,3,7} & \zeta_2^2\ \zeta_{3,3,5}   &\cr
&   &&\zeta_{3,5,5} &\zeta_2\ \zeta_3\ \zeta_{3,5} &&\zeta_{3,11} &\zeta_2\ \zeta_{3,9} && \zeta_{3,3,9}&\zeta_2\ \zeta_{3,5,5} &\zeta_2^2\ \zeta_3\ \zeta_{3,5}  &\cr
&   &&\zeta_{13} &\zeta_2\ \zeta_{11} &&\zeta_{5,9} &\zeta_2\ \zeta_3\ \zeta_9 &&\zeta_{5,3,7} &\zeta_2\ \zeta_{13} &\zeta_2^2\ \zeta_{11}   &\cr
&   &&\zeta_{3,7}\ \zeta_3 &\zeta_2\ \zeta_3^2\ \zeta_5 &&\zeta_{3,3,5}\ \zeta_3 &\zeta_2\ \zeta_5\ \zeta_7 &&\zeta_{15} &  \zeta_2\ \zeta_3\ \zeta_{3,7} &\zeta_2^2\ \zeta_3^2\ \zeta_5  &\cr
&   &&\zeta_{3,5}\ \zeta_5 &\zeta_2^2\ \zeta_3^3 &&\zeta_{3,5}\ \zeta_3^2 &\zeta_2\ \zeta_3^4 &&\zeta_{1,1,4,6}\ \zeta_3 &  \zeta_2\ \zeta_5\  \zeta_{3,5}&\zeta_2^3\ \zeta_3^3   &\cr
&   &&\zeta_3^2\ \zeta_7 &\zeta_2^2\ \zeta_9 &&\zeta_3\ \zeta_{11} &\zeta_2^2\ \zeta_{3,7} &&\zeta_{3,9}\ \zeta_3 & \zeta_2\ \zeta_3^2\ \zeta_7 &\zeta_2^3\ \zeta_9  &\cr
&   &&\zeta_3\ \zeta_5^2 &\zeta_2^3\ \zeta_7 &&\zeta_3^3\ \zeta_5 &\zeta_2^3\ \zeta_{3,5} &&\zeta_9\ \zeta_3^2 & \zeta_2\ \zeta_3\ \zeta_5^2 &\zeta_2^4\ \zeta_7  &\cr
&   && &\zeta_2^4\ \zeta_5 && \zeta_5\ \zeta_9 &\zeta_2^2\ \zeta_5^2 &&\zeta_3\ \zeta_5\ \zeta_7 &  &\zeta_2^5\ \zeta_5 &\cr
&   && &\zeta_2^5\ \zeta_3 && \zeta_7^2&\zeta_2^2\ \zeta_3\ \zeta_7 &&\zeta_3^5 &  &\zeta_2^6\ \zeta_3   & \cr
&   && & && &\zeta_2^3\ \zeta_3\ \zeta_5 && \zeta_{3,7}\ \zeta_5 &&&\cr
&   && & && &\zeta_2^4\ \zeta_3^2 &&\zeta_5^3 &&&\cr
&   && & && &\zeta_2^7 && \zeta_{3,5}\ \zeta_7 &&&\cr
\noalign{\hrule}
&d_w &&16 &  &&21 & &&28 &&   &\cr
\noalign{\hrule}}}$$
\vskip-6pt
\centerline{\noindent{\bf Table 2:}
{\sl Basis elements for $\Zc_w$, with $13\leq w\leq 15$.}}
\vskip10pt}}}

{\vbox{\ninepoint{$$
\vbox{\offinterlineskip\tabskip=0pt
\halign{\strut\vrule#
&~$#$~\hfil 
&\vrule$#$ 
&~$#$~\hfil 
&~$#$~\hfil 
&~$#$~\hfil 
&~$#$~\hfil 
&\vrule$#$\cr
\noalign{\hrule}
& w &&16 & & &     &\cr
\noalign{\hrule}
&\Zc_w&& \zeta_{1,1,6,8}    & \zeta_2\ \zeta_3\ \zeta_{3,3,5}  & \zeta_2\ \zeta_{3,3,3,5}  & \zeta_2^2\ \zeta_{1,1,4,6}   &   \cr
&     && \zeta_{3,3,3,7}    & \zeta_2\ \zeta_3^2\ \zeta_{3,5}  & \zeta_2\ \zeta_{3,11}  &    \zeta_2^2\ \zeta_{3,9} & \cr
&     && \zeta_{3,3,5,5}    & \zeta_2\ \zeta_3\ \zeta_{11}  & \zeta_2\ \zeta_{5,9}  & \zeta_2^2\ \zeta_5\ \zeta_7   &   \cr
&     && \zeta_{3,13}   & \zeta_2\ \zeta_3^3\ \zeta_5  &\zeta_2\ \zeta_5\ \zeta_9    & \zeta_2^3\ \zeta_{3,7}   &   \cr
&     &&  \zeta_{5,11}  & \zeta_2^2\ \zeta_3^4  &  \zeta_2\ \zeta_7^2&  \zeta_2^4\ \zeta_{3,5}  &   \cr
&     && \zeta_3\ \zeta_{3,3,7}     & \zeta_2^2\ \zeta_3\ \zeta_9  &  & \zeta_2^3\ \zeta_5^2   &   \cr
&     &&  \zeta_3\ \zeta_{3,5,5}   & \zeta_2^3\ \zeta_3\ \zeta_7  &   &  \zeta_2^8  &   \cr
&     && \zeta_3\ \zeta_{13}   &\zeta_2^4\ \zeta_3\ \zeta_5   &   &    &   \cr
&     && \zeta_{3,7}\ \zeta_3^2     &  \zeta_2^5\ \zeta_3^2 &   &    &   \cr
&     && \zeta_{3,5}\ \zeta_3\ \zeta_5   &   &   &    &   \cr
&     && \zeta_3^3\ \zeta_7   &   &   &    &   \cr
&     && \zeta_3^2\ \zeta_5^2   &   &   &    &   \cr
&     && \zeta_7 \ \zeta_9&   &   &    &   \cr
&     && \zeta_{3,5}^2  &   &   &    &   \cr
&     && \zeta_5\ \zeta_{11}  &   &   &    &   \cr
&     && \zeta_{3,3,5}\ \zeta_5  &   &   &    &   \cr
\noalign{\hrule}
&d_w &&37  & & &  &  \cr
\noalign{\hrule}}}$$
\vskip-6pt
\centerline{\noindent{\bf Table 3:}
{\sl Basis elements for $\Zc_{16}$.}}
\vskip10pt}}}


A slight generalization of \MPOLY\ represents the iterated integral $I_\gamma$ over a product of closed one--forms \goncharov
\eqn\Integral{
I_\gamma(a_0;a_1,\ldots,a_n;a_{n+1})=\int_{\Delta_{n,\gamma}}\fc{dz_1}{z_1-a_1}\wedge\ldots\wedge\fc{dz_n}{z_n-a_n}\ ,}
with $\gamma$ a path in $M=\IC\slash \{a_1,\ldots,a_n\}$ with endpoints $\gamma(0)=a_0\in M,\ \gamma(1)=a_{n+1}\in M$ and $\Delta_{n,\gamma}$ a simplex consisting of all ordered $n$--tuples of points $(z_1,\ldots,z_n)$ on $\gamma$.
For the map 
\eqn\map{
\rho(n_1,\ldots,n_r)=10^{n_1-1}\ldots 10^{n_r-1}\ ,}
with $n_r\geq 2$ Kontsevich observed that:
\eqn\integral{
\zeta_{n_1,\ldots,n_r}=(-1)^r\ I_\gamma(0;\rho(n_1\ldots n_r);1)\ .}
This defines an element in the category $MT(\IZ)$ of mixed Tate motives over $\IZ$. It is an Abelian tensor category, whose simple objects are the Tate motives $\IQ(n)$. The periods of $MT(\IZ)$ are $\IQ\lf[\fc{1}{2\pi i}\ri]$--linear combinations of $\zeta_{n_1,\ldots,n_r}$ \TATE.

\newsec{Open superstring amplitude}

The string $S$--matrix, which describes  string scattering processes
involving on--shell string states as external states,
comprises a perturbative expansion in the string tension $\ap$ 
and the string coupling constant $g_{\rm string}$.
From this expansion one may extract for a given order in $\ap$ and $g_{\rm string}$
the relevant interaction terms of the low--energy effective action.

Open superstring theory contains a massless vector identified as a gauge boson.
Its interactions  are studied by gluon scattering amplitudes. Geometrically, at 
tree--level (\ie at leading order in $g_{\rm string}$) the latter 
are  described by a disk with (integrated) insertions of gluon vertex operators. 
Due to the extended nature of strings the amplitudes  generically represent 
non--trivial functions of the string tension $\ap$. 
In the effective field theory description this $\ap$--dependence
gives rise to a series of infinitely many higher order gauge operators
governed by positive integer powers in $\ap$.
The classical YM term is reproduced in the zero--slope limit $\ap\ra 0$, while
its modification can be derived by studying the higher orders in $\ap$ of the 
tree--level gluon scattering amplitudes.

At string tree--level the complete open string $N$--point superstring amplitude has been computed in \refs{\MSSi,\MSSii}.
The main result is written in a strikingly compact form\foot{A very compact expression for $D=4$ maximal helicity violating 
$N$--gluon amplitudes has been derived in \AMHV.} 
\eqn\SimpleN{
\Ac(1,\ldots,N)=\sum_{\si\in S_{N-3}} A_{YM}(1,2_\si,\ldots,(N-2)_\si,N-1,N)\ 
F_{(1,\ldots,N)}^\si(\ap)\ ,}
where $A_{YM}$ represent 
$(N-3)!$ color ordered Yang--Mills (YM) subamplitudes, $F^\si(\ap)$ are generalized
Euler integrals encoding the full $\ap$--dependence of the string amplitude and $i_\si=\si(i)$. The labels $(1,\ldots,N)$ in $F_{(1,\ldots,N)}^\si$ 
are related to the integration region  of the integrals:
choosing an ordering of the vertex operator positions $z_i$ along the boundary of the disk
determines the color--ordering of the superstring subamplitude. 
The system of $(N-3)!$ multiple hypergeometric functions $F^\si$ 
appearing in \SimpleN\ are given as generalized Euler integrals (with $z_1=0,\ z_{N-1}=1$ and $z_N=\infty$) 
\eqnn\revol
$$\eqalignno{
F^{(23\ldots N-2)}_{(1,\ldots,N)}(s_{ij}) &= (-1)^{N-3}\int\limits_{z_i<z_{i+1}}  
\prod_{j=2}^{N-2} dz_j\  \lf(\prod_{i<l} |z_{il}|^{s_{il}}\ri) \ 
\left\{\ \prod_{k=2}^{N-2}  \sum_{m=1}^{k-1} \fc{ s_{mk} }{z_{mk}} \ \right\}\  ,\cr\crr
 &= (-1)^{N-3} \int\limits_{z_i<z_{i+1}}  
\prod_{j=2}^{N-2} dz_j \ \lf(\prod_{i<l} |z_{il}|^{s_{il}}\ri)  \cr\crr 
&\times\lf\{\left( \ \prod_{k=2}^{[N/2]} \  \sum_{m=1}^{k-1} \ \fc{ s_{mk}}{z_{mk}} \ \ri)\ \lf(\  
\prod_{k=[N/2]+1}^{N-2} \ \sum_{n=k+1}^{N-1} \ \fc{ s_{kn}}{z_{kn}} \ \right)\ri\}\  ,&\revol}$$
with permutations $\si\in S_{N-3}$ acting on all indices within the curly brace.
Above, $[\ldots ]$ denotes the Gauss bracket $[x] = \max_{n \in \IZ,n \leq x} n$, which picks out
the nearest integer smaller than or equal to its argument.
The $\ap$--dependence of \revol\ is encoded in the kinematic invariants
\eqn\KININV{
s_{ij}=\ap\ (k_i+k_j)^2\ ,}
with the external gluon momenta $k_i$ satisfying the on-shell
condition $k_i^2=0$. For further details we refer the reader to Refs. \refs{\MSSi,\MSSii}.

The result \SimpleN\ is valid in any space--time dimension $D$, for 
any compactification and any amount of supersymmetry. Furthermore, the expression 
\SimpleN\ does not make any reference to any kinematical or space--time helicity choices.
Hence, the same is true for our results throughout this article. 
The integrals \revol\ boil down to linear combinations of the following
generalized Euler or Selberg integrals \MSSii
\eqn\GENERIC{
B_N\lf[n\ri]=\lf(\prod_{i=1}^{N-3} \ \int^1_0 d x_i\ri) \ 
\prod_{j=1}^{N-3} \ x_j^{s_{12...j+1}+n_{j} } \ 
\prod_{l=j}^{N-3} \ \lf( \ 1 \ - \ \prod_{k=j}^l x_k \ \right)^{ s_{j+1,l + 2}+n_{jl}}\ ,}
with the set of $\h N(N-3)$ integers $n_j,n_{jl}\in\IZ$ as well as 
$s_{i\ldots l}=\ap(k_i+\ldots+k_l)^2$ 
and $s_{i,j}\equiv s_{ij}$.
The integrals $B_N$  share a very interesting mathematical structure \refs{\Dan,\MSSii}.
For a given $N$ the functions \revol\ represent integrals on the moduli space of Riemann
spheres with $N$ marked points $\Mc_{0,N}$ \refs{\Browni,\Brownii}. 
These spaces have an $N$--fold symmetry following from $N$--fold cyclic 
transformations on the disk, \cf \MSSii\ for more details.
The lowest terms of the $\ap$--expansion of the functions $F^\si$ assume the form~\MSSii
\eqn\LOW{\eqalign{
F^\si&=1+p_2^\si\ \zeta(2)+p_3^\si\ \zeta(3) +\ldots\ \ \ ,\ \ \ \si=(23\ldots N-2)\ ,\cr
F^\si&=p_2^\si\ \zeta(2)+p_3^\si\ \zeta(3)+\ldots\ \ \ ,\ \ \ \si\neq(23\ldots N-2)\ ,}}
with some polynomials $p_n^\si$ of degree $n$ in the  kinematic invariants 
$s_{ij}$ and $s_{i\ldots l}$.
Note that starting at $N\geq 7$ subsets of $F^\si$ start at even higher order in $\ap$, 
\ie $p_2^\si,\ldots,p_\nu^\si=0$ for some $\nu\geq2$.
In Refs. \refs{\Browni,\Brownii} it is proven, that at lowest order in $\ap$
 these integrals always lead to linear  $\IQ$ combinations of MZVs of weight $w\leq N-3$.
Consequently, to all orders in $\ap$ only  combinations of MZVs show up.

In the following let us discuss the cases $N=4$ and $N=5$ in more detail before
moving to the general case afterwards.
\subsec{$N=4$}

For $N=4$ \eqq \SimpleN\ becomes:
\eqn\four{
\Ac(1,2,3,4)=A_{YM}(1,2,3,4)\ F\ ,}
with the function
\eqn\funciv{
F:=F^{(2)}_{(1,2,3,4)}=s\ \int_0^1 dx\ x^{s-1}\ (1-x)^u=\fc{\Gamma(1+s)\ \Gamma(1+u)}{\Gamma(1+s+u)}\ ,}
and the two kinematic invariants $s= \alpha'(k_1+k_2)^2$ and $u= \alpha'(k_1+k_4)^2$.
With the identities
$$\eqalign{
\fc{\Gamma(1+x)}{\Gamma(1-x)}&=\exp\lf\{-2\ \sum_{n=1}^\infty \fc{x^{2n+1}}{2n+1}\ \zeta_{2n+1}\ri\}\ \exp\lf\{-2\gamma_E x\ri\}\ ,\cr
\pi\ \fc{s\ u}{s+u}\ \fc{\sin[\pi(s+u)]}{\sin(\pi s)\ \sin(\pi u)}\ 
&=\exp\lf\{2\ \sum_{n=1}^\infty\fc{\zeta_{2n}}{2n}\ [\ s^{2n}+u^{2n}-(s+u)^{2n}\ ]\ri\}\ ,}$$
we may bring \four\ into the following form
\eqn\wehave{
\Ac(1,2,3,4)=P\ \exp\lf\{\sum_{n\geq 1}\zeta_{2n+1}\ M_{2n+1}\ri\}\ A_{YM}(1,2,3,4)\ ,}
with: 
\eqn\wehaveiv{\eqalign{
P&=\exp\lf\{\sum_{n=1}^\infty\fc{\zeta_{2n}}{2n}\ [\ s^{2n}+u^{2n}-(s+u)^{2n}\ ]\ri\}\ ,
\cr\crr
M_{2n+1}&=-\fc{1}{2n+1}\ \lf[\ s^{2n+1}+u^{2n+1}-(s+u)^{2n+1}\ \ri].}}
In \wehave\ we observe  a disentanglement of Riemann zeta functions of even and odd arguments. Furthermore, no MZVs of depth  greater than one $r>1$ appear.

\subsec{$N=5$}

For $N=5$ we have a basis of two color ordered superstring amplitudes
$\Ac(1,2,3,4,5)$ and $\Ac(1,3,2,4,5)$. According to \SimpleN\ they take the form:
\eqn\five{\eqalign{
\Ac(1,2,3,4,5)=A_{YM}(1,2,3,4,5)\ F_1+A_{YM}(1,3,2,4,5)\ F_2\ ,\cr
\Ac(1,3,2,4,5)=A_{YM}(1,3,2,4,5)\ \tilde F_1+A_{YM}(1,2,3,4,5)\ \tilde F_2\ ,}}
with the hypergeometric functions \revol:
\eqnn\exv
$$\eqalignno{
F_1:=F^{(23)}_{(1,2,3,4,5)}&=s_{12}\ s_{34}\ \int_0^1dx\int_0^1 dy\  x^{s_{45}}\ y^{s_{12}-1}\ 
(1-x)^{s_{34}-1}\ (1-y)^{s_{23}}\ (1-xy)^{s_{24}}\cr
&=\fc{\Gamma(1+s_{12})\ \Gamma(1+s_{23})\ \Gamma(1+s_{34})\ \Gamma(1+s_{45})}{\Gamma(1+s_{12}+s_{23})\ \Gamma(1+s_{34}+s_{45})}\cr\crr
&\times  \ _3F_2\lf[{s_{12},\ 1+s_{45},\ -s_{24}\atop 1+s_{12}+s_{23},\ 1+s_{34}+s_{45}};1\ri]\ ,\cr
F_2:=F^{(32)}_{(1,2,3,4,5)}&=s_{13}\ s_{24}\ \int_0^1dx\int_0^1 dy\ x^{s_{45}}\ y^{s_{12}}\ 
(1-x)^{s_{34}}\ (1-y)^{s_{23}}\ (1-xy)^{s_{24}-1}\cr
&=s_{13}\ s_{24}\ \fc{\Gamma(1+s_{12})\ \Gamma(1+s_{23})\ \Gamma(1+s_{34})\ \Gamma(1+s_{45})}{\Gamma(2+s_{12}+s_{23})\ \Gamma(2+s_{34}+s_{45})}\cr\crr 
&\times \ _3F_2\lf[{1+s_{12},\ 1+s_{45},\ 1-s_{24}\atop 2+s_{12}+s_{23},\ 2+s_{34}+s_{45}};1\ri]\ .&\exv}
$$
Furthermore, we have: 
\eqn\permut{
\tilde F_1=\lf. F_1\ri|_{2\leftrightarrow3}\ \ \ ,\ \ \ 
\tilde F_2=\lf. F_2\ri|_{2\leftrightarrow3}\ .}

When investigating the $\ap$--expansions of \five\ one makes the following intriguing observation\foot{We have tested this formula up to weight $16$. Work beyond this order is in progress \BSS.}:
\eqn\VERYNICE{
\Ac=P\ Q\ :\exp\lf\{\sum\limits_{n\geq1}\zeta_{2n+1}\ M_{2n+1}\ri\}:\ A\ ,}
with the vectors
\eqn\ppp{
A=\lf(A_{YM}(1,2,3,4,5)\atop  A_{YM}(1,3,2,4,5)\ri)\ \ \ ,\ \ \ \Ac=\pmatrix{\Ac(1,2,3,4,5)\cr\Ac(1,3,2,4,5)}\ ,}
and the matrices
\eqnn\PPP
$$\eqalignno{
M_{2n+1}&=\lf.\pmatrix{F_1 & F_2\cr
\tilde F_2& \tilde F_1 }\ri|_{\zeta_{2n+1}}\ ,&\PPP\cr\crr
P&=\pmatrix{
\sum\limits_{n\geq 0} p_{2n}\ \zeta_2^n&  \sum\limits_{n\geq 0} q_{2n}\ \zeta_2^n\cr\crr
\sum\limits_{n\geq 0} \tilde q_{2n}\ \zeta_2^n&  \sum\limits_{n\geq 0} \tilde p_{2n}\ \zeta_2^n}=1+\sum_{n\geq 1} \zeta_2^n\ P_{2n}\ ,}
$$
where $\tilde p_{2n}=\lf.p_{2n}\ri|_{2\leftrightarrow3},\ 
\tilde q_{2n}=\lf.q_{2n}\ri|_{2\leftrightarrow3}$. Furthermore, we have the matrix:
\eqn\Q{
Q=1+\sum_{n\geq 8} Q_n\ ,}
with:
\eqnn\QQQ
$$\eqalignno{
Q_8&=\fc{1}{5}\ \zeta_{3,5}\ [M_5,M_3]\ \ \ ,\ \ \ Q_9=0\ ,\cr\crr
Q_{10}&=\lf\{\ \fc{3}{14}\ \zeta_5^2+\fc{1}{14}\ \zeta_{3,7}\ \ri\}\ [M_7,M_3]\ ,\cr \crr
Q_{11}&=\lf\{\ 9\ \zeta_2\ \zeta_9+\fc{6}{25}\ \zeta_2^2\ \zeta_7-\fc{4}{35}\ \zeta_2^3\ \zeta_5+\fc{1}{5}\ \zeta_{3,3,5}\ \ri\}\ [M_3,[M_5,M_3]]\ ,\cr\crr
Q_{12}&=\lf\{\ \fc{2}{9}\ \zeta_5\ \zeta_7+\fc{1}{27}\ \zeta_{3,9}\ \ri\}\ [M_9,M_3]\cr\crr
&+\fc{48}{691}\ \lf\{\ \fc{18}{35}\ \zeta_2^3\ \zeta_3^2+\fc{1}{5}\ \zeta_2^2\ \zeta_3\ \zeta_5-
10\ \zeta_2\ \zeta_3\ \zeta_7-\fc{7}{2}\ \zeta_2\ \zeta_5^2-\fc{3}{5}\ \zeta_2^2\ \zeta_{3,5}-
3\ \zeta_2\ \zeta_{3,7}\ri.\cr\crr
&\lf.-\fc{1}{12}\ \zeta_3^4-\fc{467}{108}\ \zeta_5\ \zeta_7
+\fc{799}{72}\ \zeta_3\ \zeta_9+\fc{2665}{648}\ \zeta_{3,9}+\zeta_{1,1,4,6}\ \ri\}\ 
\lf\{\ [M_9,M_3]-3\ [M_7,M_5]\ \ri\}\ ,\cr\crr
Q_{13}&=\lf\{\ \fc{11}{4}\ \zeta_2\ \zeta_{11}-\fc{2}{35}\ \zeta_2^2\ \zeta_9-\fc{16}{245}\ \zeta_2^3\ \zeta_7-\fc{3}{35}\ \zeta_{3,5,5}+\fc{1}{14}\ \zeta_{3,3,7}\ \ri\}\ [M_3,[M_7,M_3]]\cr\crr
&+\lf\{\ \fc{11}{2}\ \zeta_2\ \zeta_{11}+\fc{2}{5}\ \zeta_2^2\ \zeta_9+\fc{1}{5}\ \zeta_5\ \zeta_{3,5}+\fc{1}{25}\ \zeta_{3,5,5}\ \ri\}\ [M_5,[M_5,M_3]]\ ,\cr\crr
Q_{14}&=\lf\{\ 4\ \zeta_2\ \zeta_5\ \zeta_7+\fc{4}{175}\ \zeta_2^3\ \zeta_{3,5}-\fc{647287}{11880}\ \zeta_7^2-\fc{12775}{198}\ \zeta_5\ \zeta_9+\fc{232}{81}\ \zeta_{5,9}\ri.\cr\crr
&\lf.+\fc{2}{3}\ \zeta_2\ \zeta_{3,9}-\fc{12841}{1188}\ \zeta_{3,11}+\fc{1}{5}\ \zeta_{3,3,3,5}\ \ri\}\ [M_3,[M_3,[M_5,M_3]]]\cr\crr
&+\lf\{\ -\fc{235}{396}\ \zeta_7^2-\fc{23}{33}\ \zeta_5\ \zeta_9+\fc{1}{27}\ \zeta_{5,9}-\fc{23}{198}\ \zeta_{3,11}\ \ri\}\ [M_{11},M_3]\cr
&+\lf\{\ \fc{55}{36}\ \zeta_7^2+\fc{5}{3}\ \zeta_5\ \zeta_9+\fc{5}{18}\ \zeta_{3,11}
-\fc{2}{27}\ \zeta_{5,9}\    \ri\}\ [M_9,M_5]\ ,\cr\crr
Q_{15}&=\lf\{\ \fc{1339}{30}\ \zeta_2\ \zeta_{13}+\fc{128}{45}\ \zeta_2^2\ \zeta_{11}-
\fc{236}{4725}\ \zeta_2^3\ \zeta_9-\fc{184}{2625}\ \zeta_2^4\ \zeta_7-\fc{64}{5775}\ \zeta_2^5\ \zeta_5\ri.\cr\crr
&\lf.-\fc{2}{45}\ \zeta_5^3-\fc{1}{15}\ \zeta_7\ \zeta_{3,5}-\fc{2}{45}\ \zeta_5\ \zeta_{3,7}+\fc{1}{27}\ \zeta_{3,3,9}\ \ri\}\ [M_3,[M_9,M_3]]\cr\crr
&+\lf\{\ -\fc{143}{20}\ \zeta_2\ \zeta_{13}-\fc{11}{35}\ \zeta_2^2\ \zeta_{11}+\fc{68}{1225}\ \zeta_2^3\ \zeta_9+\fc{11}{70}\ \zeta_5^3+\fc{24}{875}\ \zeta_2^4\ \zeta_7+
\fc{48}{13475}\ \zeta_2^5\ \zeta_5\ri.\cr\crr
&\lf.+\fc{1}{5}\ \zeta_7\ \zeta_{3,5}+\fc{3}{35}\ \zeta_5\ \zeta_{3,7}-\fc{1}{70}\ \zeta_{5,3,7}\ \ri\}\ [M_5,[M_7,M_3]]+\fc{2}{15}\ \zeta_{5,3,7}\ [M_3,[M_7,M_5]]\cr\crr
&+\fc{48}{7601}\ \lf\{\ -8\ \zeta_2\ \zeta_3\ \zeta_5^2+\fc{21}{2}\ \zeta_2\ \zeta_5\ \zeta_{3,5}-\fc{14}{5}\ \zeta_2\ \zeta_{3,5,5}+2\ \zeta_2\ \zeta_{3,3,7}-26\ \zeta_2\ \zeta_3^2\ \zeta_7 \ri.\cr\crr
&-\fc{6417649}{2880}\ \zeta_2\ \zeta_{13}-6\ \zeta_2\ \zeta_3\ \zeta_{3,7}-\fc{8495287}{15120}\ \zeta_2^2\ \zeta_{11}-\fc{23}{10}\ \zeta_2^2\ \zeta_3^2\ \zeta_5-\fc{8}{5}\ \zeta_2^2\ \zeta_3\ \zeta_{3,5}\cr\crr
&+4\ \zeta_2^2\ \zeta_{3,3,5}+\fc{12}{35}\ \zeta_2^3\ \zeta_3^3 +\fc{54263011}{396900}\ \zeta_2^3\ \zeta_9+
\fc{57847}{15750}\ \zeta_2^4\ \zeta_7-\fc{1714624}{121275}\ \zeta_2^5\ \zeta_5\cr\crr
&+\fc{1451972}{716625}\ \zeta_2^6\ \zeta_3+\fc{1185701}{30240}\ \zeta_5^3-\fc{74}{3}\ \zeta_3\ \zeta_5\ \zeta_7-\fc{1}{15}\ \zeta_3^5+\fc{6775}{144}\ \zeta_3^2\ \zeta_9+\fc{2188}{945}\ \zeta_5\ \zeta_{3,7}\cr\crr
&\lf.-\fc{12199}{720}\ \zeta_7\ \zeta_{3,5}
 +\fc{29}{9}\ \zeta_3\ \zeta_{3,9}  +\zeta_3\ \zeta_{1,1,4,6}+\fc{17203}{3360}\ \zeta_{5,3,7} -\fc{853}{648}\ \zeta_{3,3,9}+\zeta_{1,1,3,4,6}\ \ri\}\cr\crr
 &\times \Big\{\ [M_3,[M_9,M_3]]-3\ [M_3,[M_7,M_5]]\ \Big\}\ ,\cr\crr
Q_{16}&=\fc{1}{50}\ \zeta_{3,5}^2\ ([M_5,M_3])^2+\lf\{\ \fc{210}{121}\ \z_9\ \z_7+
\fc{9}{11}\ \z_{11}\ \z_5-\fc{5}{242}\ \z_{5,11}+\fc{3}{22}\ 
\z_{3,13}\ \ri\}\ [M_{11},M_5]\cr\crr
&+\lf\{\ -\fc{1275}{1573}\ \z_9\ \z_7-\fc{57}{143}\ \z_{11}\ \z_5+\fc{3}{242}\ \z_{5,11}-\fc{19}{286}\ 
\z_{3,13}\ri\}\ [M_{13},M_3]\cr\crr
&+\lf\{\ \fc{24}{35}\ \z_7\ \z_5\ \z_2^2+\fc{6}{245}\ \z_5^2\ \z_2^3+\fc{2}{245}\ \z_{3,7}\ \z_2^3+
\fc{4}{35}\ \z_{3,9}\ \z_2^2+\fc{967}{56}\ \z_7^2\ \z_2\ri.\cr\crr
&+\fc{363}{14}\ \z_9\ \z_5\ \z_2-\fc{47}{42}\ \z_{5,9}\ \z_2+\fc{121}{28}\ \z_{3,11}\ \z_2
-\fc{2272973}{330330}\ \z_9\ \z_7-\fc{601677}{40040}\ \z_{11}\ \z_5\cr\crr
&\lf.+\fc{23181}{67760}\ \z_{5,11}-
\fc{200559}{80080}\ \z_{3,13}-\fc{3}{35}\ \z_{3,3,5,5}+\fc{1}{14}\ \z_{3,3,3,7}\ \ri\}\ 
[M_3,[M_3,[M_7,M_3]]]\cr\crr
&+\lf\{\ -\fc{8}{25}\ \z_7\ \z_5\ \z_2^2-\fc{2}{35}\ \z_5^2\ \z_2^3-\fc{4}{75}\ \z_{3,9}\ \z_2^2
-\fc{333}{20}\ \z_7^2\ \z_2\ri.\cr\crr
&-21\ \z_9\ \z_5\ \z_2+\z_{5,9}\ \z_2-\fc{7}{2}\ \z_{3,11}\ \z_2-\fc{299373}{7150}\ \z_9\ \z_7
-\fc{21033}{1300}\ \z_{11}\ \z_5\cr\crr
&\lf. +\fc{909}{2200}\ \z_{5,11}-\fc{7011}{2600}\ \z_{3,13}+\fc{1}{5}\ \z_5\ \z_{3,3,5}+\fc{1}{25}\ 
\z_{3,3,5,5}\ \ri\}\ [M_3,[M_5,[M_5,M_3]]]\cr\crr
&+\fc{720}{3617}\ \lf\{-\fc{21331}{525}\ \z_5\ \z_7\ \z_2^2-\fc{284}{245}\ \z_5^2\ \z_2^3+\fc{108}{875}\ \z_{3,5}\ \z_2^4-\fc{62}{245}\ \z_{3,7}\ \z_2^3\ri.\cr\crr
&-\fc{8954}{1575}\ \z_{3,9}\ \z_2^2-\fc{78201}{140}\ \z_7^2\ \z_2-\fc{12443}{14}\ \z_5\ \z_9\ \z_2
+\fc{697}{21}\ \z_{5,9}\ \z_2-\fc{1991}{14}\ \z_{3,11}\ \z_2\cr\crr
&-137\ \z_{11}\ \z_3\ \z_2-\fc{11}{7}\ \z_9\ \z_3\ \z_2^2+\fc{848}{245}\ \z_7\ \z_3\ \z_2^3+
\fc{48}{35}\ \z_5\ \z_3\ \z_2^4+\fc{408}{2695}\ \z_3^2\ \z_2^5\cr\crr
&-\fc{4}{7}\ \z_3^2\ \z_5^2-\fc{1}{3}\ \z_3^3\ \z_7+\fc{4850713}{6600}\ \z_7\ \z_9+\fc{455534}{525}\ \z_5\ \z_{11}+\fc{8497}{42}\ \z_3\ \z_{13}+\fc{1}{7}\ \z_3^2\ \z_{3,7}\cr\crr
&-\fc{114307}{7392}\ \z_{5,11}+\fc{2217053}{16800}\ \z_{3,13}-\fc{2}{5}\ \z_5\ \z_{3,3,5}
-\fc{6}{7}\ \z_3\ \z_{3,5,5}+\fc{5}{7}\ \z_{3,3,7}\ \z_3\cr\crr
&\lf.+\fc{542}{175}\ \z_{3,3,5,5}-\fc{19}{7}\ \z_{3,3,3,7}+\z_{1,1,6,8}\ri\}\ 
\lf\{\ \fc{7}{11}\ [M_{11},M_5]-\fc{2}{11}\ [M_{13},M_3]-[M_9,M_7]\ri.\cr\crr
&\lf.+\fc{6493}{9240}\ [M_3,[M_3,[M_7,M_3]]]-\fc{751}{100}\ [M_3,[M_5,[M_5,M_3]]]\ \ri\}\ .&\QQQ}
$$
Finally, in \VERYNICE\  the ordering colons 
$:\ldots :$ are defined such that matrices with larger subscript multiply matrices with smaller subscript from the left,
\eqn\order{
: \, M_{i} \ M_{j} \, : =  
\cases{M_{i} \ M_j\ , & $i \geq j\ ,$\cr 
       M_{j} \ M_i\ , &              $i<j\ .$}}
The generalization to iterated matrix products $: M_{i_1} M_{i_2} \ldots M_{i_p}:$ is straightforward. 

To illustrate the structure of the matrices $P$ and $M$, given in \PPP, 
let us display $P_2$ and $M_3$:
\eqn\EXAMP{
P_2=\pmatrix{-s_3 s_4+s_1\ (s_3-s_5)&s_{13}\ s_{24}\cr
s_1\ s_3&(s_1+s_2)\ (s_2+s_3)-s_4 s_5},\ \ \ M_3=\pmatrix{m_{11}&m_{12}\cr m_{21}&m_{22}}\ ,}
with 
\eqnn\examp{
$$\eqalignno{m_{11}&=s_3\ [\ -s_1\ (s_1+2 s_2+s_3)+s_3 s_4+s_4^2\ ]+s_1 s_5\ (s_1 +s_5)\ ,\cr
m_{12}&=-s_{13}\ s_{24}\ (s_1+s_2+s_3+s_4+s_5),\ \ \ 
m_{21}=s_1\ s_3\ [\ s_1+s_2+s_3-2\ (s_4+s_5)\ ]\ ,\cr
m_{22}&=(s_2+s_3)\ [\ (s_1+s_2) (s_1+s_3)-2\ s_1 s_4\ ]-[\ 2 s_1 s_3-s_4^2+2 s_2\ (s_3+s_4)\ ] 
s_5+s_4 s_5^2\ ,}$$
and $s_i \equiv \alpha'(k_i+k_{i+1})^2$ subject to cyclic identification $k_{i+N} 
\equiv k_i$. 
The expression \VERYNICE\ allows to conveniently extract any order in $\ap$ of the superstring amplitude by simple matrix manipulations.
{\it E.g.} at weight  $w=8$ from \VERYNICE\ we obtain the expressions
\eqnn\testa
$$\eqalignno{
\lf.\Ac\ \ri|_{\zeta_3\zeta_5}&=M_5\ M_3\ A\ ,\cr
\lf.\Ac\ \ri|_{\zeta_{3,5}}&=\fc{1}{5}\ \lf[M_5,M_3\ri]\ A\ ,\cr\crr
\lf.\Ac\ \ri|_{\zeta_2\zeta_3^2}&=\h\ P_2\ M_3\ M_3\ A\ ,\cr\crr
\lf.\Ac\ \ri|_{\zeta_2^4}&=P_8\ A\ ,&\testa}
$$
while for weight $w=10$ we get:
\eqnn\testb
$$\eqalignno{\lf.\Ac\ \ri|_{\zeta_3\zeta_7}&=M_7\ M_3\ A\ ,\cr\crr
\lf.\Ac\ \ri|_{\zeta_{3,7}}&=\fc{1}{14}\ \lf[M_7,M_3\ri]\ A\ ,\cr\crr
\lf.\Ac\ \ri|_{\zeta_5^2}&=\lf(\h\ M_5\ M_5+\fc{3}{14}\ \lf[M_7,M_3\ri]\ \ri)\ A\ ,\cr\crr
\lf.\Ac\ \ri|_{\zeta_2\zeta_3\zeta_5}&=P_2\ M_5\ M_3\ A\ ,\cr\crr
\lf.\Ac\ \ri|_{\zeta_2\zeta_{3,5}}&=\fc{1}{5}\ P_2\ \lf[M_5,M_3\ri]\ A\ ,\cr\crr
\lf.\Ac\ \ri|_{\zeta_2^2\zeta_3^2}&=\h\ P_4\ M_3\ M_3\ A\ ,\cr\crr
\lf.\Ac\ \ri|_{\zeta_2^5}&=P_{10}\ A\ .&\testb}
$$
\comment{and for weight eleven we obtain:
\eqn\testc{
\eqalign{\lf.\Ac\ \ri|_{\zeta_{3,3,5}}&=-\fc{1}{5}\ \lf[M_3,\lf[M_3,M_5\ri]\ri]\ A_{YM}\ ,\cr\crr
\lf.\Ac\ \ri|_{\zeta_{3,5}\zeta_3}&=\fc{1}{5}\ \lf[M_5,M_3\ri]\ M_3\ A_{YM}\ ,\cr\crr
\lf.\Ac\ \ri|_{\zeta_3^2\zeta_5}&=\h\ M_5\ M_3\ M_3\ A_{YM}\ ,\cr\crr
\lf.\Ac\ \ri|_{\zeta_2\zeta_3^3}&=\fc{1}{6}\ P_2\ M_3\ M_3\ M_3\ A_{YM}\ ,\cr\crr
\lf.\Ac\ \ri|_{\zeta_2\zeta_9}&=\lf(P_2\ M_9-9\ \lf[M_3,\lf[M_3,M_5\ri]\ri] \ri)\ A_{YM}\ ,\cr\crr
\lf.\Ac\ \ri|_{\zeta_2^2\zeta_7}&=\lf(P_4\ M_7-\fc{6}{25}\ \lf[M_3,\lf[M_3,M_5\ri]\ri] \ri)A_{YM}\ ,\cr\crr
\lf.\Ac\ \ri|_{\zeta_2^3\zeta_5}&=\lf(P_6\ M_5+\fc{4}{35}\ \lf[M_3,\lf[M_3,M_5\ri]\ri] \ri)A_{YM}\ ,\cr\crr
\lf.\Ac\ \ri|_{\zeta_2^4\zeta_3}&=P_8\ M_3\ A_{YM}\ .}}
}
The terms $M_5 M_3 A$ in \testa\  and $M_7 M_3 A,\ P_2 M_5 M_3A$ in \testb\ 
use the ordering prescription \order\ introduced in \VERYNICE\ for the  matrices $M_i$ 
stemming from the exponential.

\subsec{General $N$}

For generic $N$ in  \SimpleN\ we have a basis of $(N-3)!$ color ordered superstring amplitudes
$\Ac(1,2_\si,\ldots,(N-2)_\si,N-1,N)$. Putting these $(N-3)!$ amplitudes into  an 
$(N-3)!$--dimensional vector $\Ac$ according to \SimpleN\ the latter
can be expressed by an $(N-3)!\times (N-3)!$--matrix $F$ 
acting on the  vector  $A$ encoding an $(N-3)!$--dimensional YM--basis as:
\eqn\npt{
\Ac=F\ A\ .}
The matrix $F$ encodes the full $\ap$--dependence of the superstring amplitude \npt.
We conjecture, that the $\ap$--dependence of the latter assumes the same form
\VERYNICE\ as for the case $N=5$
\eqn\VERYNICEE{
F=P\ Q\  :\exp\lf\{\sum\limits_{n\geq1}\zeta_{2n+1}\ M_{2n+1}\ri\}:\ ,}
with the  matrices $P, M$ and $Q$ now being $(N-3)!\times (N-3)!$ matrices, following from
\eqnn\PP
$$\eqalignno{
M_{2n+1}&=\lf.F\ \ri|_{\zeta_{2n+1}}\ ,&\PP\cr\crr
P&=1+\sum_{n\geq 1} \zeta_2^n\ P_{2n}:=1+\sum_{n\geq 1} \zeta_2^n\ \lf.F\ \ri|_{\zeta_2^n}\ ,}
$$
with $P_{2n}=\lf.P\ri|_{\zeta_2^n}$ and $Q$ given in \Q. 
The polynomial structure of the matrices $M,P$ and $Q$ is further exhibited in \BSS.

What makes the form \VERYNICEE\ appealing is the  disentanglement of the full
$\ap$--expansion into several contributions accounting for different classes of MZVs: 
$P$ comprising  powers of $\z_2$, $M$  accounting for $\z_{2n+1}$ and powers thereof
and $Q$ encapsulating the MZVs $\z_{n_1,\ldots,n_r}$ of depth $r>1$  greater than one.
As we shall see in section 4 the specific form \VERYNICEE\ is bolstered by the decomposition of motivic MZVs. 
It is interesting to note, that in \Q\ MZVs of depth greater than one $r>1$
appear with commutators as:
\eqn\interesting{
\zeta_{n_1,\ldots,n_r}\ \ [M_{n_2},[M_{n_3},\ldots,[M_{n_r},M_{n_1}]]\ldots]\ .}
This property turns out to have a crucial impact on the closed string amplitude, \cf
section~6.

At weight $16$ in \Q\ the term $\fc{1}{50}\ \zeta_{3,5}^2\ ([M_5,M_3])^2$ gives 
rise  to speculate, that all terms in $Q$ follow from 
expanding an exponential:
\eqn\expandexp{
Q=\exp\lf\{ \fc{1}{5}\ \zeta_{3,5}\ [M_5,M_3]+\lf(\fc{3}{14}\ \zeta_5^2+\fc{1}{14}\ \zeta_{3,7}\ri)\ [M_7,M_3]+\ldots\ri\}\ .}
In fact, at weight $18$ we find the following terms\foot{Note the commutator relations:
$[M_7,M_3][M_5,M_3]=[M_5,M_3][M_7,M_3]$ and $[M_3,[M_5,[M_7,M_3]]]=[M_5,[M_3,[M_7,M_3]]]$.}
\eqnn\eightteen
$$\eqalignno{
\lf.\Ac\ri|_{\z_{3,5}\z_{3,7}}&=\fc{1}{5}\ \fc{1}{14}\ [M_7,M_3]\ [M_5,M_3]
+\fc{208926}{894845}\ [M_3,[M_3,[M_7,M_5]]]\cr\crr
&-\fc{69642}{894845}\ [M_3,[M_3,[M_9,M_3]]]\ ,&\eightteen\cr\crr
\lf.\Ac\ri|_{\z_{3,5}\z_{5}^2}&=\fc{1}{2}\ \fc{1}{5}\ [M_5,M_3]M_5^2
+\fc{1}{5}\ \fc{3}{14}\ [M_7,M_3]\ [M_5,M_3]+\fc{1}{5}\ [M_5,[M_5,M_3]]M_5\cr\crr
&+\fc{1800}{43867}\ [M_{11},M_7]-\fc{22500}{570271}\ [M_{13},M_5]+\fc{7200}{570271}\ [M_{15},M_3]\cr\crr
&-\fc{7044111243797}{6415252209080}\ [M_3,[M_3,[M_7,M_5]]]
+\fc{2792059}{5702710}\ [M_5,[M_5,[M_5,M_3]]]\cr\crr
&-\fc{2432943}{7983794}\ [M_5,[M_3,[M_7,M_3]]]-\fc{2818807834641}{6415252209080}\ [M_3,[M_3,[M_9,M_3]]]\ ,}
$$
in agreement with the Ansatz \expandexp.

Obviously, for $N=4$ in \VERYNICEE\ we have $Q=1$ as all commutators vanish for the scalars $M_{2n+1}$ given in \wehaveiv. With this information \VERYNICEE\ boils down to \wehave.
So far, for $N=6$ we have verified \VERYNICEE\ up to $\ap^8$. Further tests are in 
progress \BSS.

\subsec{Minimal depth representation with Euler sums}

The choice of basis elements may follow some minimal intrinsic representation 
guided by the minimal depth representation and the choice of positive odd indices only.
For MZVs this is achieved by also allowing for  Euler sums as basis elements:
\eqn\Euler{
\zeta(\eps_1n_1,\ldots,\eps_rn_r)=
\sum_{0<k_1<\ldots<k_r} \prod_{l=1}^r \eps_l^{k_l}\ k_l^{-n_l}\ \ \ ,\ \ \ n_l\in\IN^+\ ,\ n_r\geq2\ .}
with signs $\eps_l=\pm 1$.
For $M_{w,r}$ being the number of basis elements for MZVs when expressed in terms of Euler sums in a minimal depth representation at weight $w>2$ and depth $r$ we have $M_{12,2}=2, 
M_{12,4}=0,\ M_{15,3}=3,\ M_{15,5}=0, M_{16,2}=3$ and $M_{16,4}=2$~\Broadii.
At weight $12$ one may get rid of the basis element $\zeta_{1,1,4,6}$ with even entries at the cost of the introducing the Euler sum $\zeta_{\ov 5,\ov 7}:=\zeta(-5,-7)$ \DataMine:
\eqnn\eulerxiia
$$\eqalignno{
\zeta_{1,1,4,6}&=-\fc{5045}{648}\ \zeta_{3,9}+3\ \zeta_2\ \zeta_{3,7}+\fc{3}{5}\ 
\zeta_2^2\ \zeta_{3,5}-\fc{799}{72}\ \zeta_3\ \zeta_9-\fc{5747}{432}\ \zeta_5\ \zeta_7
+10\ \zeta_2\ \zeta_3\ \zeta_7\cr
&+\fc{7}{2}\ \zeta_2\ \zeta_5^2-\fc{1}{5}\ \zeta_2^2\ \zeta_3\ \zeta_5+\fc{1}{12}\ \zeta_3^4-\fc{18}{35}\ \zeta_2^3\ \zeta_3^2+\fc{694891}{2837835}\ \zeta_2^6-\fc{64}{27}\ \zeta_{\ov 5,\ov 7}\ .\ \ \ \ \ &\eulerxiia}
$$
Similarly, we may use the Euler sum $\zeta_{\ov 3,\ov 9}:=\zeta(-3,-9)$  to arrive at \Broadi:
\eqnn\eulerxiib
$$\eqalignno{
\zeta_{1,1,4,6}&=\fc{371}{144}\ \zeta_{3,9}+3\ \zeta_2\ \zeta_{3,7}+\fc{3}{5}\ 
\zeta_2^2\ \zeta_{3,5}-\fc{3131}{144}\ \zeta_3\ \zeta_9+\fc{107}{24}\ \zeta_5\ \zeta_7
+10\ \zeta_2\ \zeta_3\ \zeta_7\cr
&+\fc{7}{2}\ \zeta_2\ \zeta_5^2-\fc{1}{5}\ \zeta_2^2\ \zeta_3\ \zeta_5+\fc{1}{12}\ \zeta_3^4-\fc{18}{35}\ \zeta_2^3\ \zeta_3^2-\fc{117713}{2627625}\ \zeta_2^6+\fc{64}{9}\ \zeta_{\ov 3,\ov 9}\ .\ \ \ \ \ &\eulerxiib}
$$
In \DataMine\ the object $A_{5,7}$
\eqn\Axii{
A_{5,7}=\zeta_{\ov 5,\ov 7}+\zeta_{5,7}}
has been  argued to play a special status within the Euler sums, since it is 
quite similar to the MZVs.
With this \eulerxiia\ can be written:
\eqnn\eulerxiic
$$\eqalignno{
\zeta_{1,1,4,6}&=-\fc{7967}{1944}\ \zeta_{3,9}+3\ \zeta_2\ \zeta_{3,7}+\fc{3}{5}\ 
\zeta_2^2\ \zeta_{3,5}-\fc{799}{72}\ \zeta_3\ \zeta_9+\fc{11431}{1296}\ \zeta_5\ \zeta_7
+10\ \zeta_2\ \zeta_3\ \zeta_7\cr
&+\fc{7}{2}\ \zeta_2\ \zeta_5^2-\fc{1}{5}\ \zeta_2^2\ \zeta_3\ \zeta_5+\fc{1}{12}\ \zeta_3^4-\fc{18}{35}\ \zeta_2^3\ \zeta_3^2-\fc{5607853}{6081075}\ \zeta_2^6-\fc{64}{27}\ 
A_{5,7}\ .\ \ \ \ \ \ \ \ &\eulerxiic}
$$
Clearly, the above three equations \eulerxiia, \eulerxiib\ and \eulerxiic\ 
are related by the identities:
\eqn\identxii{\eqalign{
\zeta_{5,7}&=\fc{14}{9}\ \zeta_{3,9}+\fc{28}{3}\ \zeta_5\ \zeta_7-\fc{776224}{1576575}\ \zeta_2^6\ ,\cr
\zeta_{\ov 3,\ov 9}&=-\fc{1}{3}\ \zeta_{\ov 5,\ov 7}-\fc{13429}{9216}\ \zeta_{3,9}+\fc{1533}{1024}\ \zeta_3\ \zeta_9-\fc{7673}{3072}\ \zeta_5\ \zeta_7+\fc{10275263}{252252000}\ \zeta_2^6\ .}}
We can write the weight $12$ part $Q_{12}$ of \Q\ in terms of Euler 
sums in a minimal depth representation and positive odd indices only in the following 
three ways corresponding to  \eulerxiia, \eulerxiib\ and \eulerxiic, respectively;
\eqnn\QQ
$$\eqalignno{
Q_{12}&=\lf\{\fc{2}{9}\ \zeta_5\ \zeta_7+\fc{1}{27}\ \zeta_{3,9}\ri\}\ [M_9,M_3]+\fc{48}{691}\ \lf\{\ [M_9,M_3]-3\ [M_7,M_5]\ \ri\}\cr\crr
&\times\lf\{\ \fc{694891}{2837835}\ \zeta_2^6-\fc{7615}{432}\ \zeta_5\ \zeta_7-\fc{595}{162}\ \zeta_{3,9}-\fc{64}{27}\ \zeta_{\ov 5,\ov 7}\ \ri\}\cr\crr
&=\lf\{\ \fc{2}{9}\ \zeta_5\ \zeta_7+\fc{1}{27}\ \zeta_{3,9}\ri\}\ [M_9,M_3]+\fc{48}{691}\ \lf\{\ [M_9,M_3]-3\ [M_7,M_5]\ \ri\}\cr\crr
&\times\lf\{\ -\fc{117713}{2627625}\ \zeta_2^6+\fc{29}{216}\ \zeta_5\ \zeta_7-\fc{511}{48}\ \zeta_3\ \zeta_9+\fc{8669}{1296}\ \zeta_{3,9}+\fc{64}{9}\ 
\zeta_{\ov 3,\ov 9}\ \ri\}\cr\crr
&=\lf\{\fc{2}{9}\ \zeta_5\ \zeta_7+\fc{1}{27}\ \zeta_{3,9}\ri\}\ [M_9,M_3]+\fc{48}{691}\ \lf\{\ [M_9,M_3]-3\ [M_7,M_5]\ \ri\}\cr\crr
&\times\lf\{\ -\fc{5607853}{6081075}\ \zeta_2^6+\fc{5827}{1296}\ \zeta_5\ \zeta_7+
\fc{7}{486}\ \zeta_{3,9}-\fc{64}{27}\ A_{5,7}\ \ri\}\ .&\QQ}
$$
At weight $15$ in \Q\ one may get rid of the basis element $\zeta_{1,1,3,4,6}$ with even entries at the cost of the introducing the Euler sum $\zeta_{\ov3,\ov 5,\ov 7}:=\zeta(-3,-5,-7)$ \DataMine:
\eqnn\eulerxv
$$\eqalignno{
 \zeta_{1,1,3,4,6}&=
             \fc{16663}{11664}\ \zeta_{3,3,9}
       + \fc{150481}{68040}\ \zeta_{5,3,7}
       - \fc{20651486329}{4082400}\ \zeta_{15}
       + \fc{1903}{120}\ \zeta_7\ \zeta_{3,5}
       - \fc{101437}{38880}\ \zeta_5\ \zeta_{3,7}\cr
       &- \fc{1520827}{38880}\ \zeta_5^3
       + 10\ \zeta_3\ \zeta_{1,1,4,6}
       + \fc{162823}{3888}\ \zeta_3\ \zeta_{3,9}
       - \fc{93619}{1296}\ \zeta_3\ \zeta_5\ \zeta_7
       + \fc{3601}{48}\ \zeta_3^2\ \zeta_9\cr
       &- \fc{17}{20}\ \zeta_3^5
       + \fc{14}{5}\ \zeta_2\ \zeta_{3,5,5}
       - 2\ \zeta_2\ \zeta_{3,3,7}
       + \fc{31753363}{12960}\ \zeta_2 \zeta_{13}
       - \fc{21}{2}\ \zeta_2\ \zeta_5\ \zeta_{3,5}\cr
       &- 27\ \zeta_2\ \zeta_3\ \zeta_{3,7}
       - \fc{61}{2}\ \zeta_2\ \zeta_3\ \zeta_5^2
       - 84\ \zeta_2\ \zeta_3^2\ \zeta_7
       - 4\ \zeta_2^2\ \zeta_{3,3,5}
       + \fc{979621}{1701}\ \zeta_2^2\ \zeta_{11}\cr
       &- 5\ \zeta_2^2\ \zeta_3\ \zeta_{3,5}
       + \fc{9}{2}\ \zeta_2^2\ \zeta_3^2\ \zeta_5
       - \fc{490670609}{3572100}\ \zeta_2^3\ \zeta_9
       + \fc{186}{35}\ \zeta_2^3\ \zeta_3^3
       - \fc{1455253}{283500}\ \zeta_2^4\ \zeta_7\cr
       &+ \fc{4049341}{311850}\ \zeta_2^5\ \zeta_5
       + \fc{12073102}{1488375}\ \zeta_2^6\ \zeta_3
 +\fc{1408}{81}\ A_{3,5,7}\ .\ &\eulerxv}
$$
More precisely, with the relations \eulerxv\ and \eulerxiic\ the combination 
$\zeta_3\zeta_{1,1,4,6}+\zeta_{1,1,3,4,6}$ can be eliminated to cast the weight $15$ part $Q_{15}$ in terms
of Euler sums in a minimal depth representation and positive odd indices only:  
\eqnn\QQQQ
$$\eqalignno{
Q_{15}&=
\lf\{\fc{1339}{30}\ \zeta_2\ \zeta_{13}+\fc{128}{45}\ \zeta_2^2\ \zeta_{11}-
\fc{236}{4725}\ \zeta_2^3\ \zeta_9-\fc{184}{2625}\ \zeta_4^2\ \zeta_7-\fc{64}{5775}\ \zeta_2^5\ \zeta_5\ri.\cr\crr
&\lf.-\fc{2}{45}\ \zeta_5^3-\fc{1}{15}\ \zeta_7\ \zeta_{3,5}-\fc{2}{45}\ \zeta_5\ \zeta_{3,7}+\fc{1}{27}\ \zeta_{3,3,9}\ri\}\ [M_3,[M_9,M_3]]\cr\cr
&+\lf\{-\fc{143}{20}\ \zeta_2\ \zeta_{13}-\fc{11}{35}\ \zeta_2^2\ \zeta_{11}+\fc{68}{1225}\ \zeta_2^3\ \zeta_9+\fc{11}{70}\ \zeta_5^3+\fc{24}{875}\ \zeta_2^4\ \zeta_7+
\fc{48}{13475}\ \zeta_2^5\ \zeta_5\ri.\cr\crr
&\lf.+\fc{1}{5}\ \zeta_7\ \zeta_{3,5}+\fc{3}{35}\ \zeta_5\ \zeta_{3,7}-\fc{1}{70}\ \zeta_{5,3,7}\ri\}\ [M_5,[M_7,M_3]]+\fc{2}{15}\ \zeta_{5,3,7}\ [M_3,[M_7,M_5]]\cr\crr
&+\fc{48}{7601}\ \lf\{\fc{1408}{81}\  A_{3,5,7}-\fc{704}{27}\  A_{5,7}\ \zeta_3-\fc{20651486329}{4082400}\  \zeta_{15}+\fc{1149577 }{5184}\ \zeta_2\ \zeta_{13}
\ri.\cr\crr
&+\fc{1912097}{136080}\ \zeta_2^2\ \zeta_{11}-\fc{230351}{357210}\  \zeta_2^3\ \zeta_9-\fc{414007}{283500}\  \zeta_2^4\  \zeta_7-\fc{45779}{39690}\  \zeta_2^5\ \zeta_5-\fc{24257}{3869775}\  \zeta_2^6\  \zeta_3\cr\crr
&+\fc{77}{648}\ \zeta_3\ \zeta_5\ \zeta_7+\fc{77}{3888}\ \zeta_{3,9}\  \zeta_3+\fc{319}{3402}\  \zeta_5^3-\fc{15983}{54432}\  \zeta_{3,7}\ \zeta_5-\fc{781}{720}\  
\zeta_{3,5}\ \zeta_7\cr\crr
&\lf.+\fc{1995367}{272160}\  \zeta_{5,3,7}+\fc{1309}{11664}\ \zeta_{3,3,9}
\ri\}\ \Big\{\ [M_3,[M_9,M_3]]-3\ [M_3,[M_7,M_5]]\ \Big\}\ .&\QQQQ}
$$

\newsec{Motivic multiple zeta values}

In this section we want to compare our findings \VERYNICEE\ 
with the beautiful work of F. Brown on the decomposition of motivic multiple zeta values \Brown. 
For this purpose after reviewing some aspects of motivic MZVs 
we determine the decomposition of motivic MZVs for the weights $11$ until $16$.

\subsec{Motivic aspects of multiple zeta values}

An important question is to explicitly describe  the structure of the algebra $\Zc$,
which eventually allows to get a grip on all algebraic MZV identities over $\IQ$. 
For this purpose the actual MZVs \MZV\ are replaced by symbols (or motivic MZVs), 
which are elements of a certain algebra.

In this section we review some aspects of motivic MZVs \Brown.
The task is to lift the ordinary iterated integrals $I_\gamma$ given in \Integral\
to motivic versions $I^{m}$ such  that the standard relations are fulfilled. 
With an embedding $\si:F\hookrightarrow\IC$
the iterated integrals $I_\gamma$  can be upgraded to a framed mixed Tate motive
over $F$ (motivic iterated integral)
\eqn\Symbol{
I^{m}(a_0;a_1,\ldots,a_n;a_{n+1})\in\Hc(F)\ \ \ ,\ \ \ a_0,\ldots,a_{n+1}\in F\ ,}
with $p_\si(I^m(a_0;a_1,\ldots,a_n;a_{n+1}))=I(\si(a_0);\si(a_1),\ldots,\si(a_n);\si(a_{n+1}))$
\Goncharov\ and some number field $F$. The latter is a finite degree field extension of the field of rational numbers $\IQ$.
The symbols \Symbol\ are elements of a commutative graded Hopf algebra $\Hc(F)$:
\eqn\gradedalgebra{
\Hc=\bigoplus_{n\geq 0}\Hc_n\ .}
The Hopf algebra\foot{A Hopf algebra is an algebra $\Ac$ with multiplication 
$\mu: \Ac\otimes\Ac\ra\Ac$, \ie $\mu(x_1\otimes x_2)=x_1\cdot x_2$ and associativity. At the same time it is also a coalgebra with coproduct $\Delta: \Ac\ra\Ac\otimes\Ac$  and coassociativity such that the product and coproduct are compatible: $\Delta(x_1\cdot x_2)=\Delta(x_1)\cdot\Delta(x_2)$, with $x_1,x_2\in\Ac$.} $\Hc$ implies a product given by the shuffle product
\eqn\Schuffle{
I^m(x;a_1,\ldots,a_r;y)\ \cdot\ I^m(x;a_{r+1},\ldots,a_{r+s};y)=
\sum_{\si\in\Si(r,s)} I^m(x;a_{\si(1)},\ldots,a_{\si(r+s)};y) \ ,}
with $\Si(r,s)=\{\si\in\Si(r+s)\ | \ \si^{-1}(1)<\ldots<\si^{-1}(r) \cap
\si^{-1}(r+1)<\ldots<\si^{-1}(r+s) \}$ and $a_i,x,y\in\{0,1\}$
and the coproduct $\Delta$ acting on the elements $I^m$ as \Goncharov
\eqn\Coproduct{\eqalign{
\Delta\ I^m(a_0;a_1,\ldots,a_n;a_{n+1})&=\sum_{0=i_0<i_1<\ldots<i_k<i_{k+1}=n+1}
I^m(a_0;a_{i_1},\ldots,a_{i_k};a_{n+1})\cr
&\otimes\prod_{p=0}^kI^m(a_{i_p};a_{i_p+1},\ldots,a_{{i_{p+1}}-1};a_{i_{p+1}})\ ,}}
with $0\leq k\leq n$ and $a_i\in F$.
As in \integral\ by \Symbol\ with $a_i\in\{0,1\}$ we may define the motivic versions 
$\zeta^m_{n_1,\ldots,n_r}$ of the MZVs $\zeta_{n_1,\ldots,n_r}$, \ie 
by \Symbol\ the motivic MZVs are defined as
\eqn\motivicMZV{
\zeta^m_{n_1,\ldots,n_r}=(-1)^r\ I^{ m}(0;\rho(n_1,\ldots,n_r);1)\in\Hc_w(\IZ)\ ,}
with the weight $w=\sum_{l=1}^r n_l$ and $\rho$ given in \map.
Any symbol $I^{ m}(a_0;a_1,\ldots,a_n;a_{n+1})$, with $a_i\in\{0,1\}$,  can be reduced to a linear combination of elements of the form \motivicMZV, 
with  $n_i\geq 1,\ n_r\geq 2$ and $w=N$.
The dimension of the space of motivic MZVs of weight $k$ is equal to $d_k$, \ie 
$\dim_\IQ(\Hc_k)=d_k$.
The map $\Hc_k\ra\Zc_k$ is surjective, \ie $\dim_\IQ(\Zc_k)\leq \dim_\IQ(\Hc_k)=d_k$
\refs{\Deligne,\TATE}. By this certain identities between MZVs can be lifted to their motivic versions \Brown.

There is a non--canonical isomorphism\foot{Note, that in contrast to Ref. \Goncharov\ in this  setup 
$\zeta_2^m$ is non--zero.}
\eqn\isomorph{
\Hc\simeq \Ac\ \otimes_\IQ\ \IQ[\zeta^m_2]\ \ \ ,\ \ \ \Ac=\Hc\big\slash\zeta^m_2\Hc\ ,}
with the first factor graded by the weight, \ie  $\Ac=\bigoplus\limits_{n\geq 0}\Ac_n$. 

To explicitly describe the structure of $\Hc$ one  introduces the (trivial) algebra--comodule:
\eqn\introU{
\Uc=\IQ\vev{f_3,f_5,\ldots}\ \otimes_\IQ\ \IQ[f_2]\ .}
The first factor $\Uc'=\Uc\big\slash f_2\Uc$ is a cofree Hopf--algebra on the cogenerators $f_{2r+1}$ in degree $2r+1\geq 3$, whose basis consists of all non--commutative words in the $f_{2i+1}$. The multiplication on $\Uc'$ is given by the shuffle product $\shuffle$
\eqn\schuffle{
f_{i_1}\ldots f_{i_r}\shuffle f_{i_{r+1}}\ldots f_{i_{r+s}}=\sum_{\si\in\Si(r,s)}
f_{i_{\si(1)}}\ldots f_{i_{\si(r+s)}}\ ,}
with $\Si(r,s)$ given after \eqq \Schuffle. The Hopf--algebra  $\Uc'$ is 
isomorphic to the space of non--commutative polynomials in $f_{2i+1}$.
The element $f_2$ commutes with all $f_{2r+1}$. Again, there is a grading $\Uc_k$ on $\Uc$, with $\dim(\Uc_k)=d_k$.
Then, there exists a morphism $\phi$ of graded algebra--comodules
\eqn\morphism{
\phi:\ \Hc\lra\Uc\ ,}
normalized\foot{Note, that there is no canonical choice of $\phi$ and the latter depends  on the choice
of motivic generators of $\Hc$.} by:
\eqn\morphismres{
\phi\big(\zeta^m_n\big)=f_n\ \ \ ,\ \ \ n\geq 2\ .}
The map \morphism\ sends every motivic MZV to a non--commutative polynomial in the $f_i$.
Furthermore, \morphism\ respects the shuffle multiplication rule \schuffle:
\eqn\ruleshuffle{
\phi(x_1x_2)=\phi(x_1)\shuffle \phi(x_2)\ \ \ ,\ \ \ x_1,x_2\in\Hc\ .} 
It is believed, that the isomorphism $\Zc_k\simeq \Uc_k$ of graded algebras over $\IQ$ holds. 

The motivic MZVs have a hidden structure, which is revealed by  the action 
of motivic derivations. 
The latter are derived from the coaction
$\Delta: \Hc\ra \Ac\otimes_{\IQ}\Hc$ \refs{\TATE,\Brown}
\eqn\Coaction{\eqalign{
\Delta\ I^m(a_0;a_1,\ldots,a_n;a_{n+1})&=\sum_{0=i_0<i_1<\ldots<\atop<i_k<i_{k+1}=n+1}
\Pi\lf(\prod_{p=0}^kI^m(a_{i_p};a_{i_p+1},\ldots,a_{{i_{p+1}}-1};a_{i_{p+1}})\ri)\cr\crr
&\otimes I^m(a_0;a_{i_1},\ldots,a_{i_k};a_{n+1})\ ,}}
which represents a modification of the coproduct \Coproduct. Here, $\Pi$ is
the projector $\Pi:\Hc\ra\Ac$ acting on $\zeta_2^m$ as 
$\zeta^m_2{\buildrel\Pi\over\lra}0$. 
The derivations $D_r: \Hc_n\ra \Ac_r\otimes_\IQ \Hc_{n-r}{\buildrel\pi\otimes id\over\ra}\Lc_r\otimes_\IQ \Hc_{n-r}$ on $\Hc$ are defined as the infinitesimal version of the coaction \Coaction\ \Brown
\eqn\CovDer{\eqalign{
D_r\ I^m(a_0;a_1,\ldots,a_n;a_{n+1})&=\sum_{p=0}^{n-r}
\pi\lf(I^a(a_p;a_{p+1},\ldots,a_{p+r};a_{p+r+1})\ri)\cr\crr
&\otimes I^m(a_0;a_1,\ldots,a_p,a_{p+r+1},\ldots,a_n;a_{n+1})\ ,}}
with the projection $\pi:\Ac\ra\Lc$ onto the Lie coalgebra $\Lc=\fc{\Ac_{>0}}{\Ac_{>0}\Ac_{>0}}$
describing all indecomposable (irreducible) elements of $\Ac$. By this we have 
$D_{2r} I^m\equiv 0$.

\subsec{On the decomposition of motivic multi zeta values}
\def\MZ#1{\zeta^m_{#1}}

The coalgebra structure \introU\ underlying the motivic MZVs can be  used to decompose any MZV into a basis. Let us now describe the decomposition of motivic MZVs up to some weight $M\geq2$~\Brown.

We are looking for decompositions in the $\IQ$--vector space $\Hc_N,\ 2\leq N\leq M$ spanned by the symbols  \motivicMZV, with $w=N$ and $n_i\geq 1,\ n_r\geq 2$.
To check, that a (conjectural) polynomial basis $B$ of motivic MZVs 
$\bigoplus_{2\leq n\leq M} \Hc_n$ up to weight $M$ indeed represents a polynomial 
basis of motivic MZVs up to weight $M$ for $n\leq N$ for each set $B_n$ 
of elements of $B$ of weight $n$ one constructs the map \morphism:
\eqn\mapphi{
\phi:\ B_n\lra\ \Uc_n\ \ \ ,\ \ \ n\leq N\ .} 
This map  assigns to every element of our basis $B$ (of weight at most $N$) a $\IQ$--linear combination of monomials
\eqn\basisUn{
f_{2i_1+1}\ldots f_{2i_r+1}\ f_2^k\ \ \ ,\ \ \ r,k\geq 0,i_1,\ldots,i_r\geq 1\ ,\  
2\ (i_1+\ldots+i_r)+r+2k=n\ ,}
which are basis elements of the $\IQ$--vector space $\Uc_n$ supplemented by the multiplication rule $\shuffle:\ \Uc_m\times\Uc_n\ra \Uc_{n+m}$ given in \schuffle.
Actually, $\phi$ can be extended to the vector space~$\Hc_n$:
\eqn\Mapphi{
\phi:\ \Hc_n\lra\ \Uc_n\ \ \ ,\ \ \ n\leq N\ .} 

For the basis $B$ we must have: $\dim_\IQ(\vev{B}_N)=d_N,\ 2\leq N\leq M$, with 
$\vev{B}_N$ the $\IQ$--vector space spanned by monomials in the elements of $B$ of total additive weight $N$. Furthermore, we have
\eqn\basismotive{
B\supset B^0=\{\zeta^m_2\}\cup\{\zeta^m_3,\ldots,\zeta^m_{2r+1}\}\ ,}
with $r=\lfloor(M-1)/2\rfloor$. For the elements of $B^0$ the map $\phi$ is given by 
\morphismres.
For the remaining elements of $B$ the explicit construction of $\phi$ is performed inductively, \ie from \mapphi\ the case $n=N+1$ is determined.
To find $\phi(\xi)$ for a general $\xi\in B_{N+1}$, with 
$\xi=I^m(a_0;a_1,\ldots,a_{N+1};a_{N+2})$ according to \motivicMZV,
we need to compute the coefficients
\eqnn\Coeff{
$$\eqalignno{
\xi_{2r+1}&=\sum_{p=0}^{N-2r}
c_{2r+1}^\phi\lf(I^m(a_p;a_{p+1},\ldots,a_{p+2r+1};a_{p+2r+2})\ri)&\Coeff\cr \crr
&\times \phi\lf(I^m(a_0;a_1,\ldots,a_p,a_{p+2r+2},\ldots,a_{N+1};a_{N+2})\ri)
\in\Uc_{N-2r}\ \ \ ,\ \ \ 3\leq 2r+1\leq N}
$$
in the expansion:
\eqn\resphi{
\phi(\xi)=\sum_{3\leq 2r+1\leq N}f_{2r+1}\ \xi_{2r+1}\in\Uc_{N+1}\ .}
Above the operator $c_{2r+1}^\phi(\xi)$, with $\xi\in\Hc_{2r+1}$ determines the rational coefficient of $f_{2r+1}$ in the monomial $\phi(\xi)\in\Uc_{2r+1}$.
Note, that the right hand side of \Coeff\ only involves elements $I^m$ from $\Hc_{\leq N}$
for which $\phi$ has already been determined.

The above construction allows 
 to assign a $\IQ$--linear combination of monomials to every element
$\zeta^m_{n_1,\ldots,n_r}$. The map\foot{The choice of $\phi$ describes for each weight $2r+1$ the motivic derivation operators $\p_{2r+1}^\phi$  acting on the space of motivic MZVs $\p_{2r+1}^\phi:\Hc\ra\Hc$ \Brown
\eqn\motder{
\p_{2r+1}^\phi=(c_{2r+1}^\phi\otimes id) \circ D_{2r+1}\ ,}
with $D_{2n+1}$  given in \CovDer\  and the coefficient function $c_{2r+1}^\phi$, 
introduced above.
} $\phi$ sends every motivic MZV of weight less or equal
to $N$ to a non--commutative polynomial in the $f_i$'s.
Inverting this map gives the decomposition of $\zeta^m_{n_1,\ldots,n_r}$ w.r.t. the basis
$B_n$, with $n=\sum_{l=1}^rn_l$. 
In other words, the derivations \motder\  are used to detect elements 
in $\Uc$ and to decompose any motivic MZV $\xi$ into a candidate basis~$B$.

In \Brown\  the map \mapphi\ and the decomposition  are explicitly worked out
up to weight $10$.
{\it E.g.} one finds
\eqn\onefinds{
\phi(\MZ{3,5})=-5\ f_5f_3\ \ \ ,\ \ \ \phi(\MZ{3,7})=-14\ f_7f_3-6\ f_5f_5\ ,}
and  at weight 10 one has for $\xi_{10}\in\Hc_{10}$ the following decomposition
\eqnn\mdecox
$$\eqalignno{
\xi_{10}&=a_0\ (\MZ{2})^5+a_1\ (\MZ{2})^2\ (\MZ{3})^2+a_2\ \MZ{2}\ \MZ{3}\ \MZ{5}\cr
&+a_3\ (\MZ{5})^2+a_4\ \MZ{2}\ \MZ{3,5}+a_5\ \MZ{3}\ \MZ{7}+a_6\ \MZ{3,7}\ ,&\mdecox}
$$
with the operators:
\eqnn\mcoeffsx
$$\eqalignno{
a_1&=\h\ c_2^2\ \p_3^2,\ a_2=c_2\ \p_5\p_3,\ a_3=\h\ \p_5^2+
\fc{3}{14}\ [\p_7,\p_3]\ ,\cr\crr
a_4&=\fc{1}{5}\ c_2\ [\p_5,\p_3],\ a_5=\p_7\p_3,\ a_6=\fc{1}{14}\ [\p_7,\p_3]\ .&\mcoeffsx}
$$
acting on $\phi(\xi_{10})$.
The derivation operators $\p_{2n+1}:\Uc\ra\Uc$  are defined as \Brown:
\eqn\defder{
\p_{2n+1}(f_{i_1}\ldots f_{i_r})=\cases{f_{i_2}\ldots f_{i_r}\ , & $i_1=2n+1\ ,$\cr
                                        0\ , & otherwise\ ,\cr}}
with $\p_{2n+1} f_2=0$.
Furthermore, we have the product rule for the shuffle product:
\eqn\Leibniz{
\p_{2n+1}(a\shuffle b)=\p_{2n+1} a\shuffle b+a\shuffle \p_{2n+1} b\ \ \ ,\ \ \ a,b\in \Uc'\ .}
Finally, $c_2^n$ takes the coefficient of $f_2^n$.

It seems very amusing, that the coefficients \mcoeffsx\ and the commutator structure
agree exactly with  \testb. Therefore, MZVs encapsulate 
the $\ap$--expansion of the 
open superstring amplitude.

\subsec{Decomposition of motivic multi zeta values for weights $11$ through $16$}

In order to bolster this connection, in the following subsections we determine
the decompositions $\xi_w$ of any motivic MZV for the weights $11\leq w\leq 16$.

For a given weight $w$ we proceed as described in \Brown: in lines of the Tables 1--3 at weight~$w$ we first detect the new elements $B_w$  to be 
added to constitute the conjectural basis $B$ up to weight $w$.
For these new elements $B_w$ we then compute their coefficients \Coeff\ or motivic derivations $\p_{2r+1}^\phi$ by applying the relations $(R0)-(R4)$ given in section 5.1 of \Brown. Equipped with these results we then determine the map  \resphi\ by using the 
findings from the lower weights. After having derived the map \resphi\ for all $d_w$ basis elements 
of $\vev{B}_w$ we can construct the basis for $\Uc_w$ and eventually the operator $\xi_w$.

For the depth two case $\MZ{n_1,n_2}$
there exists a closed formula, which computes the map 
$\phi(\MZ{n_1,n_2})$, directly \FB. Our results for
$\phi(\MZ{3,9}),\phi(\MZ{3,11}),\phi(\MZ{5,9}),\phi(\MZ{3,13})$ and $\phi(\MZ{5,11})$
agree with what this  formula gives. However, as it will become clear in the following,
beyond depth two the computations involve new aspects and become rather involved.

\subsubsec{Decomposition at weight $11$}

At weight $11$ we take the following set of motivic MZVs
\eqn\Bxi{
B=\{\ \MZ{2},\ \MZ{3},\ \MZ{5},\ \MZ{7},\ \MZ{3,5},\ \MZ{9},\ \MZ{3,7},\ \MZ{11},\ \MZ{3,3,5}\ \}}
as independent algebra generators up to weight $11$.
In \Brown\ up to weight $n\leq 10$ 
to each element of $B$  an element of $\Uc$ is associated by the map $\phi$ given in \mapphi.
Hence, we only need to compute $\phi(\zeta_{3,3,5}^m)$, which according to \Coeff\ 
requires  the following derivatives:
\eqnn\needxi
$$\eqalign{
\p^\phi_3\MZ{3,3,5}&=0\ ,\cr\crr
\p^\phi_7\MZ{3,3,5}&=-\fc{6}{5}\ (\MZ{2})^2\ ,\cr}
\qquad 
\eqalign{
\p^\phi_5\MZ{3,3,5}&=-5\ \MZ{3,3}=-\fc{5}{2}\ (\MZ{3})^2+\fc{4}{7}\ (\MZ{2})^3\ ,\cr\crr
\p^\phi_9\MZ{3,3,5}&=-45\ \MZ{2}\ .\cr}
\eqno \hbox{\needxi}$$
From these results the expression \resphi\ gives rise to:
\eqn\verenaxi{
\phi(\zeta_{3,3,5}^m)=-\fc{5}{2}\ f_5(f_3\shuffle f_3)+\fc{4}{7}\ f_5f_2^3-
\fc{6}{5}\ f_7f_2^2-45\ f_9f_2\ .}
Gathering the information about the lower weight basis $\Uc_{k\leq 10}$ with \verenaxi\ 
we can construct  the following basis for $\Uc_{11}$:
\eqnn\basisxi
$$\eqalignno{
&-\fc{5}{2}\ f_5(f_3\shuffle f_3)+\fc{4}{7}\ f_5f_2^3-\fc{6}{5}\ f_7f_2^2-45\ f_9f_2\ ,\cr\crr 
&-5\ (f_5f_3)\shuffle f_3,\ f_{11},\ f_3\shuffle f_3\shuffle f_5,\ 
f_3\shuffle f_3\shuffle f_3 f_2\ ,\cr\crr
&f_9f_2,\ f_7f_2^2,\ f_5f_2^3,\ f_3f_2^4\ .&\basisxi}
$$
This basis gives rise to the following decomposition of any motivic MZV $\xi_{11}$ of weight $11$
\eqnn\mdecoxi
$$\eqalignno{
\xi_{11}&=a_1\ \MZ{3,3,5}+a_2\ \MZ{3,5}\ \MZ{3}+a_3\ \MZ{11}+a_4\ (\MZ{3})^2\ \MZ{5}+a_5\ \MZ{2}\ (\MZ{3})^3\cr
&+a_6\ \MZ{2}\ \MZ{9}+a_7\ (\MZ{2})^2\ \MZ{7}+a_8\ (\MZ{2})^3\ \MZ{5}+a_9\ 
(\MZ{2})^4\ \MZ{3}&\mdecoxi}
$$
with\foot{The following  relations $[\p_3,[\p_5,\p_3]]f_3\shuffle f_3\shuffle f_5=0$ and $[\p_3,[\p_5,\p_3]]f_5f_3\shuffle f_3 =0$ are useful. More generally, we have:
$[\p_a,[\p_b,\p_c]]f_a\shuffle f_b\shuffle f_c=0$ and 
$[\p_a,[\p_b,\p_a]]f_bf_a\shuffle f_a =0$.} the following operators
\eqnn\mcoeffsxi
$$\eqalignno{
a_1&=\fc{1}{5}\ [\p_3,[\p_5,\p_3]],\ a_2=\fc{1}{5}\ [\p_5,\p_3]\p_3\ ,\cr\crr
a_3&=\p_{11},\ a_4=\h\ \p_5\p_3^2,\ a_5=\fc{1}{6}\ c_2\ \p_3^3\ ,\cr
a_6&=c_2\ \p_9+9\ [\p_3,[\p_5,\p_3]],\ a_7=c_2^2\ \p_7+\fc{6}{25}\ [\p_3,[\p_5,\p_3]]\ ,\cr 
a_8&=c_2^3\ \p_5-\fc{4}{35}\ [\p_3,[\p_5,\p_3]],\ a_9=c_2^4\ \p_3\ &\mcoeffsxi}
$$
acting on $\phi(\xi_{11})$.

\subsubsec{Decomposition at weight $12$}

Next, at weight $12$ we take the set of motivic MZVs
\eqn\Bxiii{
B=\{\ \MZ{2},\ \MZ{3},\ \MZ{5},\ \MZ{7},\ \MZ{3,5},\ \MZ{9},\ \MZ{3,7},\ \MZ{11},\ \MZ{3,3,5},\ \MZ{3,9},\ \MZ{1,1,4,6}\ \}}
as independent algebra generators up to weight $12$.
We  need to compute $\phi(\zeta_{3,9}^m)$ and  $\phi(\zeta_{1,1,4,6}^m)$, which require  
the following derivatives
\eqnn\needxii
$$\eqalign{
\p^\phi_3\MZ{3,9}&=0\ ,\cr\crr
\p^\phi_5\MZ{3,9}&=-6\ \MZ{7}\ ,\cr}
\qquad 
\eqalign{
\p^\phi_7\MZ{3,9}&=-15\ \MZ{5}\ ,\cr
\p^\phi_9\MZ{3,9}&=-27\ \MZ{3}\ ,\cr}
\eqno \hbox{\needxii}$$
and 
\eqnn\needxiia
$$\eqalign{
\p^\phi_3\MZ{1,1,4,6}&=\fc{1}{3}\ (\zeta_3^m)^3-\fc{799}{72}\ \zeta^m_9+10\ \zeta^m_7\zeta^m_2
-\fc{1}{5}\ \zeta^m_5(\zeta^m_2)^2-\fc{36}{35}\ \zeta^m_3(\zeta^m_2)^3\ ,\cr\crr
\p^\phi_5\MZ{1,1,4,6}&=29\ \MZ{7}-11\ \MZ{5}\MZ{2}-\fc{16}{5}\ \MZ{3}(\MZ{2})^2\ ,\cr\crr
\p^\phi_7\MZ{1,1,4,6}&=\fc{1133}{16}\ \MZ{5}-32\ \MZ{3}\MZ{2}\ ,\cr\crr
\p^\phi_9\MZ{1,1,4,6}&=\fc{1799}{18}\ \MZ{3}\ ,\cr}
\eqno \hbox{\needxiia}$$
respectively.
With the derivatives \needxii\ and \needxiia\ we determine the following maps:
\eqn\verenaxii{\eqalign{
\phi(\zeta_{3,9}^m)&=-6\ f_5f_7-15\ f_7f_5-27\ f_9f_3\ ,\cr\crr
\phi(\zeta_{1,1,4,6}^m)&=\fc{1799}{18}\ f_9f_3-32\ f_7f_3f_2+\fc{1133}{16}\ f_7f_5+29\ f_5f_7-11\ f_5^2f_2-\fc{16}{5}\ f_5f_3f_2^2\cr\crr
&+\fc{1}{3}\ f_3 (f_3\shuffle f_3\shuffle f_3)-\fc{799}{72}\ f_3f_9+10\ f_3f_7f_2-\fc{1}{5}\ f_3f_5f_2^2-\fc{36}{35}\ f_3^2f_2^3\ .}}
Inspecting  the lower weight basis $\Uc_{k\leq 12}$ with \verenaxii\ 
we have  the following basis for $\Uc_{12}$:
\eqnn\basisxii
$$\eqalignno{
&\fc{1799}{18}\ f_9f_3-32\ f_7f_3f_2+\fc{1133}{16}\ f_7f_5+29\ f_5f_7-11\ f_5^2f_2-\fc{16}{5}\ f_5f_3f_2^2\cr\crr
&+\fc{1}{3}\ f_3 (f_3\shuffle f_3\shuffle f_3)-\fc{799}{72}\ f_3f_9+10\ f_3f_7f_2-\fc{1}{5}\ f_3f_5f_2^2-\fc{36}{35}\ f_3^2f_2^3,\cr\crr
&-6\ f_5f_7-15\ f_7f_5-27\ f_9f_3,\ f_3\shuffle f_9,\ f_5\shuffle f_7,\ f_3\shuffle f_3
\shuffle f_3\shuffle f_3,\cr\crr
&(-14f_7f_3-6f_5^2)f_2,\ -5\ f_5f_3f_2^2,\ f_5\shuffle f_5f_2,\ f_3\shuffle f_7f_2,\cr\crr&f_3\shuffle f_5f_2^2,\ f_3\shuffle f_3f_2^3,\ f_2^6\ .&\basisxii}
$$
Therefore, the decomposition of any motivic MZV $\xi_{12}$ of weight $12$ assumes the 
form
\eqnn\mdecoxii
$$\eqalignno{
\xi_{12}&=a_1\ \MZ{1,1,4,6}+a_2\ \MZ{3,9}+a_3\ \MZ{9}\ \MZ{3}+a_4\ \MZ{7}\ \MZ{5}+
a_5\ (\MZ{3})^4+a_6\ \MZ{3,7}\ \MZ{2}\cr\crr
&+a_7\ \MZ{3,5}\ (\MZ{2})^2+a_8\ (\MZ{5})^2\ \MZ{2}+a_9\ \MZ{7}\ \MZ{3}\ \MZ{2}+
a_{10}\ \MZ{5}\ \MZ{3}\ (\MZ{2})^2\cr\crr
&+a_{11}\ (\MZ{3})^2\ (\MZ{2})^3+a_{12}\ (\MZ{2})^6\ ,&\mdecoxii}
$$
with the following operators
\eqnn\mcoeffsxii
$$\eqalignno{
a_1&=\fc{48}{691}\ \lf([\p_9,\p_3]-3\ [\p_7,\p_5]\ri),\ 
a_2=\fc{1}{27}\ [\p_9,\p_3]+\fc{2665}{648}\ a_1,\cr\crr
a_3&=\p_9\p_3+\fc{799}{72}\ a_1,\ a_4=\p_7\p_5+\fc{2}{9}\ [\p_9,\p_3]-
\fc{467}{108}\ a_1,\ a_5=\fc{1}{24}\ \p_3^4-\fc{1}{12}\ a_1,\cr\crr
a_6&=\fc{1}{14}\ c_2\ [\p_7,\p_3]-3\ a_1,\ a_7=\fc{1}{5}\ c_2^2\ [\p_5,\p_3]-\fc{3}{5}\ a_1,\cr\crr
a_8&=c_2\ \lf(\h\ \p_5^2+\fc{3}{14}\ [\p_7,\p_3]\ri)-\fc{7}{2}\ a_1,\ 
a_9=c_2\ \p_7\p_3-10\ a_1,\cr\crr
a_{10}&=c_2^2\ \p_5\p_3+\fc{1}{5}\ a_1,\ a_{11}=\h\ c_2^3\ \p_3^2+\fc{18}{35}\ a_1,
\ a_{12}=c_2^6&\mcoeffsxii}
$$
acting on $\phi(\xi_{12})$.

\subsubsec{Decomposition at weight $13$}

At weight $13$  the following set of motivic MZVs
\eqn\Bxiii{
B=\{\ \MZ{2},\ \MZ{3},\ \MZ{5},\ \MZ{7},\ \MZ{3,5},\ \MZ{9},\ \MZ{3,7},\ \MZ{11},\ \MZ{3,3,5},\ \MZ{3,9},\ \MZ{1,1,4,6},\ \MZ{3,3,7},\ \MZ{3,5,5}\ \}}
represents independent algebra generators up to weight $13$.
We  need to compute $\phi(\zeta_{3,3,7}^m)$ and  $\phi(\zeta_{3,5,5}^m)$, which require  
the following derivatives
\eqnn\needxiii
$$\eqalign{
\p^\phi_3\MZ{3,3,7}&=0\ ,\cr\crr
\p^\phi_5\MZ{3,3,7}&=-6\ \MZ{3,5}\ ,\cr\crr
\p^\phi_7\MZ{3,3,7}&=-7\ (\MZ{3})^2+\fc{32}{35}\ (\MZ{2})^3\ ,\cr}
\qquad 
\eqalign{
\p^\phi_9\MZ{3,3,7}&=-\fc{56}{5}\ (\MZ{2})^2\ ,\cr\crr
\p^\phi_{11}\MZ{3,3,7}&=-\fc{407}{2}\ \MZ{2}\ ,\cr}
\eqno \hbox{\needxiii}$$
and
\eqnn\needxiiia
$$\eqalign{
\p^\phi_3\MZ{3,5,5}&=0\ ,\cr
\p^\phi_5\MZ{3,5,5}&=-5\ \MZ{3,5}\ ,\cr
\p^\phi_7\MZ{3,5,5}&=0\ ,\cr}
\qquad 
\eqalign{
\p^\phi_9\MZ{3,5,5}&=-10\ (\MZ{2})^2\ ,\cr
\p^\phi_{11}\MZ{3,5,5}&=-\fc{275}{2}\ \MZ{2}\ ,\cr}
\eqno \hbox{\needxiiia}$$
respectively.
The derivatives \needxiii\ and \needxiiia\ give rise to the maps:
\eqn\verenaxiii{\eqalign{
\phi(\zeta_{3,3,7}^m)&=30\ f_5^2f_3-7\ f_7(f_3\shuffle f_3)
+\fc{32}{35}\ f_7f_2^3-\fc{56}{5}\ f_9f_2^2-\fc{407}{2}\ f_{11}f_2\ ,\cr\crr
\phi(\zeta_{3,5,5}^m)&=25\ f_5^2f_3-10\ f_9f_2^2-\fc{275}{2}\ f_{11}f_2\ .}}
Collecting the information about the lower weight basis $\Uc_{k\leq 13}$ with \verenaxiii\ 
we have  the following basis for $\Uc_{13}$:
\eqnn\basisxiii
$$\eqalignno{
&30\ f_5^2f_3-7\ f_7(f_3\shuffle f_3)
+\fc{32}{35}\ f_7f_2^3-\fc{56}{5}\ f_9f_2^2-\fc{407}{2}\ f_{11}f_2\ ,\cr\crr 
&25\ f_5^2f_3-10\ f_9f_2^2-\fc{275}{2}\ f_{11}f_2,\ f_{13},\ (-14f_7f_3-6f_5^2)\shuffle f_3\ ,\cr\crr
&-5\ (f_5f_3)\shuffle f_5,\ f_7\shuffle f_3\shuffle f_3,\ f_5\shuffle f_5\shuffle f_5,&\basisxiii\cr\crr
&-\fc{5}{2}\ f_5(f_3\shuffle f_3)f_2+\fc{4}{7}\ f_5f_2^4-\fc{6}{5}\ f_7f_2^3-45\ f_9f_2^2,\ 
-5(f_5f_3)\shuffle f_3f_2,\cr\crr
&f_{11}f_2,\ f_5\shuffle f_3\shuffle f_3f_2,\ f_3\shuffle f_3\shuffle f_3 f_2^2,\ f_9f_2^2,\ 
f_7f_2^3,\ f_5f_2^4,\ f_3f_2^5\ .}
$$
Therefore, we have the following decomposition of any motivic MZV $\xi_{13}$ of weight $13$:
\eqnn\mdecoxiii
$$\eqalignno{
\xi_{13}&=a_1\ \MZ{3,3,7}+a_2\ \MZ{3,5,5}+a_3\ \MZ{13}+a_4\ \MZ{3,7}\ \MZ{3}+a_5\ \MZ{3,5}\ \MZ{5}+a_6\ \MZ{7}(\MZ{3})^2&\mdecoxiii\cr 
&+a_7\ (\MZ{5})^2\MZ{3}+a_8\ \MZ{3,3,5}\ \MZ{2}+a_9\ \MZ{3,5}\ \MZ{3}\ \MZ{2}+a_{10}\ \MZ{11}\ \MZ{2}+a_{11}\ \MZ{5}\ (\MZ{3})^2\ \MZ{2}\cr
&+a_{12}\ (\MZ{3})^3\ (\MZ{2})^2+a_{13}\ \MZ{9}\ (\MZ{2})^2+a_{14}\ \MZ{7}\ (\MZ{2})^3
+a_{15}\ \MZ{5}\ (\MZ{2})^4+a_{16}\ \MZ{3}\ (\MZ{2})^5\ ,}
$$
with the following operators
\eqnn\mcoeffsxiii
$$\eqalignno{
a_1&=\fc{1}{14}\ [\p_3,[\p_7,\p_3]],\ 
a_2=\fc{1}{25}\ [\p_5,[\p_5,\p_3]]-\fc{3}{35}\ [\p_3,[\p_7,\p_3]],\ a_3=\p_{13},\cr\crr
a_4&=\fc{1}{14}\ [\p_7,\p_3]\p_3,\ a_5=\fc{1}{5}\ \p_5[\p_5,\p_3],\ 
a_6=\h\p_7\p_3^2\ ,a_7=\fc{3}{14}\ [\p_7,\p_3]\p_3+\h\ \p_5^2\p_3,\cr\crr
a_8&=\fc{1}{5}\ c_2\ [\p_3,[\p_5,\p_3]],\ a_9=\fc{1}{5}\ c_2[\p_5,\p_3]\p_3,\cr\crr 
a_{10}&=c_2\p_{11}+\fc{11}{2}\ [\p_5,[\p_5,\p_3]]+\fc{11}{4}\ [\p_3,[\p_7,\p_3]],\ 
a_{11}=\h\ c_2\p_5\p_3^2,\ a_{12}=\fc{1}{6}\ c_2^2\p_3^3,\cr\crr 
a_{13}&=c_2^2\p_{9}+9\ c_2[\p_3,[\p_5,\p_3]]+\fc{2}{5}\ [\p_5,[\p_5,\p_3]]-\fc{2}{35}\ [\p_3,[\p_7,\p_3]],\cr\crr
a_{14}&=c_2^3\p_{7}+\fc{6}{25}\ c_2[\p_3,[\p_5,\p_3]]-\fc{16}{245}\ [\p_3,[\p_7,\p_3]],\cr\crr
a_{15}&=c_2^4\p_{5}-\fc{4}{35}\ c_2[\p_3,[\p_5,\p_3]],\ a_{16}=c_2^5\p_3&\mcoeffsxiii}
$$
acting on $\phi(\xi_{13})$.

\subsubsec{Decomposition at weight $14$}

At weight $14$ we  take the following set of motivic MZVs
\eqn\Bxiv{\eqalign{
B=\{\ &\MZ{2},\ \MZ{3},\ \MZ{5},\ \MZ{7},\ \MZ{3,5},\ \MZ{9},\ \MZ{3,7},\ \MZ{11},\ \MZ{3,3,5},\ \MZ{3,9},\ \MZ{1,1,4,6},\ \MZ{3,3,7},\ \MZ{3,5,5},\cr 
&\MZ{3,3,3,5},\ \MZ{3,11},\ \MZ{5,9}\ \}}}
as independent algebra generators up to weight $14$.
Hence, we only need to compute the maps $\phi(\zeta_{3,11}^m),\ \phi(\zeta^m_{5,9})$ 
and $\phi(\zeta_{3,3,3,5}^m)$, which require  the following derivatives
\eqnn\needxiv
$$\eqalign{
\p^\phi_3\MZ{3,11}&=0\ ,\cr\crr
\p^\phi_5\MZ{3,11}&=-6\ \MZ{9}\ ,\cr\crr
\p^\phi_7\MZ{3,11}&=-15\ \MZ{7}\ ,\cr}
\qquad 
\eqalign{
\p^\phi_9\MZ{3,11}&=-28\ \MZ{5}\ ,\cr\crr
\p^\phi_{11}\MZ{3,11}&=-44\ \MZ{3}\ ,\cr}
\eqno \hbox{\needxiv}$$
and
\eqnn\needxiva
$$\eqalign{
\p^\phi_3\MZ{5,9}&=0\ ,\cr
\p^\phi_5\MZ{5,9}&=0\ ,\cr
\p^\phi_7\MZ{5,9}&=-15\ \MZ{7}\ ,\cr}
\qquad 
\eqalign{
\p^\phi_9\MZ{5,9}&=-69\ \MZ{5}\ ,\cr
\p^\phi_{11}\MZ{3,5,5}&=-165\ \MZ{3}\ ,\cr}
\eqno \hbox{\needxiva}$$
and
\eqnn\needxivb
$$\eqalign{
\p^\phi_3\MZ{3,3,3,5}&=0\ ,\cr\crr
\p^\phi_5\MZ{3,3,3,5}&=-\fc{5}{6}\ (\MZ{3})^3-\fc{5}{3}\ \MZ{9}+\fc{4}{7}\ \MZ{3}\ (\MZ{2})^3\ ,\cr\crr
\p^\phi_7\MZ{3,3,3,5}&=-51\ \MZ{7}+30\ \MZ{5}\ \MZ{2}\ ,\cr}
\qquad 
\eqalign{
\p^\phi_9\MZ{3,3,3,5}&=-\fc{405}{2}\ \MZ{5}+90\ \MZ{3}\ \MZ{2}\ ,\cr\crr
\p^\phi_{11}\MZ{3,3,3,5}&=-15\ \MZ{3}\ ,\cr}
\eqno \hbox{\needxivb}$$
respectively.
These derivatives give rise to :
\eqn\verenaxiv{\eqalign{
\phi(\zeta^m_{3,11})&=-6\ f_5f_9-15\ f_7^2-28\ f_9f_5-44\ f_{11}\ f_3\ ,\cr
\phi(\zeta^m_{5,9})&=-15\ f_7^2-69\ f_9f_5-165\ f_{11}\ f_3\ ,\cr
\phi(\zeta_{3,3,3,5}^m)&=-\fc{5}{6}\ f_5\ (f_3\shuffle f_3\shuffle f_3)-\fc{5}{3}\ f_5f_9+\fc{4}{7}\ f_5f_3f_2^3-51 f_7^2\cr
&+30\ f_7f_5f_2-\fc{405}{2}\ f_9f_5+90\ f_9f_3f_2-15\ f_{11}f_3\ ,}}
respectively.
Gathering the information about the lower weight basis $\Uc_{k\leq 13}$ with \verenaxiv\ 
we can construct the basis for $\Uc_{14}$ displayed in (A.1).
This basis (A.1) gives rise to the following decomposition of any motivic MZV $\xi_{14}$ of weight $14$
\eqnn\mdecoxiv
$$\eqalignno{
\xi_{14}&=a_1\ \MZ{3,3,3,5}+a_2\ \MZ{3,11}+a_3\ \MZ{5,9}+a_4\ \MZ{3,3,5}\ \MZ{3}+a_5\ 
\MZ{3,5}\ (\MZ{3})^2+a_6\ \MZ{3}\ \MZ{11}\cr
&+a_7\ (\MZ{3})^3\ \MZ{5}+a_8\ \MZ{5}\ \MZ{9}+a_9\ (\MZ{7})^2+a_{10}\ \MZ{1,1,4,6}\ \MZ{2}
+a_{11}\ \MZ{3,9}\ \MZ{2}\cr
&+a_{12}\ \MZ{3}\ \MZ{9}\ \MZ{2}+a_{13}\ \MZ{5}\ \MZ{7}\ \MZ{2}+a_{14}\ (\MZ{3})^4\ \MZ{2}
+a_{15}\ \MZ{3,7}\ (\MZ{2})^2\cr
&+a_{16}\ \MZ{3,5}\ (\MZ{2})^3+a_{17}\ (\MZ{5})^2\ (\MZ{2})^2+a_{18}\ \MZ{7}\ \MZ{3}\ (\MZ{2})^2+a_{19}\ \MZ{5}\ \MZ{3}\ (\MZ{2})^3\cr
&+a_{20}\ (\MZ{3})^2\ (\MZ{2})^4+a_{21}\ (\MZ{2})^7&\mdecoxiv}
$$
with the operators $a_i$ acting on $\phi(\xi_{14})$ and given in (A.2).

\subsubsec{Decomposition at weight $15$}

At weight $15$ we have the following set of motivic MZVs
\eqn\Bxv{\eqalign{
B=\{\ &\MZ{2},\ \MZ{3},\ \MZ{5},\ \MZ{7},\ \MZ{3,5},\ \MZ{9},\ \MZ{3,7},\ \MZ{11},\ \MZ{3,3,5},\ \MZ{3,9},\ \MZ{1,1,4,6},\ \MZ{3,3,7},\ \MZ{3,5,5},\cr 
&\MZ{3,3,3,5},\ \MZ{3,11},\ \MZ{5,9},\ \MZ{5,3,7},\ \MZ{3,3,9},\ \MZ{1,1,3,4,6} \}}}
as independent algebra generators up to weight $15$.
Hence, we only need to compute the maps $\phi(\zeta_{5,3,7}^m),\ \phi(\zeta^m_{3,3,9})$ 
and $\phi(\zeta_{1,1,3,4,6}^m)$, which require  the following derivatives
\eqnn\needxv
$$\eqalign{
\p^\phi_3\MZ{5,3,7}&=0\ ,\cr\crr
\p^\phi_5\MZ{5,3,7}&=-3\ (\MZ{5})^2+\fc{96}{385}\ (\MZ{2})^5+6\ \MZ{3,7}\ ,\cr\crr
\p^\phi_7\MZ{5,3,7}&=-14\ \MZ{5,3}=-14\ \MZ{3}\ \MZ{5}+14\ \MZ{3,5}
+\fc{48}{25}\ (\MZ{2})^4\ ,\cr}
\qquad 
\eqalign{
\p^\phi_9\MZ{5,3,7}&=\fc{136}{35}\ (\MZ{2})^3\ ,\cr\crr
\p^\phi_{11}\MZ{5,3,7}&=-22\ (\MZ{2})^2\ ,\cr\crr
\p^\phi_{13}\MZ{5,3,7}&=-\fc{1001}{2}\ \MZ{2}}
\eqno \hbox{\needxv}$$
and
\eqnn\needxva
$$\eqalign{
\p^\phi_3\MZ{3,3,9}&=0\ ,\cr\crr
\p^\phi_5\MZ{3,3,9}&=-6\ \MZ{3,7}\ ,\cr\crr
\p^\phi_7\MZ{3,3,9}&=-\fc{72}{175}\ (\MZ{2})^4-15\ \MZ{3,5}\ ,\cr\crr}
\qquad 
\eqalign{
\p^\phi_9\MZ{3,3,9}&=-\fc{27}{2}\ (\MZ{3})^2-\fc{116}{35}\ (\MZ{2})^3\ ,\cr\crr
\p^\phi_{11}\MZ{3,3,9}&=-\fc{252}{5}\ (\MZ{2})^2\ ,\cr\crr
\p^\phi_{13}\MZ{3,3,9}&=-\fc{1209}{2}\ \MZ{2}\ ,\cr\crr}
\eqno \hbox{\needxva}$$
and
\eqnn\needxvb
$$\eqalignno{
\p^\phi_3\MZ{1,1,3,4,6}&=\fc{74}{3}\ \MZ{5}\MZ{7}-83\ \MZ{3}\MZ{9}-\fc{29}{9}\ \MZ{3,9}-\MZ{1,1,4,6}+6\ \MZ{3,7}\ \MZ{2}+\fc{8}{5}\ \MZ{3,5}\ (\MZ{2})^2\cr\crr
&+8\ (\MZ{5})^2\ \MZ{2}+42\ \MZ{3}\MZ{7}\MZ{2}+\fc{24}{5}\ \MZ{3}\MZ{5}(\MZ{2})^2-
\fc{1451972}{716625}\ (\MZ{2})^6\cr\crr
\p^\phi_5\MZ{1,1,3,4,6}&=-\fc{12263}{112}\ (\MZ{5})^2-\fc{245}{2}\ \MZ{3}\MZ{7}+\fc{145}{112}\ \MZ{3,7}-\fc{25}{2}\ \MZ{3,5}\MZ{2}+\fc{87}{2}\ \MZ{3}\MZ{5}\MZ{2}\cr\crr
&+\fc{15}{2}\ (\MZ{3})^2(\MZ{2})^2+\fc{19939}{1617}\ (\MZ{2})^5 \ ,\cr\crr
\p^\phi_7\MZ{1,1,3,4,6}&=\fc{31}{4}\ \MZ{3}\MZ{5}+\fc{481}{20}\ \MZ{3,5}
-12\ (\MZ{3})^2\MZ{2}+\fc{6404}{2625}\ (\MZ{2})^4\ ,\cr\crr
\p^\phi_9\MZ{1,1,3,4,6}&=-\fc{5599}{72}\ (\MZ{3})^2+\fc{25687}{630}\ (\MZ{2})^3\ ,\cr\crr
\p^\phi_{11}\MZ{1,1,3,4,6}&=\fc{28519}{60}\ (\MZ{2})^2\ ,\cr\crr
\p^\phi_{13}\MZ{1,1,3,4,6}&=\fc{56717}{120}\ \MZ{2}\ ,&\needxvb}
$$
respectively.
These derivatives give rise to :
\eqnn\verenaxv{
$$\eqalignno{
\phi(\zeta^m_{5,3,7})&=-3\ f_5\ (f_5\shuffle f_5)+\fc{96}{385}\ f_5f_2^5-6\ f_5\ (14f_7f_3+6f_5^2)-14\ f_7(f_3\shuffle f_5)\cr\crr
&-70\ f_7f_5f_3+\fc{48}{25}\ f_7f_2^4+\fc{136}{35}\ f_9f_2^3-22\ f_{11}f_2^2-
\fc{1001}{2}\ f_{13}f_2\ ,\cr\crr
\phi(\zeta^m_{3,3,9})&=6\ f_5\ (14f_7f_3+6f_5^2)-\fc{72}{175}\ f_7f_2^4+75\ f_7f_5f_3-\fc{27}{2}\ f_9\ (f_3\shuffle f_3)\cr\crr
&-\fc{116}{35}\ f_9f_2^3-\fc{252}{5}\ f_{11}f_2^2-\fc{1209}{2}\ f_{13}f_2\ ,\cr\crr
\phi(\zeta_{1,1,3,4,6}^m)&=-\fc{29}{9}\ f_3\ \phi(\MZ{3,9})-f_3\ \phi(\MZ{1,1,4,6})+
\fc{74}{3}\ f_3(f_5\shuffle f_7)-83\ f_3(f_3\shuffle f_9)\cr\crr
&-6\ f_3(14f_7f_3+6f_5^2)f_2-8\ f_3f_5 f_3f_2^2
+8\ f_3(f_5\shuffle f_5)f_2\cr\crr
&+42\ f_3(f_3\shuffle f_7)f_2+\fc{24}{5}\ 
f_3(f_3\shuffle f_5)f_2^2-\fc{1451972}{716625}\ f_3f_2^6\cr\crr
&-\fc{12263}{112}\ f_5(f_5\shuffle f_5)-\fc{245}{2}\ f_5(f_3\shuffle f_7)-\fc{145}{112}\ f_5(14f_7f_3+6f_5^2)\cr\crr
&+\fc{125}{2}\ f_5^2f_3f_2+\fc{87}{2}\ f_5(f_3\shuffle f_5)f_2+\fc{15}{2}\ f_5(f_3\shuffle f_3)f_2^2+\fc{19939}{1617}\ f_5f_2^5 \ ,\cr\crr
&+\fc{31}{4}\ f_7(f_3\shuffle f_5)-\fc{481}{4}\ f_7f_5f_3
-12\ f_7(f_3\shuffle f_3)f_2+\fc{6404}{2625}\ f_7f_2^4\ ,\cr\crr
&-\fc{5599}{72}\ f_9(f_3\shuffle f_3)+\fc{25687}{630}\ f_9f_2^3+\fc{28519}{60}\ f_{11}\ f_2^2+\fc{56717}{120}\ f_{13}f_2\ ,&\verenaxv}
$$
respectively. The maps $\phi(\MZ{3,9})$ and $\phi(\MZ{1,1,4,6})$ are given in \verenaxii.
With the information about the lower weight basis $\Uc_{k\leq 14}$ with \verenaxv\ 
we can construct the  basis for $\Uc_{15}$ shown in (A.6).
This basis (A.6) gives rise to the following decomposition of any motivic MZV $\xi_{15}$ of weight $15$
\eqnn\mdecoxv
$$\eqalignno{
\xi_{15}&=a_1\ \MZ{1,1,3,4,6}+a_2\ \MZ{3,3,9}+a_3\ \MZ{5,3,7}+a_4\ \MZ{15}+a_5\ \MZ{1,1,4,6}\ \MZ{3}
+a_6\ \MZ{3,9}\ \MZ{3}\cr
&+a_7\ \MZ{9}\ (\MZ{3})^2+a_8\ \MZ{3}\ \MZ{5}\ \MZ{7}+a_9\ (\MZ{3})^5+
a_{10}\ \MZ{3,7}\ \MZ{5}+a_{11}\ (\MZ{5})^3+a_{12}\ \MZ{3,5}\ \MZ{7}\cr
&+a_{13}\ \MZ{2}\ \MZ{3,3,7}+a_{14}\ \MZ{2}\ \MZ{3,5,5}
+a_{15}\ \MZ{2}\ \MZ{13}+a_{16}\ \MZ{2}\ \MZ{3}\ \MZ{3,7}+a_{17}\ \MZ{2}\ \MZ{5}\ \MZ{3,5}\cr
&+a_{18}\ \MZ{2}\ (\MZ{3})^2\ \MZ{7}+a_{19}\ \MZ{2}\ \MZ{3}\ (\MZ{5})^2
+a_{20}\ (\MZ{2})^2\ \MZ{3,3,5}+a_{21}\ (\MZ{2})^2\ \MZ{3}\ \MZ{3,5}\cr
&+a_{22}\ (\MZ{2})^2\ \MZ{11}+a_{23}\ \MZ{5}\ (\MZ{2})^2\ (\MZ{3})^2+a_{24}\ (\MZ{2})^3\ (\MZ{3})^3
+a_{25}\ (\MZ{2})^3\ \MZ{9}\cr
&+a_{26}\ (\MZ{2})^4\ \MZ{7}+a_{27}\ (\MZ{2})^5\ \MZ{5}+a_{28}\ (\MZ{2})^6\ \MZ{3}\ ,&\mdecoxv}
$$
with the  operators $a_i$ acting on $\phi(\xi_{15})$ and collected in (A.7).

\subsubsec{Decomposition at weight $16$}

Finally, at weight $16$ we have the following set of motivic MZVs
\eqn\Bxvi{\eqalign{
B=\{\ &\MZ{2},\ \MZ{3},\ \MZ{5},\ \MZ{7},\ \MZ{3,5},\ \MZ{9},\ \MZ{3,7},\ \MZ{11},\ \MZ{3,3,5},\ \MZ{3,9},\ \MZ{1,1,4,6},\ \MZ{3,3,7},\ \MZ{3,5,5},\ \MZ{3,3,3,5},\cr 
&\MZ{3,11},\ \MZ{5,9},\ \MZ{5,3,7},\ \MZ{3,3,9},\ \MZ{1,1,3,4,6},\ \MZ{3,3,3,7},\ \MZ{3,3,5,5},\ \MZ{3,13},\ \MZ{5,11},\ \MZ{1,1,6,8} \}}}
as independent algebra generators up to weight $16$.
Hence, we only need to compute the maps $\phi(\zeta_{3,3,3,7}^m),\ 
\phi(\zeta^m_{3,3,5,5}),\ \phi(\MZ{3,13}),\ \phi(\MZ{5,11})$ 
and $\phi(\zeta_{1,1,6,8}^m)$, which require  the following derivatives
\eqnn\needxvi
$$\eqalign{
\p^\phi_3\MZ{3,3,3,7}&=0,\ \ \ \p^\phi_5\MZ{3,3,3,7}=-6\ \MZ{3,3,5}\ ,\cr\crr
\p^\phi_7\MZ{3,3,3,7}&=-\fc{775}{6}\ \MZ{9}-\fc{7}{3}\ (\MZ{3})^3+63\ \MZ{7}\MZ{2}+\fc{36}{5}\ \MZ{5}(\MZ{2})^2+\fc{8}{5}\ \MZ{3}(\MZ{2})^3\ ,\cr\crr
\p^\phi_9\MZ{3,3,3,7}&=-476\ \MZ{7}+280\ \MZ{5}\MZ{2}\ ,\cr\crr
\p^\phi_{11}\MZ{3,3,3,7}&=-\fc{3723}{4}\ \MZ{5}+407\ \MZ{3}\MZ{2},\ \ \ 
\p^\phi_{13}\MZ{3,3,3,7}=-165\ \MZ{3}\ ,}
\eqno \hbox{\needxvi}$$
\eqnn\needxvia
$$\eqalign{
\p^\phi_3\MZ{3,3,5,5}&=0,\ \ \ \p^\phi_5\MZ{3,3,5,5}=-5\ \MZ{3,3,5}\ ,\cr
\p^\phi_7\MZ{3,3,5,5}&=\fc{381}{2}\ \MZ{9}-105\ \MZ{7}\MZ{2}-6\ \MZ{5}(\MZ{2})^2,\cr\crr
\p^\phi_9\MZ{3,3,5,5}&=70\ \MZ{7}+25\ \MZ{5}\MZ{2}-36\ \MZ{3}(\MZ{2})^2\ ,\cr\crr
\p^\phi_{11}\MZ{3,3,5,5}&=-\fc{1881}{4}\ \MZ{5}+275\ \MZ{3}\MZ{2},\ \ \ 
\p^\phi_{13}\MZ{3,3,5,5}=\fc{99}{2}\ \MZ{3}\ ,}
\eqno \hbox{\needxvia}$$
\eqnn\needxvib
$$\eqalign{
\p^\phi_3\MZ{3,13}&=0\ ,\cr
\p^\phi_5\MZ{3,13}&=-6\ \MZ{11}\ ,\cr
\p^\phi_7\MZ{3,13}&=-15\ \MZ{9}\ ,\cr}
\qquad 
\eqalign{
\p^\phi_9\MZ{3,13}&=-28\ \MZ{7}\ ,\cr
\p^\phi_{11}\MZ{3,13}&=-45\ \MZ{5}\ ,\cr
\p^\phi_{13}\MZ{3,13}&=-65\ \MZ{3}\ ,}
\eqno \hbox{\needxvib}$$
\eqnn\needxvic
$$\eqalign{
\p^\phi_3\MZ{5,11}&=0\ ,\cr
\p^\phi_5\MZ{5,11}&=0\ ,\cr
\p^\phi_7\MZ{5,11}&=-15\ \MZ{9}\ ,\cr}
\qquad 
\eqalign{
\p^\phi_9\MZ{5,11}&=-70\ \MZ{7}\ ,\cr
\p^\phi_{11}\MZ{5,11}&=-209\ \MZ{5}\ ,\cr
\p^\phi_{13}\MZ{5,11}&=-429\ \MZ{3}\ ,}
\eqno \hbox{\needxvic}$$
and
\eqnn\needxvid
$$\eqalign{
\p^\phi_3\MZ{1,1,6,8}&=-\fc{5}{7}\ \MZ{3,3,7}+\fc{6}{7}\ \MZ{3,5,5}-\fc{2}{7}\ \MZ{3}\MZ{3,7}
-\fc{8497}{42}\ \MZ{13}+\fc{8}{7}\ \MZ{3}(\MZ{5})^2+(\MZ{3})^2\MZ{7}\cr\crr
&+137\ \MZ{11}\MZ{2}+\fc{11}{7}\ \MZ{9}(\MZ{2})^2-\fc{848}{245}\ \MZ{7}(\MZ{2})^3-\fc{48}{35}\ \MZ{5}(\MZ{2})^4-\fc{816}{2695}\ \MZ{3}(\MZ{2})^5 \ ,\cr\crr
\p^\phi_5\MZ{1,1,6,8}&=-\fc{2}{5}\ \MZ{3,3,5}-\fc{18211}{240}\ \MZ{11} +(\MZ{3})^2\MZ{5}
+\fc{71}{2}\ \MZ{9}\MZ{2}+\fc{163}{25}\ \MZ{7}(\MZ{2})^2\cr\crr
&+\fc{36}{35}\ \MZ{5}(\MZ{2})^3-\fc{132}{175}\ \MZ{3}(\MZ{2})^4\ ,\cr\crr
\p^\phi_7\MZ{1,1,6,8}&=72\ \MZ{9}+(\MZ{3})^3-22\ \MZ{7}\MZ{2}-7\ \MZ{5}(\MZ{2})^2-
\fc{116}{35}\ \MZ{3}(\MZ{2})^3\ ,\cr\crr
\p^\phi_9\MZ{1,1,6,8}&=\fc{26921}{72}\ \MZ{7}-\fc{277}{2}\ \MZ{5}\MZ{2}-41\ 
\MZ{3}(\MZ{2})^2\ ,\cr\crr
\p^\phi_{11}\MZ{1,1,6,8}&=\fc{11536}{15}\ \MZ{5}-\fc{727}{2}\ \MZ{3}\MZ{2},\ \ \ \p^\phi_{13}\MZ{1,1,6,8}=\fc{28513}{25}\ \MZ{3}\ ,}
\eqno \hbox{\needxvid}$$
respectively.
These derivatives give rise to :
\eqnn\verenaxvi{
$$\eqalignno{
\phi(\zeta^m_{3,3,3,7})&=
-6\ f_5\ \phi(\MZ{3,3,5})-\fc{775}{6}\ f_7f_9-\fc{7}{3}\ f_7(f_3\shuffle f_3\shuffle f_3)+63\ f_7^2f_2+\fc{36}{5}\ f_7f_5f_2^2\cr\crr
&+\fc{8}{5}\ f_7f_3f_2^3-476\ f_9f_7+280\ f_9f_5f_2-\fc{3723}{4}\ f_{11}f_5+407\ f_{11}f_3f_2-165\ f_{13}f_3,
\cr\crr
\phi(\zeta^m_{3,3,5,5})&=
-5\ f_5\ \phi(\MZ{3,3,5})+\fc{381}{2}\ f_7f_9-105\ f_7^2f_2-6\ f_7f_5f_2^2+
70\ f_9f_7+25\ f_9f_5f_2\cr\crr
&-36\ f_9f_3f_2^2-\fc{1881}{4}\ f_{11}f_5+275\ f_{11}f_3f_2+\fc{99}{2}\ f_{13}f_3\ ,\cr\crr
\phi(\MZ{3,13})&=-6\ f_5f_{11}-15\ f_7f_9-28\ f_9f_7-45\ f_{11}f_5-65\ f_{13}f_3\ ,
&\verenaxvi\cr
\phi(\MZ{5,11})&=-15\ f_7f_9-70\ f_9f_7-209\ f_{11}f_5-429\ f_{13}f_3\ ,\cr\crr
\phi(\zeta_{1,1,6,8}^m)&=-\fc{5}{7}\ f_3\ \phi(\MZ{3,3,7})+\fc{6}{7}\ f_3\ \phi(\MZ{3,5,5})+\fc{2}{7}\ f_3[f_3\shuffle(14f_7f_3+6f_5^2)]
-\fc{8497}{42}\ f_3f_{13}\cr\crr
&+\fc{8}{7}\ f_3(f_3\shuffle f_5\shuffle f_5)+f_3(f_3\shuffle f_3\shuffle f_7)+137\ f_3f_{11}f_2+\fc{11}{7}\ f_3f_9f_2^2\cr\crr
&-\fc{848}{245}\ f_3f_7f_2^3-\fc{48}{35}\ f_3f_5f_2^4-\fc{816}{2695}\ f_3f_3f_2^5-\fc{2}{5}\ f_5\ \phi(\MZ{3,3,5})-\fc{18211}{240}\ f_5f_{11} \cr\crr
&+f_5(f_3\shuffle f_3
\shuffle f_5)+\fc{71}{2}\ f_5f_9f_2+\fc{163}{25}\ f_5f_7f_2^2+\fc{36}{35}\ f_5f_5f_2^3-\fc{132}{175}\ f_5f_3f_2^4\ ,\cr\crr
&+72\ f_7f_9+f_7(f_3\shuffle f_3\shuffle f_3)-22\ f_7f_7f_2-7\ f_7f_5f_2^2-
\fc{116}{35}\ f_7f_3f_2^3+\fc{26921}{72}\ f_9f_7\cr\crr
&-\fc{277}{2}\ f_9f_5f_2-41\ f_9f_3f_2^2+\fc{11536}{15}\ f_{11}f_5-\fc{727}{2}\ f_{11}f_3f_2+\fc{28513}{25}\ f_{13}f_3,}
$$
respectively. The maps $\phi(\MZ{3,3,5}), \phi(\MZ{3,3,7})$ and 
 $\phi(\MZ{3,5,5})$ are given in \verenaxi\  and \verenaxiii, respectively.
Gathering the information about the lower weight basis $\Uc_{k\leq 15}$ with \verenaxvi\ 
we can construct  the basis for $\Uc_{16}$ shown in (A.8).
This basis (A.8) gives rise to the following decomposition of any motivic MZV $\xi_{16}$ of weight $16$
\eqnn\mdecoxvi
$$\eqalignno{
\xi_{16}&=a_1\ \MZ{1,1,6,8}+a_2\ \MZ{3,3,3,7}+a_3\ \MZ{3,3,5,5}+a_4\ \MZ{3,13}+a_5\ \MZ{5,11}
+a_6\ \MZ{3}\ \MZ{3,3,7}\cr
&+a_7\ \MZ{3}\ \MZ{3,5,5}+a_8\ \MZ{13}\ \MZ{3}+a_9\ \MZ{3,7}\ (\MZ{3})^2
+a_{10}\ \MZ{3,5}\ \MZ{3}\ \MZ{5}+a_{11}\ \MZ{7}(\MZ{3})^3\cr
&+a_{12}\ (\MZ{5})^2\ (\MZ{3})^2+a_{13}\ \MZ{9}\ \MZ{7}+a_{14}\ (\MZ{3,5})^2+
a_{15}\ \MZ{11}\ \MZ{5}+a_{16}\ \MZ{3,3,5}\ \MZ{5}\cr
&+a_{17}\ \MZ{2}\ \MZ{3}\ \MZ{3,3,5}+a_{18}\ \MZ{3,5}\ (\MZ{3})^2\ \MZ{2}+
a_{19}\ \MZ{11}\ \MZ{3}\ \MZ{2}+a_{20}\ \MZ{5}\ (\MZ{3})^3\ \MZ{2}\cr
&+a_{21}\ (\MZ{3})^4\ (\MZ{2})^2+a_{22}\ \MZ{9}\ \MZ{3}\ (\MZ{2})^2
+a_{23}\ \MZ{7}\ \MZ{3}\ (\MZ{2})^3+a_{24}\ \MZ{5}\ \MZ{3}\ (\MZ{2})^4\cr
&+a_{25}\ (\MZ{3})^2\ (\MZ{2})^5+a_{26}\ \MZ{3,3,3,5}\ \MZ{2}+
a_{27}\ \MZ{3,11}\ \MZ{2}+a_{28}\ \MZ{5,9}\ \MZ{2}+a_{29}\ \MZ{9}\ \MZ{5}\ \MZ{2}\cr
&+a_{30}\ (\MZ{7})^2\ \MZ{2}+a_{31}\ \MZ{1,1,4,6}\ (\MZ{2})^2+
a_{32}\ \MZ{3,9}\ (\MZ{2})^2+a_{33}\ \MZ{7}\ \MZ{5}\ (\MZ{2})^2\cr
&+a_{34}\ \MZ{3,7}\ (\MZ{2})^3+a_{35}\ \MZ{3,5}\ (\MZ{2})^4+
a_{36}\ (\MZ{5})^2\ (\MZ{2})^3+a_{37}\ (\MZ{2})^8\ ,&\mdecoxvi}
$$
with the  operators $a_i$ acting on $\phi(\xi_{16})$ and listed in (A.9).

\subsubsec{Comments on regularizing the coproduct and the map $\phi$}

Some terms in the sum of the coproduct \Coproduct\ may imply divergences \refs{\goncharov,\Goncharov,\Claude}. 
Divergences of multiple polylogarithms are end--point divergences, \ie
the poles in the integrand \Integral\  coincide with the endpoints of the path $\gamma$. 
A canonical regularization has been introduced in \goncharov\ by
shifting the endpoints by a small parameter~$\eps$:
\eqn\regular{
I^m(0;a_1,\ldots,a_n;1)\ra I^m(\eps;a_1,\ldots,a_n;1-\eps)\ .}
Expanding the latter w.r.t. small $\eps$ gives a polynomial in $\ln \eps$.
Its constant term defines the regularized value $\hat{I}^m(0;a_1,\ldots,a_n;1)$.
The coproduct in the non--generic case is defined by replacing in the sum of \Coproduct\ 
every multiple polylogarithm $I^m(0;a_1,\ldots,a_n;1)$ by its regularized value 
$\hat{I}^m(0;a_1,\ldots,a_n;1)$ \refs{\goncharov,\Goncharov}.

Also the coaction \Coaction\ and therefore \CovDer\ and  \Coeff\ may be plagued by
divergences. We have regularized the terms in the sum \Coeff\ in the same way as described above for the coproduct \Coproduct. The problem, which  only affects the first
factor $c_{2r+1}^\phi(\ldots)$ of the terms in \Coeff,  occurs only in  the computation 
of the maps $\phi(\MZ{1,1,4,6}), \phi(\MZ{1,1,3,4,6})$ and $\phi(\MZ{1,1,6,8})$.
In addition, in the above three cases $c_{2r+1}^\phi(\ldots)$ computes the coefficient of $\zeta^m_{2r+1}$, 
which does not depend on the regularization, \ie it is independent on $\eps$.

Let us demonstrate the regularization at the computation of $\p^\phi_3(\MZ{1,1,6,8})$,
whose result is given in \needxvid.
With $\MZ{1,1,6,8}=I^m(0;1,1,1,0,0,0,0,0,1,0,0,0,0,0,0,0;1)$ computing \Coeff\ for $r=1$ 
yields:
\eqn\REGU{
\eqalign{
\xi_3&=c_3^\phi\lf[I^m(0;1,1,0;1)+I^m(0;1,0,1;1)\ri]\ I^m(0;1,0,0,0,0,1,0,0,0,0,0,0,0;1)\cr
&-c_3^\phi\lf[I^m(0;0,0,1;1)\ri]\ I^m(0;1,1,0,0,0,1,0,0,0,0,0,0,0;1)\ .}}
Above, the integral $I^m(0;1,0,1;1)$ has to be replaced by its regularized value
$\hat{I}^m(0;1,0,1;1)$. The latter is computed from expanding
\eqn\regi{\eqalign{
I^m(\eps;1,0,1;1-\eps)&\simeq\int_\eps^{1-\eps}\fc{dt_3}{1-t_3}\ \int_\eps^{t_3}\fc{dt_2}{t_2}\ \int_\eps^{t_2}\fc{dt_1}{1-t_1}\cr\crr
&=-\z_2^m\ \ln\eps-2\ \z_3^m+\lf[2+\z_2^m-(\ln\eps)^2\ri]\ \eps+\Oc(\eps^2)}}
w.r.t. small $\eps$. Hence, we have\foot{With this result \eqq \REGU\ becomes: $\xi_3=c_3^\phi\lf(\zeta^m_{1,2}-2\z^m_3\ri)\ \phi(\MZ{5,8})+c_3^\phi\lf(-\zeta^m_3\ri)\ \phi(\MZ{1,4,8})=-\phi(\MZ{5,8})-\phi(\MZ{1,4,8})$.}:
\eqn\henceregu{
\hat{I}^m(0;1,0,1;1)=-2\ \z^m_3\ .}
Note, that this agrees, with what one would obtain by applying the shuffle rule \Schuffle
\eqn\schuff{
I^m(0;1,0;1)\ I^m(0;1;1)=I^m(0;1,0,1;1)+2\ I^m(0;1,1,0;1)\ ,}
from which we obtain:
\eqn\regii{
I^m(0;1,0,1;1)=I^m(0;1,0;1)\ I^m(0;1;1)-2\ I^m(0;1,1,0;1)\ .}
With $I^m(0;1,1,0;1)=\MZ{1,2}=\MZ{3}$ 
the two expressions \regi\ and \regii\ give the same finite piece.
This is a consequence of the fact, that the shuffle relation also holds for the canonical 
regularization of multiple polylogarithms~\goncharov. 
An other way to arrive at the conclusion \henceregu\ follows from
simply identifying $I^m(a_0;a_1;a_2)\simeq 0$ for $a_i\in\{0,1\}$ in the shuffle relation \schuff, \cf \Brown.

\subsec{Motivic decomposition  operators and $\ap$--expansion}

By comparing the decomposition operators $\xi_l$ given for $l=10,\ldots,16$ in
\mcoeffsx, \mcoeffsxi, \mcoeffsxii, \mcoeffsxiii, (A.2), (A.7) and (A.9), respectively 
with the corresponding order $\ap^l$ in the expansion of \VERYNICEE\ 
(with the operators \PP\ and \QQQ) we see an exact match in the coefficient and 
commutator structure by identifying the motivic derivation operators \motder\ and the 
matrix operators \PP
\eqn\identification{
\p_{2n+1}\simeq M_{2n+1}\ ,}
and the coefficient operator $c_2$ with the matrix operators \PP:
\eqn\identificationi{
c_2^k\simeq P_{2k}\ \ \ ,\ \ \ k\geq1\ .}

We can further strengthen this connection. 
Let $\Lc'=\IQ\vev{e_3,e_5,\ldots}$ 
be the free graded Lie algebra (some vector space  over $\IQ$) 
freely generated by the generators $e_{2r+1}$ of degree $-(2r+1)$ with the Lie--bracket $(e_i,e_j)\mapsto [e_i,e_j]$ and the Jacobi relations:
\eqn\jacobi{
[e_i,[e_j,e_k]]+[e_j,[e_k,e_i]]+[e_k,[e_i,e_j]]=0\ .}
With $\Lc=\IQ[e_2]\oplus\Lc'$ the 
underlying graded vector space over $\IQ$ is generated by the following elements \Schneps:
\eqn\spanE{
e_2,\ e_3,\ e_5,\ e_7,\ [e_3,e_5],\ e_9,\ [e_3,e_7],\ e_{11},\ [e_3,[e_5,e_3]],\ [e_3,e_9],\ 
[e_5,e_7],\ \ldots\ .}
E.g. at weight $11$  the elements $e_{11}$ and $[e_3,[e_3,e_5]]$ generate $\Lc'_{11}$.
For $f_3,f_5,\ldots$ being the functionals on the vector space generated by the 
vectors $e_3,e_5,\ldots$ such that $\vev{f_i,e_j}=\delta_{ij}$ the dual to the universal enveloping algebra $U(\Lc)$ is isomorphic to the space $\Uc$ of non--commutative polynomials in $f_{2n+1}$ with the shuffle product \refs{\goncharov,\ABG}.

In fact, the Lie algebra $\Lc$ generators $e_i$ can be identified with the matrices  $M_{2n+1}$ and $P_2$ introduced in \PP, \ie
\eqn\isomorphic{\eqalign{
&e_{2n+1}\simeq M_{2n+1}\ ,\cr
&e_2\simeq P_2\ ,}}
and of course  the matrices $M_{2n+1}$ fulfill the Jacobi identity \jacobi:
\eqn\Jacobi{
[M_i,[M_j,M_k]]+[M_j,[M_k,M_i]]+[M_k,[M_i,M_j]]=0\ .}

To conclude, motivic MZVs encapsulate the $\ap$--expansion of the 
open superstring amplitude.

\newsec{Motivic structure  of the open superstring amplitude}

The symbol of a transcendental function represents a motivic road map encoding all the 
relevant information about the function without further specifying the latter explicitly 
in terms of multiple polylogarithms \refs{\Symbols,\GSVV,\SpradlinWP}. In particular, the various relations among different multiple polylogarithms become simple algebraic identities
in the corresponding tensor algebra.
In this section we show, that the isomorphism  $\phi$, which is induced by the coaction
\Coaction, encapsulates all the relevant  information of the $\ap$--expansion of the 
open superstring amplitude without further specifying the latter explicitly 
in terms of MZVs. 
By passing from the MZVs $\zeta\in\Zc$ to their motivic versions $\zeta^m\in\Hc$ and then mapping the latter to elements $\phi(\zeta^m)$ of the Hopf algebra $\Uc$
the map  $\phi$ endows the superstring amplitude with its motivic structure:
it maps the $\ap$--expansion  into a
very short and intriguing form \summ\ in terms of the non--commutative Hopf algebra $\Uc$.
In particular, the various relations among different MZVs become simple algebraic 
identities in the Hopf algebra $\Uc$. 
Moreover, in this writing the final result \summ\ for the superstring expansion does not depend on the choice of a specific set\foot{For instance instead of taking a basis containing
the depth one elements $\zeta^m_{2n+1}$ 
one also could choose the set of Lyndon words in the Hoffman elements
$\zeta^m_{n_1,\ldots,n_r}$, with $n_i=2,3$  \refs{\TATE,\DataMine}.}
 of MZVs as basis elements.

In this section we apply the isomorphism $\phi$ to the motivic version $\Ac^m$ of the open superstring amplitude expression \npt
\eqn\PHIMAP{
\phi(\Ac^m)=\phi(F^m)\ A\ ,}
with 
\eqn\Fmotivic{\eqalign{
F^m&=P^m\ Q^m\ :\exp\lf\{\sum_{n\geq 1}\zeta_{2n+1}^m\ M_{2n+1}\ri\}:\ ,\cr\crr
P^m&=\lf.P\ri|_{\zeta_2\ra \z_2^m}\ \ \ ,\ \ \ Q^m=\lf.Q\ri|_{\zeta_{n_1,\ldots,n_r}\ra\zeta^m_{n_1,\ldots,n_r}}\ ,}}
and the matrices $P,M$ and $Q$ defined in \PP\ and \QQQ, respectively.
The action \morphism\ of $\phi$ on the motivic MZVs is explained in the previous section. 

\subsec{Motivic structure up to weight 16}

The first hint of a simplification under $\phi$ occurs in \testa\ at weight $w=8$, 
where the  commutator term $[M_5,M_3]$ together with the 
prefactor ${1\over 5}\zeta_{3,5}^m$ conspires\foot{Note the useful relation $\phi(Q_8^m)=f_5f_3\ [M_3,M_5]$ for $Q_8^m={1\over 5}\zeta_{3,5}^m \ [M_5,M_3]$.}} into:
\eqn\testx{
\phi\lf(\ \zeta_3^m \zeta_5^m \ M_5 M_3  + Q_8^m\ \ri) 
=  f_3 f_5 \ M_5 M_3 \ + \ f_5 f_3 \ M_3 M_5\ .}
The right hand side obviously treats the objects $f_3,M_3$ and $f_5,M_5$ in a democratic way.
The effect of the map $\phi$ becomes even more drastic\foot{We use the identity:
$\phi(Q_{11}^m)=f_5f_3^2\ [M_3,[M_3,M_5]]$.}  at weight $w=11$ at the permutations of $M_3 M_3 M_5$:
\eqnn\testy
$$\eqalignno{
\phi\lf(\lf.F^m\ri|_{w=11}\ri)&= \phi\Big(\ Q_{11}^m  + 
Q_8^m\ \z_3^m\ M_3 +  {1\over 2} \ (\zeta_3^m)^2 \zeta_5^m\  M_5 M_3^2 +  \zeta_{11}^m \ M_{11}+  {1 \over 6} \ (\zeta_3^m)^3 \zeta_2^m \ P_2 M_3^3  \cr\crr
&+ \zeta_9^m \zeta_2^m \  P_2 M_9 + \zeta_7^m (\zeta_2^m)^2 \  P_4 M_7  +  \zeta_5^m (\zeta_2^m)^3\  P_6 M_5 + \zeta_3^m (\zeta_2^m)^4\  P_8 M_3  \  \Big)  
\cr\crr
&= f_{11}\  M_{11}  +  f_{3}^2 f_5\  M_5 M_3^2  +  f_3 f_5 f_3\  M_3 M_5 M_3  +  f_5 f_3^2\  M_3^2 M_5 \cr
&  +  P_2  f_2\  \big( f_9\  M_9 \ + \ f_3^3\  M_3^3 \big)   +  P_4  f_2^2  f_7\  M_7
+  P_6  f_2^3  f_5\  M_5 +  P_8  f_2^4  f_3 \ M_3  \ .&\testy}$$
{}From \testx\ and \testy\ we observe, that in the Hopf algebra ${\cal U}$, every non--commutative word 
of odd letters $f_{2k+1}$ multiplies the  associated 
reverse product of matrices $M_{2k+1}$. 
Powers $f_2^k$ of the commuting generator $f_2$ are accompanied by  $P_{2k}$, which multiplies all the 
 operators  $M_{2k+1}$  from the left. 
Most notably, in contrast to the representation in terms of motivic MZVs, 
the numerical factors become unity, \ie all the rational numbers in \Q\ drop out. 
Our explicit results confirm, that the  beautiful structure
with the combination of 
operators $M_{i_p} \ldots M_{i_2} M_{i_1}$ accompanying the word $f_{i_1} f_{i_2} \ldots f_{i_p}$,
continues to hold through weight $w=16$:
\eqnn\testw
$$\eqalignno{
\phi \lf(F^m\ri) &= \lf(\   1  +  f_2 P_2  +  f_2^2 P_4   +   f_2^3 P_6  +  f_2^4 P_8  +   f_2^5 P_{10}  +   f_2^6 P_{12}  +   f_2^7 P_{14}  +   f_2^8 P_{16}+\ldots  \  \ri) \cr
&  \times  \lf(\  1  +  f_3\ M_3  +  f_5\ M_5  +  f_3^2\ M_3^2  +  f_7\ M_7  +  f_3 f_5\ M_5 M_3  +  f_5 f_3\ M_3 M_5  \ri.\cr
& +  f_9\ M_9+  f_3^3\ M_3^3  +  f_5^2\ M_5^2  +  f_3 f_7\ M_7 M_3  +  f_7 f_3\ M_3 M_7+  f_{11}\ M_{11}    \cr
&  +  f_3^2 f_5\ M_5 M_3^2  +  f_3 f_5 f_3\ M_3 M_5 M_3  +  f_5 f_3^2\ M_3^2 M_5 +  f_3^4\ M_3^4  +  f_3 f_9\ M_9 M_3    \cr
&  +  f_9 f_3\ M_3 M_9  +  f_5 f_7\ M_7 M_5   +  f_7 f_5\ M_5 M_7  +  f_{13}\ M_{13}  +  f_3^2 f_7\ M_7 M_3^2 \cr
&  +  f_3 f_7 f_3\ M_3 M_7 M_3    +  f_7 f_3^2\ M_3^2 M_7   +  f_3 f_5^2\ M_5^2 M_3  +  f_5f_3 f_5\ M_5 M_3 M_5 \cr
&    +  f_5^2 f_3\ M_3 M_5^2  +  f_7^2\ M_7^2  +  f_3 f_{11}\ M_{11} M_3 +  f_{11} f_3\ M_3 M_{11}  +  f_5 f_9\ M_9 M_5  \cr
& +  f_9f_5\ M_5 M_9  +  f_3^3 f_5\ M_5 M_3^3   +  f_3^2 f_5 f_3\ M_3 M_5 M_3^2  +  f_3 f_5 f_3^2\ M_3^2 M_5 M_3  \cr
&  +  f_5 f_3^3\ M_3^3 M_5   +  f_{15}\ M_{15}  +  f_5^3\ M_5^3  +  f_3^5\ M_3^5  +  f_3^2 f_9\ M_9 M_3^2   +   f_3 f_9 f_3\ M_3 M_9 M_3  \cr
& +  f_9 f_3^2\ M_3^2 M_9  +  f_3 f_5 f_7\ M_7 M_5 M_3    +  f_3 f_7 f_5\ M_5 M_7 M_3  +  f_5 f_3 f_7\ M_7 M_3 M_5  \cr
&  +  f_5 f_7 f_3\ M_3 M_7 M_5  +  f_7f_3 f_5\ M_5 M_3 M_7 +  f_7 f_5 f_3\ M_3 M_5 M_7   +  f_7 f_9\ M_9 M_7 \cr
&  +  f_9f_7\ M_7 M_9  +  
f_{11} f_5\ M_5 M_{11}  +  f_{5} f_{11}\ M_{11} M_5  +  f_3 f_{13}\ M_{13}M_3   +  f_{13}f_3\ M_3 M_{13}  \cr
&  +  f_3^2 f_5^2\ M_5^2 M_3^2  +  
f_5^2 f_3^2\ M_3^2 M_5^2+f_3 f_5^2 f_3\ M_3 M_5^2 M_3 +  f_5 f_3^2 f_5\ M_5 M_3^2 M_5  \cr
&  +  f_3 f_5 f_3 f_5\ M_5 M_3 M_5 M_3 +  f_5 f_3 f_5 f_3\ M_3 M_5 M_3 M_5 +   
f_3^3 f_7\ M_7 M_3^3  \cr
&\lf. +   f_3^2 f_7 f_3\ M_3 M_7 M_3^2 +   f_3 f_7 f_3^2\  M_3^2 M_7 M_3 +  f_7 f_3^3\ M_3^3 M_7 +\ldots\ \ri)\ .&\testw
} $$
Writing the amplitude \PHIMAP\ in terms of elements of the algebra comodule $\Uc$, with 
$\phi(F^m)$ given above encodes  all the information contained in \QQQ. 

\subsec{Motivic structure at general weight}

Motivated by the observation, that every non--commutative word constructed from odd generators $f_{2k+1}$ shows up in \testw\  
we write down the following formula 
\eqn\summtwo{
\phi(F^m) =\left( \sum_{k=0}^{\infty} f_2^k\ P_{2k} \right)\ \lf( \sum_{p=0}^{\infty} \sum_{ i_1,\ldots, i_p \atop  \in 2 \IN^+ + 1}
f_{i_1} f_{i_2}\ldots f_{i_p}\  M_{i_p} \ldots M_{i_2} M_{i_1}\ \ri) ,}
for the image\foot{Note, that a different normalization \morphismres\ or choice of $\phi$ (\cf footnote 6) can be compensated by an appropriate  modification of the definition of the matrices $M_{2n+1}$ such that the form of  \summtwo\ stays unchanged.} $\phi(F^m)$ valid for any weight.
In \summtwo\ the sum over the combinations $ f_{i_1} f_{i_2}\ldots f_{i_p} M_{i_p} \ldots M_{i_2} M_{i_1}$ includes 
{\it all} possible 
non--commutative words $f_{i_1} f_{i_2} \ldots f_{i_p}$ with coefficients 
$M_{i_p} \ldots M_{i_2} M_{i_1}$ graded by their length $p$. 
Matrices $P_{2k}$ associated with the  powers $f_2^k$ always act by left multiplication. The commutative nature of $f_2$ w.r.t. the odd 
generators $f_{2k+1}$ ties in with the fact that in the matrix ordering the  matrices $P_{2k}$ have the well--defined place left of all matrices $M_{2k+1}$. With \testw\ one  easily checks that \summtwo\ is compatible through weights less or equal to $16$. 
Combining \eqqs \summtwo\ and \PHIMAP\ gives the final result\foot{The  combinations $ f_{i_1} f_{i_2}\ldots f_{i_p} M_{i_p} \ldots M_{i_2} M_{i_1}$ in \summ\ reflect the agreement of the coefficients in the $\ap$ expansion $\lf.\Ac\ri|_w$ 
of the superstring amplitude with those in the motivic decomposition operators $\xi_w$ observed in subsection 4.4.}
 \summ:
\eqn\summ{
\phi(\Ac^m)=\lf
( \sum_{k=0}^{\infty} f_2^k\ P_{2k} \ri) \ \lf\{\  \sum_{p=0}^{\infty} 
\ \sum_{ i_1,\ldots, i_p \atop\in 2 \IN^+ + 1}
f_{i_1} f_{i_2}\ldots f_{i_p}\ M_{i_p} \ldots M_{i_2} M_{i_1}\  \ri\}\  A\ .}

In the following we shall give further evidence, that the validity extends to higher weights.
In subsection 4.4 we have already pointed out, that the decomposition formula $\xi_w$ for  
MZVs  of weight $w$ exactly matches the corresponding $\ap^w$--part of the 
superstring amplitude subject to the replacements \identification\ and \identificationi. 
If  this mapping  holds to arbitrary weight, then the simplicity of our final result \summtwo\ reflects the role of $\phi(\xi_w)$ being the unit operator projected to weight $w$, \eg
\eqnn\testw
$$\eqalignno{
\phi(\xi_{10})&= \ f_2^5 c_2^5 \ + \ f_2^2 f_3^2 \, c_2^2 \partial_3^2 \ + \ f_2 \, \big( f_3 f_5 \, \partial_5 \partial_3 \ + \ f_5 f_3 \, \partial_3 \partial_5 \big) \, c_2 \cr
&+ f_5^2 \, \partial_5^2 \ + \ f_7 f_3 \, \partial_3 \partial_7 \ + \ f_3 f_7 \, \partial_7 \partial_3 \ = \  id \, \big|_{w=10}
}$$
maps any non--commutative weight ten polynomial in $f_2,f_3,f_5,f_7,f_9$ to itself.
More generally, the differential operator $c_2^k \partial_{i_p} \ldots \partial_{i_2} \partial_{i_1}$ annihilates all ${\cal U}$ elements except for $f_2^k f_{i_1} f_{i_2}\ldots f_{i_p}$. Hence, the weight $w$ identity operator is given by
\eqnn \idw
$$\eqalignno{
\phi(\xi_{w})&  \ = \ \sum_{k=0}^{\infty} f_2^k c_2^k \ \sum_{p=0}^{\infty} \sum_{ i_1,\ldots, i_p \atop  \in 2 \IN^+ + 1}
f_{i_1} f_{i_2}\ldots f_{i_p} \, \partial_{i_p} \ldots \partial_{i_2} \partial_{i_1} \cr&\times\de \big( i_1+i_2+\ldots +i_p \,  + \, 2k - w\big) = \  id \, \big|_{w}\ , &\idw
}$$
where the $\de(\ldots)$ function makes sure that the correct weight is picked up. Clearly, \idw\ maps to the weight $w$ contributions of \summtwo\ under the replacements \identification\ and \identificationi. In this sense, the image under $\phi$ of the disk amplitude at weight $w$ is closely related to the identity operator in the algebra comodule $\Uc$, restricted to weight $w$.

\newsec{Closed superstring amplitude}

The string world--sheet describing the tree--level string $S$--matrix of $N$ 
gravitons has the topology of a complex sphere with $N$ (integrated) insertions
of graviton vertex operators.
One of the key properties of graviton amplitudes in string theory is that 
at tree--level they can be expressed as sum over squares of (color ordered) 
gauge amplitudes in the left-- and right--moving sectors. 
This map, known as Kawai--Lewellen--Tye (KLT) relations  \KawaiXQ, 
gives a relation between 
a closed string tree--level amplitude $\Mc$ involving $N$ closed strings and a sum of squares of 
(partial ordered) open string tree--level amplitudes.  
We may write these relations in matrix notation as follows
\eqn\KLT{
\Mc(1,\ldots,N)=\Ac^t\ S\ \Ac\ ,}
with the vector $\Ac$ encoding the $(N-3)!$ independent color ordered open string subamplitudes 
and some $(N-3)!\times(N-3)!$ matrix $S$. 
The latter encodes the $\sin$--factors from the KLT relations \KawaiXQ\ and the contributions 
from the monodromy relations \refs{\StiebergerHQ,\BjerrumBohrRD} to express both 
left-- and right--movers in terms of the same open string basis $\Ac$. 
Hence, in superstring theory the tree--level computation of graviton amplitudes 
boils down to considering squares of tree--level gauge amplitudes $\Ac$ given in \SimpleN.
For this sector the explicit expressions \npt\ and \VERYNICEE\ and subsequent
results from the previous sections can be used.
The KLT relations  are insensitive
to the compactification details or the amount of supersymmetries of the superstring background. 
Hence, the following discussions and results are completely general.

In the sequel we shall discuss the implication of \VERYNICEE\ 
to the closed string amplitude \KLT.
Especially, we shall be interested in the structure of its $\ap$--expansion.
The latter has been already investigated up to the order $\ap^8$ for the cases $N=4,5$ and $N=6$ with the remarkable observation, that the graviton amplitudes do not allow
for powers of $\zeta_2$ in their $\ap$--expansions up to the order $\ap^8$ \GRAV.
With the explicit expression \VERYNICEE\ for the open superstring amplitude 
we are now able to reveal the pattern  and more general framework 
behind these findings.

\subsec{$N=4$}

For $N=4$ the KLT relation \KLT\ can be written as:
\eqn\closedd{\Mc(1,2,3,4)=\Ac^t\ S\ \Ac\ ,}
with the basis $\Ac=\Ac(1,2,3,4)$ of open string amplitudes \wehave\ 
and the scalar:
\eqn\Sfactor{
S=\sin(\pi s)\ \fc{\sin(\pi u)}{\sin(\pi t)}\ .}
With \wehave\ and 
\eqn\Pvier{
P=\lf\{\pi\ \fc{s\ u}{s+u}\ \fc{\sin[\pi(s+u)]}{\sin(\pi s)\ \sin(\pi u)}\ri\}^{1/2}\ ,}
Eq. \closedd\ yields:
\eqn\graviv{
\Mc(1,2,3,4)=\pi\  \fc{su}{s+u}\  \exp\lf\{2\sum_{n\geq 1}\zeta_{2n+1}\ M_{2n+1}\ri\}\ |A|^2\ ,}
with the YM subamplitude $A=A_{YM}(1,2,3,4)$ and $M_{2n+1}$ given in \wehaveiv. Obviously, in the 
four--graviton amplitude \graviv, not any  Riemann zeta function with even entries shows up.

The field--theory contribution from \closedd\  arises from $P=1$
and $\Ac=A$, \ie
\eqn\ftgrav{
\lf.\Mc(1,2,3,4)\ri|_{FT}=A^t\ S_0\ A\ ,}
with
\eqn\SSS{
S_0\equiv \lf.S\ri|_{FT}=\pi\ s\ \fc{u}{t}\ .}
We observe, that:
\eqn\Wehave{
P^tSP=S_0\ .}
This equation guarantees the absence of powers of $\z_2$  in \graviv.
Stated differently, the absence of powers of $\z_2$ in \closedd\ allows to determine the scalar 
$P=P^t$ from the equation \Wehave\  as:
\eqn\invert{
P=S_0^{1/2}\ (S^{-1})^{1/2}\ .}

\subsec{$N=5$}

For $N=5$ the closed string amplitude \KLT\ can be cast into
\eqn\closed{
\Mc(1,2,3,4,5)=\Ac^t\ S\ \Ac\ ,}
with the basis $\Ac$ of open string amplitudes given in \ppp\ 
and the symmetric matrix $S$ encoding the diagonal matrix 
${\rm diag}\{\sin(\pi s_{12})\sin(\pi s_{34}),\ \sin(\pi s_{13})\sin(\pi s_{24})\}$
from the KLT relation \KawaiXQ\ and further $\sin$--factors from the monodromy relations \refs{\StiebergerHQ,\BjerrumBohrRD} expressing the string amplitudes $\Ac(2,1,4,3,5)$ and $\Ac(3,1,4,2,5)$ in terms of the basis elements 
$\Ac(1,2,3,4,5)$ and $\Ac(1,3,2,4,5)$. More precisely, we have
\eqn\SFactorv{
S=\lf[\sin(\pi s_{35})\ \sin(\pi s_{25})\ \sin(\pi s_{14})\ri]^{-1}\ \pmatrix{\Si_{11}&\Si_{12}\cr 
         \Si_{12}&\Si_{22}}\ ,}
with:
\eqnn\Sfactorv
$$\eqalignno{
\Si_{11}&=\fc{1}{4}\ \sin(\pi s_1)\ \sin(\pi s_3)\ \lf[\ \sin\pi(s_1-s_2-s_3)-
\sin\pi(s_1+s_2-s_3)\ri.\cr
&+\sin\pi(s_1+s_2+s_3)+\sin\pi(s_1+s_2-s_3-2 s_4)\cr
&\lf.+\sin\pi(-s_1+s_2+s_3-2s_5)-\sin\pi(s_1+s_2+s_3-2s_4-2s_5)\ \ri]\ ,\cr
\Si_{12}&=-\sin(\pi s_1)\ \sin(\pi s_3)\ \sin(\pi s_{13})\ \sin(\pi s_{24})\  
\sin\pi(s_4+s_5)\ ,\cr
\Si_{22}&=\fc{1}{4}\ \sin(\pi s_{13})\ \sin(\pi s_{24})\  \cr
&\times\lf[\ \sin\pi(s_1+s_2-s_3-s_4-s_5)-\sin\pi(s_1+s_2-s_3-s_4+s_5)\ri.\cr
&-\sin\pi(s_1+s_2+s_3-s_4-s_5)+\sin\pi(s_1-s_2-s_3-s_4+s_5)\cr
&\lf.-\sin\pi(s_1-s_2-s_3+s_4+s_5)+\sin\pi(s_1+s_2+s_3+s_4+s_5)\ \ri]\ .&\Sfactorv}$$

The field--theory contribution from \closed\  arises from $P=\pmatrix{1&0\cr 0&1}$
and $\Ac=A$, with the YM basis vector $A$ given in \ppp, \ie
\eqn\ftgrav{
\lf.\Mc(1,2,3,4,5)\ri|_{FT}=A^t\ S_0\ A\ ,}
with
\eqn\sFactorv{
S_0\equiv \lf.S\ri|_{FT}=\pi^2\ (s_{25}\ s_{35}\ s_{14})^{-1}\ 
\pmatrix{\si_{11}&\si_{12}\cr 
         \si_{12}&\si_{22}}\ ,}
and:
\eqnn\sfactorv
$$\eqalignno{
\si_{11}&=s_1s_3\ [\ s_4(s_3-s_5)(-s_2+s_4+s_5)+s_1(-s_3(s_4+s_5)+s_5(-s_2+s_4+s_5))\ ]\ ,\cr
\si_{12}&=-s_1s_3\ s_{13}\ s_{24}\ (s_4+s_5)\ ,\cr
\si_{22}&=-s_{13}\ s_{24}\ [\ s_1s_4(s_2+s_3)+s_1s_3s_5+s_2s_5(s_3+s_4)\ ]\ .&\sfactorv}$$

By considering the closed superstring amplitude \closed\ and analyzing 
its $\ap$--expansion  \GRAV\ we find, that the following matrix equation holds:
\eqn\vanishi{
P^t\ S P=S_0\ .}
We have checked the validity of \vanishi\ up to the order $\ap^{18}$.
As a consequence of the relation \vanishi, 
the contribution of the matrix $P$ 
stemming from the open superstring amplitudes  \VERYNICE\ and accounting for powers of $\zeta_2$
drops out of the $\ap$--expansion of \closed.
In addition, we find the relation
\eqn\newstuff{
M_l^t \ S_0 = S_0\ M_l\ ,
}
which we have verified up to weight $l=19$. For commutators $\Qc_{(r)}$ of $M_l$, \eqq \newstuff\ implies
\eqn\vanishii{\eqalign{
S_0\ \Qc_{(2)}+\Qc_{(2)}^t\ S_0&=0\ \ \ ,\ \ \ \Qc_{(2)}=[M_l,M_m]\ ,\cr
S_0\ \Qc_{(3)}-\Qc_{(3)}^t\ S_0&=0\ \ \ ,\ \ \ \Qc_{(3)}=[M_l,[M_m,M_n]]\ ,\cr
S_0\ \Qc_{(4)}+\Qc_{(4)}^t\ S_0&=0\ \ \ ,\ \ \ \Qc_{(4)}=[M_k,[M_l,[M_m,M_n]]]\ }}
generalizing to
\eqn\vanishiii{
S_0\ \Qc_{(r)} +(-1)^r\ \Qc^t_{(r)} S_0=0 \ \ \ , \ \ \ \Qc_{(r)}=[M_{n_2},[M_{n_3},\ldots,[M_{n_r},M_{n_1}]]\ldots] }
for nested commutators $\Qc_{(r)}$ of generic depth $r$.
In contrast to  \vanishiii\ in the closed string amplitude the nested commutators $\Qc_{(r)}$
show up in the combination $S_0 \Qc_{(r)} +\Qc^t_{(r)} S_0$, which only vanishes for commutators of even depth 
$r \in  2\IN$. Assuming for $Q$ the exponential form \expandexp\ the relation
\eqn\vanishiv{
S_0\ e^{\Qc_{(r)}} = e^{(-1)^{r+1} \Qc^t_{(r)}}\ S_0}
following from \vanishiii\ guarantees the decoupling of any power of nested commutators $\Qc_{(r)}$ of even depth 
$r \in  2\IN$ in \closed. 

On the basis of \vanishi\ and \newstuff, we obtain the final form 
\eqn\gravfivept{\eqalign{
\Mc(1,2,3,4,5) &=A^t\ \lf( : \exp\lf\{\sum_{r\in 2\IN^+ +1} \! \! \!  \zeta_{r}\ M_{r}\ri\} : \ri)^t\ 
Q^t \,  S_0\, Q \ :\exp\lf\{\sum_{s\in 2\IN^+ +1} \! \! \! \zeta_{s}\ M_{s}\ri\} : \ A \cr
&= A^t  \
 S_0\ \lf( : \exp\lf\{\sum_{r\in 2\IN^+ +1} \! \! \!  \zeta_{r}\ M_{r}^t \ri\} : \ri)^t 
 \tilde Q \ Q \ :\exp\lf\{\sum_{s \in 2\IN^++1} \! \! \!  \zeta_{s}\ M_{s}\ri\} : \ A
\ ,}}
where the ordering colons enclosing the exponentials\foot{Note that the transpositions involved in the expression $\lf( : \exp\lf\{\sum_{r}  \zeta_{r}\ M_{r}^t \ri\} : \ri)^t$ lead to a reversal of the matrix multiplication order compared to the ordered product $:\exp\lf\{\sum_{s}  \zeta_{s}\ M_{s}\ri\} :$ without transposition, \ie:
$\lf( : \exp\lf\{\sum_{r\in 2\IN^+ +1}   \zeta_{r}\ M_{r}^t \ri\} : \ri)^t= 1+\zeta_3 M_3 + \zeta_5 M_5 + {1 \over 2} \zeta_3^2 M_3^2 + \zeta_7 M_7 + \zeta_3 \zeta_5 M_3 M_5  
+{1 \over 6} \zeta_3^3 M_3^3+ \zeta_9 M_9 
 + {1 \over 2} \zeta_5^2 M_5^2 + \zeta_3 \zeta_7 M_3 M_7 + {1 \over 2} \zeta_3^2 \zeta_5 M_3^2 M_5 + \zeta_{11} M_{11} + \ldots \ .$} are defined in \order\ and 
the matrix $\tilde Q$ is obtained from $Q$ by  replacing commutators $\Qc_{(r)}$ as follows:
\eqn\obtaintilde{
\tilde Q = \lf.Q\ \ri|_{\Qc_{(r)} \rightarrow (-1)^{r+1}\Qc_{(r)}} \ .} 
As a consequence  terms with commutator factors ${\Qc_{(2n)}}$ of even depth do not show up in the product\foot{The exponential form \expandexp\ leads us to expect even weight contributions to $\tilde Q Q$ starting at weight 22, e.g. $Q_{22} = {1\over 2} Q_{11}^2 + \ldots$ such that 
$\tilde Q  \lf.Q\ri|_{w=22} = 2\ Q_{11}^2$.}:
\eqn\onlyterms{
\tilde Q\ Q = 1 + 2\ Q_{11} + 2\ Q_{13}+2\ Q_{15}+\ldots \ .}
Hence, we observe, that in \gravfivept\ MZVs of even weight or depth $\geq 2$ only enter through the product \onlyterms\ starting at weight $w=11$.
This result
is in agreement with the observation made in \GRAV. We now have verified this observation
through weight $18$. 
Let us display the expansion of \gravfivept\ through the  order $\ap^{14}$:
\comment{\eqn\gravlowmom{\eqalign{
\Mc(1,2,3,4,5) \ = \ A^t\ 
&S_0\ \big( \ 1+2 \zeta_3 M_3 + 2 \zeta_5 M_5 + 2 \zeta_3^2 M_3^2 + 2 \zeta_7 M_7 + 2 \zeta_3 \zeta_5 (M_3 M_5 + M_5 M_3) + 2 \zeta_9 M_9 + {4 \over 3} \zeta_3^3 M_3^3 + 2 \zeta_5^2 M_5^2 + 2 \zeta_3 \zeta_7 (M_3 M_7 + M_7 M_3) + 2 Q_{11} + 2 \zeta_{11} M_{11} + \zeta_3^2 \zeta_5 ( M_3^2 M_5 + 2 M_3 M_5 M_3 + M_5 M_3^2 ) + {2 \over 3} \zeta_3^4 M_3^4 + 2 \zeta_3 \zeta_9 (M_3 M_9 + M_9 M_3) + 2 \zeta_5 \zeta_7 (M_5 M_7 + M_7 M_5) + 2 Q_{13} + \zeta_{13} M_{13} + \zeta_3^2 \zeta_7 ( M_3^2 M_7 + 2 M_3 M_7 M_3 + M_7 M_3^2 ) + 2 \zeta_3 \zeta_5^2 (M_3 M_5^2 + M_5^2 M_3) + 2 \zeta_3 (M_3 Q_{11}+Q_{11}M_3) + 2 \zeta_7^2 M_7^2 + 2 \zeta_3 \zeta_{11} (M_3 M_{11} + M_{11} M_3) + 2 \zeta_5 \zeta_9 (M_5 M_9 + M_9 M_5) + \zeta_3^3 \zeta_5 ( {1\over 3} M_3^3 M_5 + M_3^2 M_5 M_3 + M_3 M_5 M_3^2 + {1 \over 3} M_5 M_3^3) + \ldots \ \big) \ A
\ ,}}}
\eqnn\gravlowmom{
$$\eqalignno{
\Mc(1,2,3,4,5)  &=  A^t\
 S_0\ \big( \ 1+2\ \zeta_3 M_3 + 2\ \zeta_5 M_5 + 2\ \zeta_3^2 M_3^2 + 2\ \zeta_7 M_7+ 2\ \zeta_3 \zeta_5\ \{M_3, M_5\} \cr
& + 2\ \zeta_9 M_9 + {4 \over 3}\ \zeta_3^3 M_3^3 + 2\ \zeta_5^2 M_5^2 
  + 2\ \zeta_3 \zeta_7\ \{M_3 ,M_7\}+ 2\ Q_{11} + 2\ \zeta_{11} M_{11} \cr 
&  + \zeta_3^2 \zeta_5\ \{ M_3,\{M_3,M_5\}\} + {2 \over 3}\ \zeta_3^4 M_3^4 
+ 2\ \zeta_3 \zeta_9\ \{M_3, M_9 \}+ 2\ \zeta_5 \zeta_7\ \{M_5, M_7\}  \cr
&+ 2\ Q_{13} + 2\ \zeta_{13} M_{13} + \zeta_3^2 \zeta_7\ \{ M_3,\{M_3, M_7\}\} 
 + 2\ \zeta_3 \zeta_5^2\ \{M_3, M_5^2\} \cr
 &+ 2\ \zeta_3\ \{M_3 ,Q_{11}\}+ 2\ \zeta_7^2 M_7^2
  + 2\ \zeta_3 \zeta_{11}\ \{M_3, M_{11}\}+ 2\ \zeta_5 \zeta_9\ \{M_5 ,M_9\}\cr
  &  
+{1\over 3}\  \zeta_3^3 \zeta_5\ \{M_3,\{M_3,\{M_3,M_5\}\}\} + \ldots \ \big)\  A\ .&\gravlowmom}$$}
Up to the  order shown,  MZVs of depth $r\geq 2$ enter through the objects $Q_{11}, Q_{13}$ and $\{M_3, Q_{11}\}$. In the single zeta sector, the coefficient of the general power $(\zeta_{2k+1} M_{2k+1})^p$ is given by $2^p/p!$.


\subsec{General $N$}

Let us now phrase the observation from above for general multiplicities $N$.
The general form of the $N$--point closed string amplitude is given in \KLT, 
\eqn\Closed{
\Mc(1,\ldots,N)=\Ac^t\ S\ \Ac\ ,}
with the $(N-3)!\times(N-3)!$ matrix $S$ specified above and the vector $\Ac$ 
encoding the $(N-3)!$ open string subamplitudes \npt. Just as in the five--point case, the relations
\eqn\vanishi{
P^t\ S P=S_0\ ,}
\eqn\newnewstuff{
M_l^t \ S_0 = S_0\ M_l \ ,}
\eqn\vanishii{
S_0\ \Qc_{(r)} +(-1)^r\ \Qc^t_{(r)} S_0=0\ ,}
with $S_0\equiv \lf.S\ri|_{FT}$ and $\Qc_{(r)}=[M_{n_2},[M_{n_3},\ldots,[M_{n_r},M_{n_1}]]\ldots]$ imply that from  \VERYNICEE\ both the matrix $P$ and the part of $Q$ with admixtures of even depth commutators $\Qc_{(2n)}$ are cancelled in the $N$--point closed string amplitude \Closed. With the informations \vanishi\  and \newnewstuff\ the closed superstring \Closed\ amplitude for any number $N$ of external states takes 
the generic form
 %
 \eqn\gravN{
\Mc(1,\ldots,N) = A^t  \
 S_0\ \lf( : \exp\lf\{\sum_{r\in 2\IN^+ +1} \! \! \!  \zeta_{r}\ M_{r}^t \ri\} : \ri)^t \tilde Q\ Q : \exp\lf\{\sum_{s \in 2\IN^++1} \! \! \!  \zeta_{s}\ M_{s}\ri\} : \ A
 \ ,}
with the $(N-3)!$ dimensional vector $A$ specifying a YM basis $A\equiv A_{YM}$,
the $(N-3)!\times(N-3)!$ matrix $S_0$
introduced above and the $(N-3)!\times(N-3)!$ matrices $M_{2n+1}$ defined in \PP.
The ordering colons enclosing the exponentials are defined in \order\ and the matrix 
$\tilde Q$ is obtained from $Q$ according to \obtaintilde. Due to \vanishii\ and the exponential form \expandexp, the product 
\eqn\nochmal{
\tilde Q\ Q = 1 + 2\ Q_{11} + 2\ Q_{13}+2\ Q_{15}+\ldots 
}
 in \gravN\ is free of even depth commutators $\Qc_{(2n)}$.
Finally, the $\ap$--expansion of the $N$--point amplitude \gravN\ assumes the same form as  \gravlowmom, with the matrices $M_{2n+1}$ given in \PP.

We would like to mention two final remarks:
Similarly as for the $N=4$ case \invert\ one can constrain $P$ from the matrix equation \vanishi.
Moreover, \eqq \newnewstuff\ provides restrictive relations between entries of the matrices $M_{2k+1}$. Their information content on the polynomial structure of 
$P$ and $M_{2k+1}$ is further investigated and exhibited in more details in \BSS.
Of course, with the explicit expression for $P$ and $M$ the relations \vanishi\ and 
\newnewstuff\ and hence \gravN\ can be verified to all orders.

\subsec{Motivic structure of the closed superstring amplitude}

Experiencing the simplicity in the open string sector suggests to also investigate the image under $\phi$ of the gravity amplitude \Closed. 
We insert the result \summtwo\ for $\phi({\cal A}^m)$ into  \Closed.
The multiplication rule \ruleshuffle\ of the isomorphism $\phi$ yields:
\eqnn\gravmotiv
$$\eqalignno{
\phi( {\cal M}^m ) \ = \ A^t \
&\bigg\{ \sum_{p=0}^{\infty} \sum_{ i_1,\ldots, i_p \atop  \in 2 \IN^+ + 1}
f_{i_1} f_{i_2}\ldots f_{i_p}\ M_{i_p} \ldots M_{i_2} M_{i_1} \bigg\}^t \cr 
&\shuffle \ S_0 \ \bigg\{ \sum_{q=0}^{\infty} \sum_{ j_1,\ldots, j_q \atop  \in 2 \IN^+ + 1}
f_{j_1} f_{j_2}\ldots f_{j_q}\ M_{j_q} \ldots M_{j_2} M_{j_1} \bigg\} \ A\ .
&\gravmotiv
}$$
The sum over $f_2^k P_{2k}$ in the open string amplitudes $\Ac^t,\Ac$ has already been dropped 
taking into account the motivic version of the relation \vanishi. As a result the commutative Hopf algebra element $f_2$ is absent in $\phi( {\cal M}^m )$.

In order to simplify \gravmotiv\ we can make use of \newnewstuff\ to convert all the $M_i^t$ from the left moving open string amplitude to $M_i$ factors multiplying $S_0$ from the right:
%
\eqnn\gravsimple
$$\eqalignno{
\phi( {\cal M}^m ) \ = \ A^t \ S_0 \ \bigg\{ \ \sum_{p=0}^{\infty} \sum_{ i_1,\ldots, i_p \atop  \in 2 \IN^+ + 1}&
f_{i_1} f_{i_2}\ldots f_{i_p} \ M_{i_1}  M_{i_2} \ldots M_{i_p} \cr
\shuffle \ &\sum_{q=0}^{\infty} \sum_{ j_1,\ldots, j_q \atop  \in 2 \IN^+ + 1}  f_{j_1} f_{j_2}\ldots f_{j_q} \ 
 M_{j_q} \ldots M_{j_2} M_{j_1} \ \bigg\} \ A  &\gravsimple \cr
 = \ A^t \ S_0 \ \bigg\{\ \sum_{p=0}^{\infty} \sum_{ i_1,\ldots, i_p \atop  \in 2 \IN^+ + 1}&  \ M_{i_1}  M_{i_2} \ldots M_{i_p} 
\ \sum_{k=0}^p  \ f_{i_1} f_{i_2} \ldots f_{i_k}  \shuffle  f_{i_p} f_{i_{p-1}} \ldots f_{i_{k+1}}  \ \bigg\} \ A
}$$
On the way to the last line of \gravsimple, the double sum over non-commutative words in $f_i$ has been rearranged to identify the overall coefficient of $A^t S_0 M_{i_1}  M_{i_2} \ldots M_{i_p}A$. Symmetry of the shuffle product implies that each string of matrices $M_{i_1}  M_{i_2} \ldots M_{i_p}$ multiplies the same $f_i$ polynomials as its reverse $M_{i_p}  \ldots M_{i_2} M_{i_1}$. In particular, this assigns the symmetric coefficient $2 f_i \shuffle f_j = \phi( 2 \zeta^m_i \zeta^m_j)$ to matrix products $M_i M_j$ of length two, reflecting the absence of the first double zetas along with $\Qc_{(2)}$.

Let us present the momentum expansion of \gravsimple\ up to weight $14$:
\eqnn\gravexpa
$$\eqalignno{
\phi( {\cal M}^m ) & = A^t \ S_0 \ \big( \ 1 + 2\ f_3 M_3 + 2\ f_5 M_5 + 4\ f_3^2 M_3^2 + 2\ f_7 M_7 + 2\ f_3 \shuffle f_5 \ \{M_3,M_5\}  \cr
&+  2\ f_9 M_9 + 8 f_3^3 M_3^3  + 4\ f_5^2 M_5^2 + 2f_3 \shuffle f_7\  \{M_3, M_7\} + 2 f_{11} M_{11} \cr
& + f_3 \shuffle f_3 \shuffle f_5\ \{M_3,\{M_3,M_5\}\}+ 2\ f_5 f_3^2\ [M_3,[M_3, M_5]]  + 16 f_3^4 M_3^4 \cr
&+ 2\ f_3 \shuffle f_9\  \{M_3, M_9\}  + 2 f_5 \shuffle f_7\  \{M_5, M_7\} + 2 f_{13} M_{13} \cr
& + f_3 \shuffle f_3 \shuffle f_7\ \{M_3,\{M_3,M_7\}\}+ 2 f_7 f_3^2\ [M_3,[M_3, M_7 ]] \cr
& + f_5 \shuffle f_5 \shuffle f_3\ \{M_5,\{M_5,M_3\}\}+ 2 f_3 f_5^2\ [M_5,[ M_5 ,M_3]] \cr
& + 4\ f_7^2 M_7^2+ 2 f_{11} \shuffle f_3\ \{M_3, M_{11}\}+ 2 f_{9} \shuffle f_5\ \{M_5,M_9\}\cr
&+ {1 \over 3}\ f_3 \shuffle f_3 \shuffle f_3 \shuffle f_5\ \{M_3,\{M_3,\{M_3,M_5\}\}\} \cr
&+ 2\ f_{3} \shuffle (f_5 f_3^2)\ \{M_3, [M_3,[M_3,M_5]]\}+ \ldots  \ \big) \ A\ .
 &\gravexpa
}$$
Starting from weight eleven, some of the $f_i$ polynomials cannot be represented as a pure shuffle product $f_{i_1} \shuffle f_{i_2} \shuffle \ldots \shuffle f_{i_p}$ reflecting the presence of depth $\geq 2$ MZVs in \gravN\ due to $\tilde Q Q$. As expected from $Q_{11}, Q_{13}, \ldots$ given in \QQQ, they multiply nested $M_i$ commutators $\Qc_{(3)}, \Qc_{(5)},\ldots$ of odd depth, see e.g. $\ldots + 2 f_5 f_3^2 [M_3,[M_3, M_5]] + \ldots$ or the last line of \gravexpa.

To summarize, we have shown that also the closed string tree amplitude has an $\ap$ expansion whose beautiful motivic structure is revealed through the $\phi$ isomorphism.

\comment{Note, that the representation \gravmotiv\ for $\phi({\cal M}^m)$ has the shortcoming, that it hides 
the cancellation of the part $\sum_{n\geq 4}  Q^m_{2n}$ of $Q^m$ due to \vanishii. }

\comment{At even weights $\leq 12$, these commutation relations are required to show, that in \gravmotiv\ all the ordered products  $f_{i_1} f_{i_2} \ldots f_{i_p}$  can effectively be replaced by shuffle products. The latter can be identified as the image of single zetas under $\phi$:
\eqn\singlezeta{
f_{i_1} \shuffle f_{i_2}\shuffle \ldots \shuffle f_{i_p} \ = \ \phi \Big( \prod_{j=1}^p \zeta^m_{i_j} \Big) \ .
}
An alternative approach is to start from \gravN, where the absence of depth $\geq 2$ MZV at weights $w \leq 10$ and $w=12$ is already manifest\foot{When directly applying the $\phi$ map to \gravN, the leading terms of $\phi(Q_{odd})$ are given by
$$\eqalignno{\phi(Q_{11}^m+Q_{13}^m + Q_{15}^m+\ldots)=&f_5 f_3^2 [M_3,[M_3,M_5]] + f_7 f_3^2 [M_3,[M_3,M_7]] - f_5(f_5 f_3+f_3 f_5) [M_5,[M_5,M_3]] + f_9 f_3^2 [M_3,[M_3,M_9]] + f_7 (f_5 f_3 + f_3 f_5) [M_5,[M_3,M_7]] + f_5 (f_3 f_7 + f_7 f_3) [M_7,[M_3,M_5]] + \ldots}$$
}.
The resulting representation for $\phi( {\cal M}^m)$ appears to differ from \gravmotiv. However, the difference between the two expressions is composed of nested 
commutators of $M_{2k+1}$ of even depth, which ultimately drop out thanks to the relations \vanishii.
}

\newsec{Conclusion}

In this work  we have investigated the structure of the $\ap$--expansion of the open and closed
superstring amplitude at tree--level with particular emphasis on their transcendentality properties. 
The strict matching of powers $\ap^w$ with their associated MZV prefactors of weight $w$ 
constituting a well--confirmed pattern has been considerably refined. 

The main point is to replace the $\IC$ valued MZVs $\zeta$ by more abstract versions thereof, the 
so--called motivic MZVs $\zeta^m$, which are endowed by a Hopf algebra structure.
Furthermore, through the isomorphism $\phi$ the  motivic MZVs are mapped into an algebra comodule 
generated by the non--commutative words in generators $f_3,f_5,f_7,\ldots$ and an additional commutative element $f_2$. 
In the same way as the symbol conveniently captures patterns of field theory amplitudes the isomorphism $\phi$ 
yields a strikingly simple and compact expression \summ\ 
for the open superstring disk amplitude: the systematics of the $\ap$--dependence 
is written in closed and short form to all weights.
In contrast to the symbol, the map $\phi$ does not lose any information and can be inverted 
to recover the tree amplitude in terms of motivic MZVs.

In the closed superstring sector the properties of the matrix $P$ encoded in \vanishi\
and the commutation relations \newnewstuff\ between matrices $M_{2r+1}$ and $S_0$
result in the compact form \gravN, where
MZVs of even weight or depth $\geq 2$ only enter through \nochmal\ starting at weight $w=11$.
On the other hand, after applying the map $\phi$ this result turns into \gravmotiv, in which
the element $f_2$ is absent and all matrices $M_{2r+1}$ and Hopf--algebra generators $f_{2s+1}$ are treated democratically 
without the necessity for the ordering prescription \order\ in \gravN.

The polynomial structure of the matrices $M$ and $P$ and various other aspects of 
$\ap$--expansions are further elaborated in \BSS.

\vskip2cm
\goodbreak
\centerline{\noindent{\bf Acknowledgments}}

St.St. is indebted to Francis Brown for very helpful discussions and useful comments.
Moreover,  St.St. is very thankful to Francis Brown for urging him to go up to weight sixteen
in the $\ap$--expansion of the open superstring amplitude.

This work has  crucially benefited by the great deal of  
computer based achievements exhibited in \refs{\HuberYG,\Harmpol,\DataMine} 
and \refs{\MochZR,\MochUC}.
In particular, St.St. wishes to thank  Thomas~Hahn, 
Tobias Huber, Daniel Ma\^{\i}tre, Sven--Olaf Moch, and especially 
Jos Vermaseren for very helpful correspondence
and support. Moreover, St.St. is very thankful to Jos Vermaseren for streamlining 
and improving his FORM code to be more efficient for even higher weights than presented here.
A portion of the computations have been performed on the SGE cluster
of the Arnold~Sommerfeld~Center for Theoretical Physics at M\"unchen.

O.S. thanks James Drummond and Michael Green for giving motivation to revisit 
\vanishiii\ and to further explore the consequences of the alternating sign therein.
We wish to thank Johannes Br\"odel for useful discussions. 
O.S.  is grateful to the Werner--Heisenberg--Institut in M\"unchen for 
hospitality and financial support during early and late stages of this work.
St.St. thanks the Albert--Einstein--Institut in Potsdam  for hospitality and financial 
support during preparation and completion of this work.

\appendix{A}{Decomposition of motivic multi zeta values}

\subsec{Decomposition at weight $14$}

Gathering the information about the lower weight basis $\Uc_{k\leq 13}$ with \verenaxiv\ 
we can construct  the following basis for $\Uc_{14}$:
\eqnn\basisxiv
$$\eqalignno{
&-\fc{5}{6}\ f_5\ (f_3\shuffle f_3\shuffle f_3)-\fc{5}{3}\ f_5f_9+\fc{4}{7}\ f_5f_3f_2^3-51 f_7^2+30\ f_7f_5f_2-\fc{405}{2}\ f_9f_5\cr\crr
&+90\ f_9f_3f_2-15\ f_{11}f_3\ ,\cr
&-6\ f_5f_9-15\ f_7^2-28\ f_9f_5-44\ f_{11}\ f_3,\ -15\ f_7^2-69\ f_9f_5-165\ f_{11}\ f_3,\cr\crr 
&\lf(-\fc{5}{2}\ f_5(f_3\shuffle f_3)+\fc{4}{7}\ f_5f_2^3-\fc{6}{5}\ f_7f_2^2-45\ f_9f_2\ri)\shuffle f_3,\ -5\ (f_5f_3)\shuffle f_3\shuffle f_3,\cr\crr
&f_{11}\shuffle f_3,\ f_3\shuffle f_3\ \shuffle f_5\ \shuffle f_3,\ f_9\shuffle f_5,\ f_7\shuffle f_7,\ &\basisxiv\cr\crr
&\lf(\fc{1799}{18}\ f_9f_3-32\ f_7f_3f_2+\fc{1133}{16}\ f_7f_5+29\ f_5f_7-11\ f_5^2f_2-\fc{16}{5}\ f_5f_3f_2^2\ri.\cr\crr
&\lf.+\fc{1}{3}\ f_3 (f_3\shuffle f_3\shuffle f_3)-\fc{799}{72}\ f_3f_9+10\ f_3f_7f_2-\fc{1}{5}\ f_3f_5f_2^2-\fc{36}{35}\ f_3^2f_2^3\ri)f_2,\cr\crr
&\lf(-6\ f_5f_7-15\ f_7f_5-27\ f_9f_3\ri)f_2,\ f_9\shuffle f_3f_2,\ f_7\shuffle f_5f_2,\ f_3\shuffle f_3\shuffle f_3\shuffle f_3f_2,\cr\crr
&(-14f_7f_3-6f_5^2)f_2^2,\ -5\ f_5f_3f_2^3,\ f_5\shuffle f_5f_2^2,\ f_3\shuffle f_7f_2^2,\ f_3\shuffle f_5f_2^3,\ f_3\shuffle f_3f_2^4,\ f_2^7\ .}
$$
The operators $a_i$ of the decomposition \mdecoxiv\ are:
\eqnn\mcoeffsxiv
$$\eqalignno{
a_1&=\fc{1}{5}\ [\p_3,[\p_3,[\p_5,\p_3]]],\ a_2=-\fc{23}{198}\ [\p_{11},\p_3]+\fc{5}{18}\ [\p_9,\p_5]-\fc{12841}{1188}\ [\p_3,[\p_3,[\p_5,\p_3]]],\cr\crr
a_3&=-\fc{2}{27}\ [\p_9,\p_5]+\fc{1}{27}\ [\p_{11},\p_3]+\fc{232}{81}\  
[\p_3,[\p_3,[\p_5,\p_3]]],\cr\crr
a_4&=\fc{1}{5}\ [\p_3,[\p_5,\p_3]]\p_3,\ 
a_5=\fc{1}{10}\ [\p_5,\p_3]\p_3^2,\ a_6=\p_{11}\p_3,\ a_7=\fc{1}{6}\ \p_5\p_3^3,\cr\crr
a_8&=\p_9\p_5-\fc{23}{33}\ [\p_{11},\p_3]+\fc{5}{3}\ [\p_9,\p_5]-\fc{12775}{198}\ [\p_3,[\p_3,[\p_5,\p_3]]]\ ,\cr\crr 
a_9&=\h\ \p_7^2-\fc{235}{396}\ [\p_{11},\p_3]+\fc{55}{36}\ [\p_9,\p_5]-\fc{647287}{11880}\ [\p_3,[\p_3,[\p_5,\p_3]]]\cr\crr
a_{10}&=c_2a_0,\ a_{11}=c_2\lf(\fc{1}{27}\ [\p_9,\p_3]+\fc{2665}{648}\ a_0\ri)+\fc{2}{3}\ [\p_3,[\p_3,[\p_5,\p_3]]],\ \cr
a_{12}&=c_2\lf(\p_9\p_3+\fc{799}{72}\ a_0\ri)+9\ [\p_3,[\p_5,\p_3]]\p_3,\cr\crr
a_{13}&=c_2\lf(\p_7\p_5+\fc{2}{9}\ [\p_9,\p_3]-\fc{467}{108}\ a_0\ri)+4\ [\p_3,[\p_3,[\p_5,\p_3]]],\cr\crr
a_{14}&=c_2\ \lf(\fc{1}{24}\p_3^4-\fc{1}{12}\ a_0\ri),\ a_{15}=\fc{1}{14}\ c_2^2\ [\p_7,\p_3]-3\ c_2a_0,\cr\crr
a_{16}&=\fc{1}{5}c_2^3\ [\p_5,\p_3]-\fc{3}{5}\ c_2\ a_0+\fc{4}{175}\ [\p_3,[\p_3,[\p_5,\p_3]]],\cr\crr 
a_{17}&=\h\ c_2^2\lf(\p_5^2+\fc{3}{7}\ [\p_7,\p_3]\ri)-\fc{7}{2}\ c_2a_0,\cr\crr
a_{18}&=c_2^2\ \p_7\p_3-10\ c_2a_0+\fc{6}{25}\ [\p_3,[\p_5,\p_3]]\p_3,\cr\crr 
a_{19}&=c_2^3\ \p_5\p_3+\fc{1}{5}\ c_2a_0-\fc{4}{35}\ [\p_3,[\p_5,\p_3]]\p_3,\cr\crr
a_{20}&=\h\ c_2^4\p_3^2+\fc{18}{35}\ c_2a_0,\ a_{21}=c_2^7&\mcoeffsxiv}
$$
acting on $\phi(\xi_{14})$. Above we have introduced the operator:
\eqn\defop{
a_0=\fc{48}{691}\ \lf(\ [\p_9,\p_3]-3\ [\p_7,\p_5]\ \ri)\ .}
Furthermore, we have used some useful formulae exhibited in the following. 
Nested  commutators involving the derivatives $\partial_3$ and $\partial_5$ acting on various products of $f_3$ and $f_5$ have a ``diagonal'' structure:
\eqnn\partialdiag
$$\eqalignno{
[\partial_3,[\partial_3,[\partial_3,\partial_5]]] \, f_5f_3f_3f_3 &= 1\ , \cr
[\partial_3,[\partial_3,\partial_5]] \, \partial_3 \, (f_5f_3f_3) \shuffle f_3 & =  1\ , 
\cr
[\partial_3,\partial_5] \, \partial_3^2 \, (f_5f_3) \shuffle f_3 \shuffle f_3 & =  2\ , \cr
\partial_5 \, \partial_3^3 \, f_5 \shuffle f_3 \shuffle f_3 \shuffle f_3 &= 6\ .&\partialdiag}
$$
On the other hand, all the other combinations of differential operators 
$$[\partial_3,[\partial_3,[\partial_3,\partial_5]]],\ [\partial_3,[\partial_3,\partial_5]] \partial_3,\ [\partial_3,\partial_5] \partial_3^2, \ \partial_5 \, \partial_3^3 $$ 
acting on the products $\{ f_5f_3^3 , \ (f_5f_3f_3) \shuffle f_3, \  (f_5f_3) \shuffle f_3 \shuffle f_3,$
$f_5 \shuffle f_3 \shuffle f_3 \shuffle f_3 \}$ vanish. {\it E.g.}  $[\partial_3,[\partial_3,[\partial_3,\partial_5]]]$ annihilates all of $(f_5f_3f_3) \shuffle f_3, \  (f_5f_3) \shuffle f_3 \shuffle f_3, \ f_5 \shuffle f_3 \shuffle f_3 \shuffle f_3$. 
More generally, we have:
\eqn\padiag{
\underbrace{[\partial_3, [\partial_3, [ \ldots , [\partial_3,\partial_5] \ldots ]]] }_{(k-p) {\rm -fold \ commutator} }\ \partial_3^p \, (f_5 f_3^{k-q}) \, (\shuffle f_3)^q \ = \ p! \   \delta_{p,q} \ .}

\subsec{Decomposition at weight $15$}
\def\commm#1#2#3{[\p_{#1},[\p_{#2},\p_{#3}]]}

At weight $15$ 
we collect the information about the lower weight basis $\Uc_{k\leq 14}$ and with \verenaxv\ we can construct  the following basis for $\Uc_{15}$:
\eqnn\basisxv
$$\eqalignno{
&\phi(\zeta_{1,1,3,4,6}^m),\ \phi(\zeta^m_{3,3,9}),\ \phi(\zeta^m_{5,3,7}),\ f_{15},\ 
f_3\shuffle\phi(\MZ{1,1,4,6}),\ f_3\shuffle\phi(\MZ{3,9}),\cr 
&f_9\shuffle f_3\shuffle f_3,\ f_3\shuffle f_5\shuffle f_7,\ f_3\shuffle f_3\shuffle f_3\shuffle f_3\shuffle f_3,\cr 
&(-14f_7f_3-6f_5^2)\shuffle f_5,\ f_5\shuffle f_5\shuffle f_5,\ (-5f_5f_3)\shuffle f_7,\cr
&\phi(\MZ{3,3,7})f_2,\ \phi(\MZ{3,5,5})f_2,\ f_{13}f_2,\ (-14f_7f_3-6f_5^2)\shuffle f_3f_2,\ (-5f_5f_3)\shuffle f_5f_2,\cr 
&f_7\shuffle f_3\shuffle f_3f_2,\ f_5\shuffle f_5\shuffle f_3f_2,\cr
&\phi(\MZ{3,3,5})f_2^2,\ (-5f_5f_3)\shuffle f_3f_2^2,\ f_{11}f_2^2,\ f_5\shuffle f_3\shuffle f_3f_2^2,\ f_3\shuffle f_3\shuffle f_3f_2^3,\cr
&f_9f_2^3,\ f_7f_2^4,\ f_5f_2^5,\ f_3f_2^6\ ,&\basisxv}
$$
with $\phi(\MZ{3,3,5}),\phi(\MZ{3,9}), \phi(\MZ{1,1,4,6}),\phi(\MZ{3,3,7}),\phi(\MZ{3,5,5}),\phi(\zeta_{1,1,3,4,6}^m),\phi(\zeta^m_{3,3,9})$ and $\phi(\zeta^m_{5,3,7})$ 
given in \verenaxi, \verenaxii, \verenaxiii\ and \verenaxv, respectively.
The operators $a_i$ of the decomposition \mdecoxv\ are:
\eqnn\mcoeffsxv
$$\eqalignno{
a_1&=\fc{48}{7601}\ \lf(\commm{3}{9}{3}-3\ \commm{3}{7}{5}\ri),\ a_2=\fc{1}{27}\ \commm{3}{9}{3}-
\fc{853}{648}\ a_1,\cr\crr
a_3&=\fc{2}{15}\ \commm{3}{7}{5}-\fc{1}{70}\ \commm{5}{7}{3}+\fc{17203}{3360}\ a_1,\ a_4=\p_{15},\cr\crr
a_5&=a_1+a_0\ \p_3,\ a_6=\fc{1}{27}\ [\p_9,\p_3]\p_3+\fc{2665}{648}\ a_0\p_3+\fc{29}{9}\ a_1\ ,\cr\crr
a_7&=\h\ \p_9\p_3^2+\fc{799}{72}\ a_0\p_3+\fc{6775}{144}\ a_1,\cr\crr 
a_8&=\p_7\p_5\p_3+\fc{2}{9}\ [\p_9,\p_3]\p_3-\fc{467}{108}\ a_0\p_3-\fc{74}{3}\ a_1,\cr\crr
a_9&=\fc{1}{5!}\ \p_3^5-\fc{1}{12}\ a_0\p_3-\fc{1}{15}\ a_1,\cr\crr
a_{10}&=\fc{1}{14}\ [\p_7,\p_3]\p_5+\fc{2188}{945}\ a_1+\fc{3}{35}\ \commm{5}{7}{3}-\fc{2}{45}\ \commm{3}{9}{3},\cr\crr
a_{11}&=\fc{1}{6}\ \p_5^3+\fc{3}{14}\ [\p_7,\p_3]\p_5+\fc{1185701}{30240}\ a_1+\fc{11}{70}\ \commm{5}{7}{3}
-\fc{2}{45}\ \commm{3}{9}{3}\cr\crr
a_{12}&=\fc{1}{5}\ [\p_5,\p_3]\p_7-\fc{12199}{720}\ a_1+\fc{1}{5}\ \commm{5}{7}{3}
-\fc{1}{15}\ \commm{3}{9}{3}\cr\crr
a_{13}&=\fc{1}{14}\ c_2\ \commm{3}{7}{3}+2\ a_1,\cr\crr
a_{14}&=-\fc{3}{35}\ c_2\ \commm{3}{7}{3}+\fc{1}{25}\ c_2\ \commm{5}{5}{3}-\fc{14}{5}\ a_1,\cr\crr
a_{15}&=c_2\ \p_{13}-\fc{6417649}{2880}\ a_1-\fc{143}{20}\ \commm{5}{7}{3}+
\fc{1339}{30}\ \commm{3}{9}{3},\cr\crr
a_{16}&=\fc{1}{14}\ c_2\ [\p_7,\p_3]\p_3-3\ a_0\p_3-6\ a_1,\cr\crr
a_{17}&=\fc{1}{5}\ c_2\ [\p_5,\p_3]\p_5+\fc{1}{5}\ c_2\ \commm{5}{5}{3}+\fc{21}{2}\ a_1,\cr\crr
a_{18}&=\h\ c_2\ \p_7\p_3^2-10\ a_0\p_3-26\ a_1,\cr\crr
a_{19}&=\h\ c_2\ \ \p_5^2\p_3+\fc{3}{14}\ c_2\ [\p_7,\p_3]\p_3-\fc{7}{2}\ a_0\p_3-8\ a_1,\cr\crr
a_{20}&=\fc{1}{5}\ c_2^2\ \commm{3}{5}{3}+4\ a_1,\ 
a_{21}=\fc{1}{5}\ c_2^2\ [\p_5,\p_3]\p_3-\fc{3}{5}\ a_0\p_3-\fc{8}{5}\ a_1,\cr\crr
a_{22}&=c_2^2\ \p_{11}+\fc{11}{4}\ c_2\ \commm{3}{7}{3}+\fc{11}{2}\ c_2\ \commm{5}{5}{3}-
\fc{8495287}{15120}\ a_1\cr\crr
&-\fc{11}{35}\ \commm{5}{7}{3}+\fc{128}{45}\ \commm{3}{9}{3},\ a_{23}=\h\ c_2^2\ \p_5\p_3^2+\fc{1}{5}\ a_0\p_3-\fc{23}{10}\ a_1,\cr\crr
a_{24}&=\fc{1}{6}\ c_2^3\ \p_3^3+\fc{18}{35}\ a_0\p_3+\fc{12}{35}\ a_1,\cr\crr 
a_{25}&=c_2^3\ \p_9+9\ c_2^2\ \commm{3}{5}{3}-\fc{2}{35}\ c_2\ \commm{3}{7}{3}+
\fc{2}{5}\ c_2\ \commm{5}{5}{3}\cr\crr
&+\fc{54263011}{396900}\ a_1+\fc{68}{1225}\ \commm{5}{7}{3}-\fc{236}{4725}\ \commm{3}{9}{3},\cr\crr
a_{26}&=c_2^4\ \p_7+\fc{6}{25}\ c_2^2\ \commm{3}{5}{3}-\fc{16}{245}\ c_2\ \commm{3}{7}{3}+
\fc{57847}{15750}\ a_1\cr\crr
&+\fc{24}{875}\ \commm{5}{7}{3}-\fc{184}{2625}\ \commm{3}{9}{3},\cr\crr
a_{27}&=c_2^5\ \p_5-\fc{4}{35}\ c_2^2\ \commm{3}{5}{3}-\fc{1714624}{121275}\ a_1
+\fc{48}{13475}\ \commm{5}{7}{3},\cr\crr 
&-\fc{64}{5775}\ \commm{3}{9}{3},\ a_{28}=c_2^6\ \p_3+\fc{1451972}{716625}\ a_1&\mcoeffsxv}
$$
acting on $\phi(\xi_{15})$. Above we have used the operator $a_0$ defined in \defop.

\subsec{Decomposition at weight $16$}

\def\opi{[\p_3,[\p_5,[\p_5,\p_3]]]}
\def\opii{[\p_3,[\p_3,[\p_7,\p_3]]]}
\def\opa{[\p_3,[\p_3,[\p_5,\p_3]]]}

Gathering the information about the lower weight basis $\Uc_{k\leq 15}$ with \verenaxv\ 
we can construct  the following basis for $\Uc_{16}$:
\eqnn\basisxvi
$$\eqalignno{
&\phi(\MZ{1,1,6,8}),\ \phi(\MZ{3,3,3,7}),\ \phi(\MZ{3,3,5,5}),\ \phi(\MZ{3,13}),\ 
\phi(\MZ{5,11})\cr
&f_3\shuffle\phi(\MZ{3,3,7}),\ f_3\shuffle\phi(\MZ{3,5,5}),\ f_3\shuffle f_{13},\ 
(-14f_7f_3-6f_5^2)\shuffle f_3\shuffle f_3,\cr
& (-5f_5f_3)\shuffle f_3\shuffle f_5,\   f_3\shuffle f_3\shuffle f_3\shuffle f_7,\ f_3\shuffle f_3\shuffle  f_5\shuffle f_5,\ f_7\shuffle f_9,\cr
&25\ (f_5f_3)\shuffle(f_5f_3),\ f_5\shuffle f_{11},\ f_5\shuffle \phi(\MZ{3,3,5}),\cr
&f_3\shuffle \phi(\MZ{3,3,5})f_2,\ f_3\shuffle f_3\shuffle(-5f_5f_3)f_2,\ f_3\shuffle f_{11}f_2,\ f_3\shuffle f_3\shuffle f_3\shuffle f_5f_2\cr
&f_3\shuffle f_3\shuffle f_3\shuffle f_3f_2^2,\ f_3\shuffle f_9f_2^2,\ f_3\shuffle f_7f_2^3,\ f_3\shuffle f_5f_2^4,\ f_3\shuffle f_3f_2^5,\cr
&\phi(\MZ{3,3,3,5})f_2,\ \phi(\MZ{3,11})f_2,\  \phi(\MZ{5,9})f_2,\ f_5\shuffle f_9f_2,\ 
f_7\shuffle f_7f_2\ ,\cr
&\phi(\MZ{1,1,4,6})f_2^2,\ \phi(\MZ{3,9})f_2^2,\ f_5\shuffle f_7f_2^2,\ (-14f_7f_3-6f_5^2)f_2^3,-5f_5f_3f_2^4,\cr 
&f_5\shuffle f_5f_2^3,\ f_2^8\ ,&\basisxvi}
$$
with the maps $\phi(\MZ{3,3,5}),\ \phi(\MZ{3,9}), \phi(\MZ{1,1,4,6}),\ \phi(\MZ{3,3,7}),\phi(\MZ{3,5,5}),\ \phi(\MZ{3,3,3,5}),\phi(\MZ{3,11}),\phi(\MZ{5,9})$
$\phi(\MZ{3,3,3,7}),\phi(\MZ{3,3,5,5}),\phi(\MZ{3,13}),\phi(\MZ{5,11})$ and $\phi(\MZ{1,1,6,8})$ 
given in \verenaxi, \verenaxii, \verenaxiii, \verenaxiv\  and \verenaxvi, respectively.
The operators $a_i$ of the decomposition \mdecoxvi\ are:
\eqnn\mcoeffsxvi
$$\eqalignno{
a_1&=\fc{720}{3617}   \lf\{\ \fc{7}{11}\ [\p_{11},\p_5]-\fc{2}{11}\ [\p_{13},\p_3]-[\p_9,\p_7]+\fc{6493}{9240}\ \opii\ri.\cr\crr
&\lf.-\fc{751}{100}\ \opi\ \ri\},\ a_2=-\fc{19}{7}\ a_1+\fc{1}{14}\ \opii,\cr \crr
a_3&=\fc{542}{175}\ a_1-\fc{3}{35}\ \opii+\fc{1}{25}\ \opi,\cr\crr
a_4&=-\fc{19}{286}\ [\p_{13},\p_3]+\fc{3}{22}\ [\p_{11},\p_5]+\fc{2217053}{16800}\ a_1
-\fc{200559}{80080}\ \opii\cr\crr
&-\fc{7011}{2600}\ \opi\cr\crr
a_5&=\fc{3}{242}\ [\p_{13},\p_3]-\fc{5}{242}\ [\p_{11},\p_5]-\fc{114307}{7392}\ a_1
+\fc{23181}{67760}\ \opii\cr\crr
&+\fc{909}{2200}\ \opi,\ a_6=-\fc{1}{14}\ \commm{3}{3}{7}\p_3+\fc{5}{7}\ a_1,\cr\crr
a_7&=-\fc{1}{25}\ \commm{5}{3}{5}\p_3+\fc{3}{35}\ \commm{3}{3}{7}\p_3-\fc{6}{7}\ a_1,\ 
a_8=\p_{13}\p_3+\fc{8497}{42}\ a_1,\cr\crr
a_9&=\fc{1}{28}\ [\p_7,\p_3]\p_3^2+\fc{1}{7}\ a_1,\ a_{10}=\fc{1}{5}\ [\p_5,\p_3]\p_5\p_3+\fc{1}{5}\ \commm{5}{5}{3}\p_3,\cr\crr
a_{11}&=\fc{1}{3!}\ \p_7\p_3^3-\fc{1}{3}\ a_1,\ a_{12}=\h\ \lf(\h\ \p_5^2+\fc{3}{14}\ [\p_7,\p_3]\ri)\p_3^2-\fc{4}{7}\ a_1,\cr\crr
a_{13}&=\p_9\p_7+\fc{4850713}{6600}\ a_1-\fc{2272973}{330330}\ \opii-\fc{299373}{7150}\ \opi\cr\crr
&-\fc{1275}{1573}\ [\p_{13},\p_3]+\fc{210}{121}\ [\p_{11},\p_5],\ a_{14}=\fc{1}{50}\ [\p_5,\p_3]^2,\cr\crr
a_{15}&=\p_{11}\p_5+\fc{455534}{525}\ a_1-\fc{601677}{40040}\ \opii-\fc{21033}{1300}\ \opi\cr\crr
&-\fc{57}{143}\ [\p_{13},\p_3]+\fc{9}{11}\ [\p_{11},\p_5],\cr\crr
a_{16}&=\fc{1}{5}\ \commm{3}{5}{3}\p_5+\fc{1}{5}\ \opi-\fc{2}{5}\ a_1,\cr\crr
a_{17}&=\fc{1}{5}\ c_2\ \commm{3}{5}{3}\p_3,\ a_{18}=\fc{1}{10}\ c_2\ [\p_5,\p_3]\p_3^2,\cr\crr
a_{19}&=c_2\ \p_{11}\p_3-\fc{11}{4}\ \commm{3}{3}{7}\p_3-\fc{11}{2}\ \commm{5}{3}{5}\p_3-
137\ a_1,\cr\crr
a_{20}&=\fc{1}{3!}\ c_2\ \p_5\p_3^3,\ a_{21}=\fc{1}{4!}\ c_2^2\ \p_3^4-\fc{1}{12}\ c_2^2\ a_0,\cr\crr
a_{22}&=c_2^2\ \p_9\p_3+9\ c_2\ \commm{3}{5}{3}\p_3+\fc{799}{72}\ c_2^2\ a_0-\fc{2}{35}\ \commm{3}{7}{3}\p_3\cr\crr
&+\fc{2}{5}\ \commm{5}{5}{3}\p_3-\fc{11}{7}\ a_1,\cr\crr
a_{23}&=c_2^3\ \p_7\p_3+\fc{6}{25}\ c_2\ \commm{3}{5}{3}\p_3-10\ c_2^2\ a_0-\fc{16}{245}\ \commm{3}{7}{3}\p_3+\fc{848}{245}\ a_1,\cr\crr
a_{24}&=c_2^4\ \p_5\p_3-\fc{4}{35}\ c_2\ \commm{3}{5}{3}\p_3+\fc{1}{5}\ c_2^2\ a_0+\fc{48}{35}\ a_1,\cr\crr
a_{25}&=\h\ c_2^5\ \p_3^2+\fc{18}{35}\ c_2^2\ a_0+\fc{408}{2695}\ a_1,\ 
a_{26}=\fc{1}{5}\ c_2\ \opa,\cr\crr
a_{27}&=-\fc{23}{198}\ c_2\ [\p_{11},\p_3]+\fc{5}{18}\ c_2\ [\p_{9},\p_5]-\fc{12841}{1188}\ c_2\ \opa-\fc{1991}{14}\ a_1\cr\crr
&+\fc{121}{28}\ \opii-\fc{7}{2}\ \opi,\cr\crr
a_{28}&=\fc{1}{27}\ c_2\ [\p_{11},\p_3]-\fc{2}{27}\ c_2\ [\p_{9},\p_5]+\fc{232}{81}\ c_2\ \opa+\fc{697}{21}\ a_1\cr\crr
&-\fc{47}{42}\ \opii+\ \opi,\cr\crr
a_{29}&=c_2\ \p_9\p_5+9\ \commm{3}{5}{3}\p_5-\fc{23}{33}\ c_2\ [\p_{11},\p_3]
+\fc{5}{3}\ c_2\ [\p_9,\p_5]-\fc{12443}{14}\ a_1\cr\crr
&-\fc{12775}{198}\ c_2\ \opa+\fc{363}{14}\ \opii-21\ \opi,\cr\crr
a_{30}&=\h\ c_2\ \p_7^2-\fc{235}{396}\ c_2\ [\p_{11},\p_3]
+\fc{55}{36}\ c_2\ [\p_9,\p_5]-\fc{647287}{11880}\ c_2\ \opa\cr\crr
&-\fc{78201}{140}\ a_1+\fc{967}{56}\ \opii-\fc{333}{20}\ \opi,\ a_{31}=c_2^2\ a_0,\cr\crr
a_{32}&=\fc{1}{27}\ c_2^2\ [\p_9,\p_3]+\fc{2665}{648}\ c_2^2\ a_0+\fc{2}{3}\ c_2\ \opa-\fc{8954}{1575}\ a_1\cr\crr
&+\fc{4}{35}\ \opii-\fc{4}{75}\ \opi,\cr\crr
a_{33}&=c_2^2\ \p_7\p_5+\fc{2}{9}\ c_2^2\ [\p_9,\p_3]+\fc{6}{25}\ \commm{3}{5}{3}\p_5-\fc{467}{108}\ c_2^2\ a_0+4\ c_2\ \opa\cr\crr
&-\fc{21331}{525}\ a_1+\fc{24}{35}\ \opii-\fc{8}{25}\ \opi,\cr\crr
a_{34}&=\fc{1}{14}\ c_2^3\ [\p_7,\p_3]-3\ c_2^2\ a_0-\fc{62}{245}\ a_1+\fc{2}{245}\ \opii,\cr\crr
a_{35}&=\fc{1}{5}\ c_2^4\ [\p_5,\p_3]-\fc{3}{5}\ c_2^2\ a_0+\fc{4}{175}\ c_2\ \opa+\fc{108}{875}\ a_1,\cr\crr
a_{36}&=c_2^3\ \lf(\h\ \p_5^2+\fc{3}{14}\ [\p_7,\p_3]\ri)-\fc{4}{35}\ \commm{3}{5}{3}\p_5-\fc{7}{2}\ c_2^2\ a_0\cr\crr
&-\fc{284}{245}\ a_1+\fc{6}{245}\ \opii-\fc{2}{35}\ \opi,\ a_{37}=c_2^8\ ,&\mcoeffsxvi}
$$
acting on $\phi(\xi_{16})$. Again, we have used the operator $a_0$ defined in \defop.

\listrefs

\end